\documentclass[preprint2]{emulateapj}

\usepackage{subfigure}

\def\p{\partial}
\def\nab{\mbox{\boldmath $\nabla$}}

\def\rb{\bar{\rho}}

\def\tb{\bar{T}}
\def\sb{\bar{S}}

\newcommand{\DD}{\mbox{\boldmath ${\cal D}$}}

\newcommand{\degree}{\ensuremath{^\circ}}

\shorttitle{Hunting for giant cells in deep stellar convective zones}

\shortauthors{Bessolaz \& Brun}

\begin{document}


\title{Hunting for giant cells in deep stellar convective zones using
wavelet analysis}


\author{Nicolas BESSOLAZ and Allan Sacha BRUN}
\affil{Laboratoire AIM Paris-Saclay, CEA/Irfu Universit\'e Paris-Diderot \\
        CNRS/INSU, 91191 Gif-sur-Yvette Cedex, France}



\begin{abstract}

We study the influence of stratification on stellar turbulent convection near
the stellar surface and in depth by carrying out 3D high resolution hydrodynamic
simulations with the ASH code. Four simulations with different radial density contrast 
corresponding to different aspect ratio for the same underlying 4 Myr 
$0.7 M_\odot$ pre-main sequence star model are thus performed. 
We highlight the existence of giant cells which are embedded in the complex 
surface convective patterns using a wavelet and time correlation analysis. Next, we study their
properties such as their lifetime, aspect ratio and spatial extension in the different 
models both in latitude and depth according to the density contrast. 
We find that these giants cells have a lifetime larger than the stellar period
with a typical longitudinal width of 490 Mm and a latitudinal extension
increasing with the radial density contrast overpassing 50\degree of latitude
 for the thickest convective zone. Their rotation rate is much larger than the
 local differential rotation rate increasing also with the radial density
 contrast. However, their spatial coherence as a function of depth
 decrease with the density contrast due to the stronger shear present in these
 more stratified cases.

\end{abstract}





\section{Introduction}

Most low-mass stars have a deep convective envelope. Understanding its
dynamical properties is of fundamental interest to explain the
evolution and redistribution of heat, energy and angular momentum transport in
stars. It is also a key step towards a global comprehension of the generation 
and organisation of magnetic fields within these stars. 
Little is known about the spectral energy distribution as a function of depth
within stellar convective zones. For instance, is there a clear separation 
between different spatial scales in convective patterns or can they genuily 
manifest themselves at the stellar surface ? 
Unfortunately, the Sun is the only star
 where we can observe with enough spatial resolution the multi-scales 
 properties of surface convection going from granulation with 1-2 Mm diameter
 cells and typical lifetime of 5 minutes to possible giant cells of few 100 Mm
 with lifetime of several weeks passing through supergranulation 
 cells of 20-50 Mm diameter with lifetime of the order of a day 
 (Norlund, Stein \& Asplund 2009, Rieutord \& Rincon 2010).
 
 The origin of such a large range of spatio-temporal
 scales and the physical processes behind these distinctive scales including
 possibly also mesogranulation has been debated for decades. Simon \& Weiss
 (1968) have proposed the location of H, He and metals recombination layers to explain the 
 granules, supergranules and giant cells existence. Rast (2003) has proposed an
 advection/fragmentation process in the upper layers of the convective zone to 
 explain the mesogranulation and supergranulation from an initial random distribution of
 downflow plumes. However, a turbulent origin 
 for these convective patterns is much convincing nowadays. Whereas granulation
and supergranulation have clear evidence of existence via respectively G-band 
or white light intensity observations and time averaged dopplergrams over 30 
minutes or magnetograms, the intermediate mesogranulation is still controversial.
Rieutord et al. (2000) stresses the spatial and time filtering biases in the
correlation tracking techniques (see Meunier et al. 2006 for a
detailed discussion). As far as the giant cells are concerned, they initially
have been predicted by Simon \& Weiss (1968) as an efficient way of 
transporting heat over several density scale height thanks to the lower 
superadiabatic temperature gradient necessary for convection over such larger
 scales. But they are very difficult to measure since they are on the one hand 
 merged within stronger signals like granulation or supergranulation which have
 to be removed correctly and on the other hand, looking for such a large scale
 signal imply to substract both the global differential rotation of the Sun and the limb effect and to disentangle possible line shifts due to the solar magnetic 
 field in active regions.  Evidence for an underlying organisation of the convective flows have been 
advocated by Lisle et al. (2004), who found a north-south alignment of 
supergranulation patterns possibly due to the presence of giant cells. Spectroscopic studies (LaBonte et al.
 1981, Scherrer et al. 1986) found cells with a longitudinal extension around
 45\degree and a typical velocity lower than 10 m/s but the use of magnetically 
 sensitive FeI line could tune down these conclusions. 
 Giant cells might also have been detected indirectly in velocity 
spectra from SOHO/MDI time averaged data by Hathaway et al. (2000) with a 
significant power at low spherical harmonics degrees ($l<30$) and a proper rotation rate similar to the
Sun. Another approach consists at developping full sphere simulations
resolving only the largest scales of turbulent convection down to
supergranulation with mean flows. They can be used as a tool to probe 
these giant cells and useful to develop specific proxies to help their 
observational detection. Much local cartesian simulations including a
higher turbulent degree and also radiation processes are able to probe the
properties of granulation (Stein \& Nordlund 1998, Vogler 2005). 
From the simulations and
the observational data, there is a recent trend to bring together 
the different scales of convection contrary to the previous paradigm which
identified specific discrete scales for granulation, mesogranulation and
supergranulation as different features of convection. This is well highlighted
by looking at the velocity spectrum $\sqrt{ k |FFT(V)|^2}$
increasing linearly with the wave number k over this broad range of spatial 
scales presented in the Nordlung, Stein \& Asplung (2009) review on Figure 22.

In the perspective of future work dedicated to low mass young stars presenting
much deeper convective zones than the Sun and thus possibly more extended 
giant cells, we try to quantify their own identity in our simulations, if 
their properties depend on the size of the convective zone and how 
they might modify the dynamics of convection by advecting smaller scale motions.
Actually, as the local density scale height diminishes much less towards 
surface
in low mass pre-main sequence stars than within the Sun, we can expect lower 
horizontal expansion rate of giant cells near the surface. This could imply a 
potential larger lifetime due to their weaker distortions by the mean flows. Indeed, these giant cells should also be important in the global
stellar activity cycle, particularly to transport magnetic flux from the bottom of the convective zone up to the stellar surface. Finally, it is also important to notice that
the equatorial modes associated to the linear growth of the convection
instability and discussed by Busse \& Cuong (1977) and Gilman (1975) could 
be candidates for these giant cells if they survive in the non linear regime
when turbulence is fully developed.   

In this paper, we present a set of high resolution 3D simulations of a subpart
 of the whole convective zone for a low-mass young star assuming various 
radial density contrasts.
The initial 3D  stellar models implemented in our simulations are presented in 
Section 2 with the numerical methods and boundary conditions used to compute 
them. Next, the wavelet analysis pipeline developed specifically for this study 
is detailed in Section 3 with the main methods used to probe the existence of stellar giant 
cells and their characteristics as the main goal of this paper. Then, we 
present our results showing the properties of convection in the different
models in Section 4, hunt for giant cells and their global
 properties in Section 5 and conclude in Section 6.     

\section{The 3D stellar models setup}


Four different models with an e-folding radial density contrast from 13 to 272  
are computed. The different main characteristics of the four models computed
are presented in Table \ref{sim_prop} and discussed in Section 4. Typical run
for these simulations lasts 3000 days, i.e. several convective overturning time.
The outer boundary condition is common to all models and is
located at 0.98$R_*$ below the stellar surface to keep the anelastic treatment
of the ASH code correct. The inner boundary location for each model is
determined to have the chosen density contrast. 
We restrict here to very slow rotators equal to the mean solar rotation rate.
However, common low mass YSO such as CTTS have mean stellar angular velocity as high as 5
$\Omega_\odot$ and have been already studied by Ballot et al (2007). Other
spectral type of YSO have been studied by Brown et al. (2008, 2010).

\subsection{Equations and boundary conditions}

The simulations described here were performed with the Anelastic Spherical
Harmonic (ASH) code. ASH solves the three-dimensional anelastic equations
of motion in a rotating spherical geometry using a pseudospectral
semi-implicit approach \citep[e.g.,][]{Clune99,BMT04}.  
These equations are fully nonlinear in velocity variables and linearized in
thermodynamic variables with respect to a spherically symmetric mean state.
This mean state is taken to have density $\bar{\rho}$, pressure $\bar{P}$,
temperature $\bar{T}$, specific entropy $\bar{S}$; perturbations about this
mean state are written as $\rho$, $P$, $T$, and $S$.  Conservation of mass,
momentum, and energy in the rotating reference frame are therefore written as
\begin{equation}
\nab\cdot(\rb {\bf v}) = 0,
\end{equation}
\vspace{-0.5cm}
\begin{eqnarray}\label{eq:momentum}
\rb \left(\frac{\p {\bf v}}{\p t}+({\bf v}\cdot\nab){\bf v}+2{\bf
\Omega_o}\times{\bf v}\right)
 &=& -\nab P \nonumber + \rho {\bf g} \nonumber \\ 
 &-& [\nab\bar{P}-\rb{\bf g}] \nonumber \\
 &-& \nab\cdot\mbox{\boldmath $\cal D$},
\end{eqnarray}
\begin{eqnarray}\label{eq:energy}
\rb \tb \frac{\p S}{\p t}&=&\nab\cdot[\kappa_r \rb c_p \nab
(\tb+T)+\kappa \rb \tb \nab (\sb+S)] \nonumber \\ &-&\rb \tb{\bf v}\cdot\nab
(\sb+S) \nonumber \\ &+& 2\rb\nu\left[e_{ij}e_{ij}-1/3(\nab\cdot{\bf v})^2\right]+\rb\epsilon
\end{eqnarray}
where $c_p$ is the specific heat at constant pressure, ${\bf
v}=(v_r,v_{\theta},v_{\phi})$ is the local velocity in spherical
geometry in the rotating frame of constant angular velocity ${\bf
\Omega_o} = \Omega_0 {\bf \hat{e}_z}$, ${\bf g}$ is the gravitational acceleration, $\kappa_r$ is the
radiative diffusivity,$\epsilon$ is the heating rate per unit mass due to gravitational contraction, 
and $\DD$ is the viscous stress tensor, with components
\begin{eqnarray}\label{eq:viscoustensor}
{\cal D}_{ij}=-2\rb\nu[e_{ij}-1/3(\nab\cdot{\bf v})\delta_{ij}],
\end{eqnarray}
where $e_{ij}$ is the strain rate tensor.  Here $\nu$ and $\kappa$ are
effective
eddy diffusivities for vorticity and entropy.  To close the set of equations,
linearized relations for the thermodynamic fluctuations are taken as
\begin{equation}\label{eos}
\frac{\rho}{\rb}=\frac{P}{\bar{P}}-\frac{T}{\tb}=\frac{P}{\gamma\bar{P}}
-\frac{S}{c_p},
\end{equation}
assuming the ideal gas law
\begin{equation}\label{eqn: gp}
\bar{P}={\cal R} \rb \tb ,
\end{equation}

\noindent where ${\cal R}$ is the gas constant.  The effects of
compressibility on the convection are taken into account by means of the
anelastic approximation, which filters out sound waves that would otherwise
severely limit the time steps allowed by the simulation.

For boundary conditions at the top and bottom of the domain, we impose:
\begin{enumerate}
\item impenetrable walls : 
\[
v_r=0|_{r=r_{\rm bot},r_{\rm top}},
\]
\item stress free conditions: 
\[
\frac{\p}{\p r}\left(\frac{v_{\theta}}{r}\right)=\frac{\p}{\p r}\left(\frac{v_{\phi}}{r}\right)=0|_{r=r_{\rm bot},r_{\rm top}},
\]
\item and constant entropy gradient : 
\[
\frac{\p \sb}{\p r}=cst|_{r=r_{\rm bot},r_{\rm top}} .
\]
\end{enumerate}

Convection in stellar environments occurs over a large range of scales.
Numerical simulations cannot, with present computing technology, consider
all these scales simultaneously.  We therefore seek to resolve the largest
scales of the nonlinear flows since we are interested in the global transport
 of heat, energy and angular momentum in
convective envelopes and particularly to the existence of giant cells.  
We do so within a large-eddy simulation (LES)
formulation, which explicitly follows larger scale flows while employing
subgrid-scale (SGS) descriptions for the effects of the unresolved
motions.  Here, those unresolved motions are treated as enhancements to the
viscosity and thermal diffusivity ($\nu$ and $\kappa$), which are thus
effective eddy viscosities and diffusivities.  For simplicity, we have
taken these to be functions of radius alone, and to scale as the inverse of
the square root of the mean density. However, the deepest model $\Delta 
\rho =272$ only depends on the inverse of the
cubic root of the mean density to avoid to increase too much the horizontal
resolution in order to fully resolve the convective patterns near the 
base of the convective zone. We emphasize that currently tractable 
simulations are still many orders of magnitude away in parameter space from the highly
turbulent conditions likely to be found in real stellar convection zones.
These large-eddy simulations should therefore be viewed only as indicators of the properties of the real flows.

\subsection{ASH numerical methods}

To solve numerically with ASH the anelastic equations of motion we use 
a pseudo spectral method (Glatzmaier 1984, Canuto 1995, Fornberg 1999). 
The velocity and thermodynamic variables within ASH are expanded in spherical harmonics
$Y^m_{\ell}(\theta,\phi)$ in the horizontal directions and in Chebyshev
polynomials $T_n (r)$ in the radial direction. Spatial resolution is thus uniform
everywhere on a sphere when a complete set of spherical harmonics of degree
$\ell$ is used, retaining all azimuthal orders $m$.  We truncate our
expansion at degree $\ell=\ell_{max}$, which is related to the number of
latitudinal mesh points $N_{\theta}$ (here $\ell_{max}=(2N_{\theta}-1)/3$),
take $N_{\phi}=2 N_{\theta}$ azimuthal mesh points, and utilize $N_r$
collocation points for the projection onto the Chebyshev polynomials.
A semi-implicit, second-order Crank-Nicolson
scheme is used in determining the time evolution of the linear terms,
whereas an explicit second-order Adams-Bashforth scheme is employed for the
advective and Coriolis terms.  The ASH code has been optimized to run
efficiently on massively parallel supercomputers such as the IBM SP-6, SGI Altix
or IBM Blue Gene/P, and has demonstrated excellent scalability on such machines.




\subsection{Initial conditions deduced from a 1D stellar structure model
\label{model}}

\begin{figure}[!h]
\centering
   \includegraphics[width=5.8cm,angle=90]{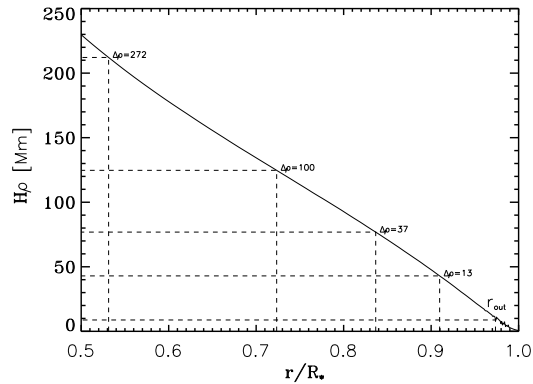}
      \caption{Local density scale height of our $0.7M_{\odot}$ YSO model at 4 Myr 
      old. The different inner boundaries for the convection zone of our set 
      of simulations correspond to
      the vertical dashed lines. The outer boundary for all models corresponds
      to $H_{\rho}$=10 Mm.}
       \label{Hrho}
\end{figure}

\begin{figure}[!h]
\centering
   \includegraphics[width=5.8cm,angle=90]{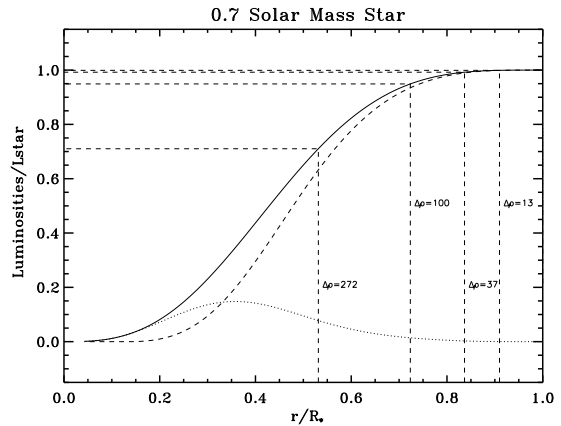}
      \caption{Total (solid curve), convective(dashed curve) and radiative
      (doted line) luminosities from the
      CESAM calculations of our $0.7M_{\odot}$ YSO model at 4 Myr old. These
      luminosities are normalized with respect to the stellar luminosity. The different inner
      boundaries for the convection zone of our set of simulations correspond to
      the vertical dashed lines.}
       \label{lum}
\end{figure}

The CESAM code (Morel 1997) is used to calculate the internal stellar structure of a low mass
$0.7M_{\odot}$ PMS young star. Such stars are fully convective until they reach
$3.76$ Myr. We choose the initial condition for our models just after the
appearance of the radiative core at 4 Myr. At this time, the stellar radius is
$1.224 R_\odot$, the base of the convection zone is at $0.18 R_\odot$ and the stellar luminosity
is $0.4 L_\odot$. Moreover, the mass stratification within the star is really
different from the Sun since 87 \% (vs 2.8 \%) of the mass is localized within the convective zone and
the central density is only 3.29 $g.cm^{-3}$ (vs 160 $g.cm^{-3}$). This is also stressed by the small variation of the local density scale height as function of depth with very different domain 
aspect ratio for the different density contrast considered (see Figure \ref{Hrho}). 
The radiative transport is
really weak and reaches a maximum of $0.148 L_*$ at $r=0.35 R_*$ (see Figure
\ref{lum}). $L_{rad}$ is equal to
$0.079 L_*$ at the inner boundary $r=0.53 R_*$ for the stronger density 
contrast case and thus is nearly negligible for all models.     

The main source of energy in these young stars is the energy release from the
gravitational contraction. In CESAM, this source term is written as $\epsilon=cT$ where
T is the local temperature within the star and c is the constant of contraction characterizing the
quasi-static state of the star (see Iben 1965 for more details). A typical value
for c is c=0.02 which gives an initial central temperature for the stellar core 
of $10^5$K. After an evolution of 1 Myr on the PMS phase, the initial guess
 for c is not so important since the relative difference on the surface stellar
luminosity and the position of the base of the convection zone is only 2\% for an 
initial value more than two order of magnitudes smaller. This energy source term is much less localized than in main sequence stars where the nuclear 
reactions are mainly confined within the deep central region translating into a source term depending on a high power of temperature (vs a power of one in our study). Here, the coefficient 
of proportionality c in $\epsilon$ is adjusted such that we obtain the
stellar luminosity at the surface after integration of the energy source term over radius. In order to have consistent models for each density
contrast case, the luminosity at the inner boundary needs to match the 1D
stellar luminosity curve. From Figure \ref{lum}, we deduce the luminosity imposed at the inner edge. For 
instance, the models for cases $\Delta \rho=100$ and $\Delta \rho=272$ will 
have an inner luminosity of $0.95 L_*$ and $0.71 L_*$ respectively.   
As a direct consequence, a diffusive flux must be imposed in order to carry the flux 
that would have otherwise been transported by convection within 
the unsolved central region (see Figure \ref{flux_bal}). The radial component of this diffusive flux is then  chosen
to have a swift decrease within the part of the convective zone studied to let
the resolved convection operate self-consistently. 

\begin{figure*}
  \centering
\subfigure{\includegraphics[width=4.8cm,angle=90]{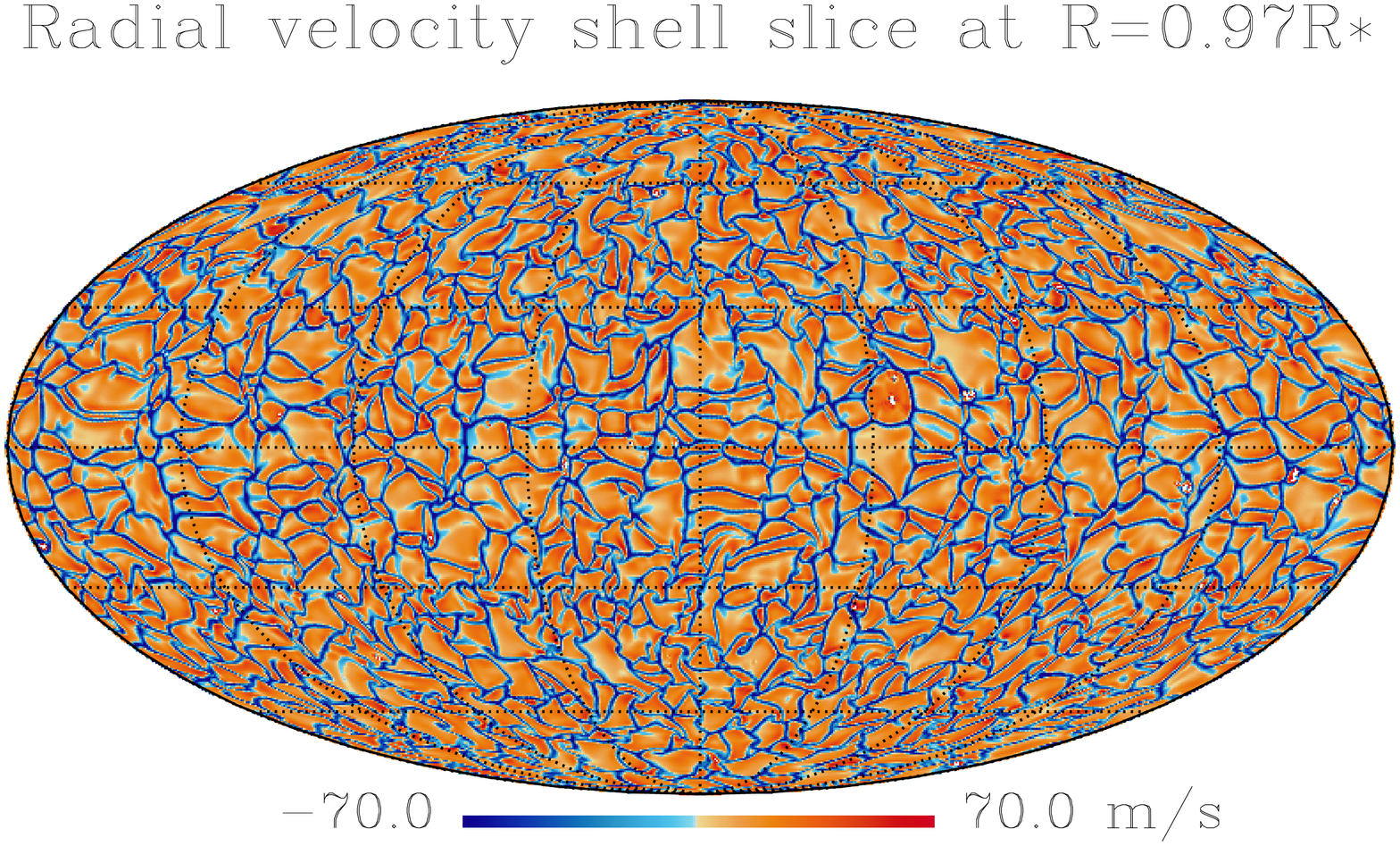}} 
\subfigure{\includegraphics[width=4.8cm,angle=90]{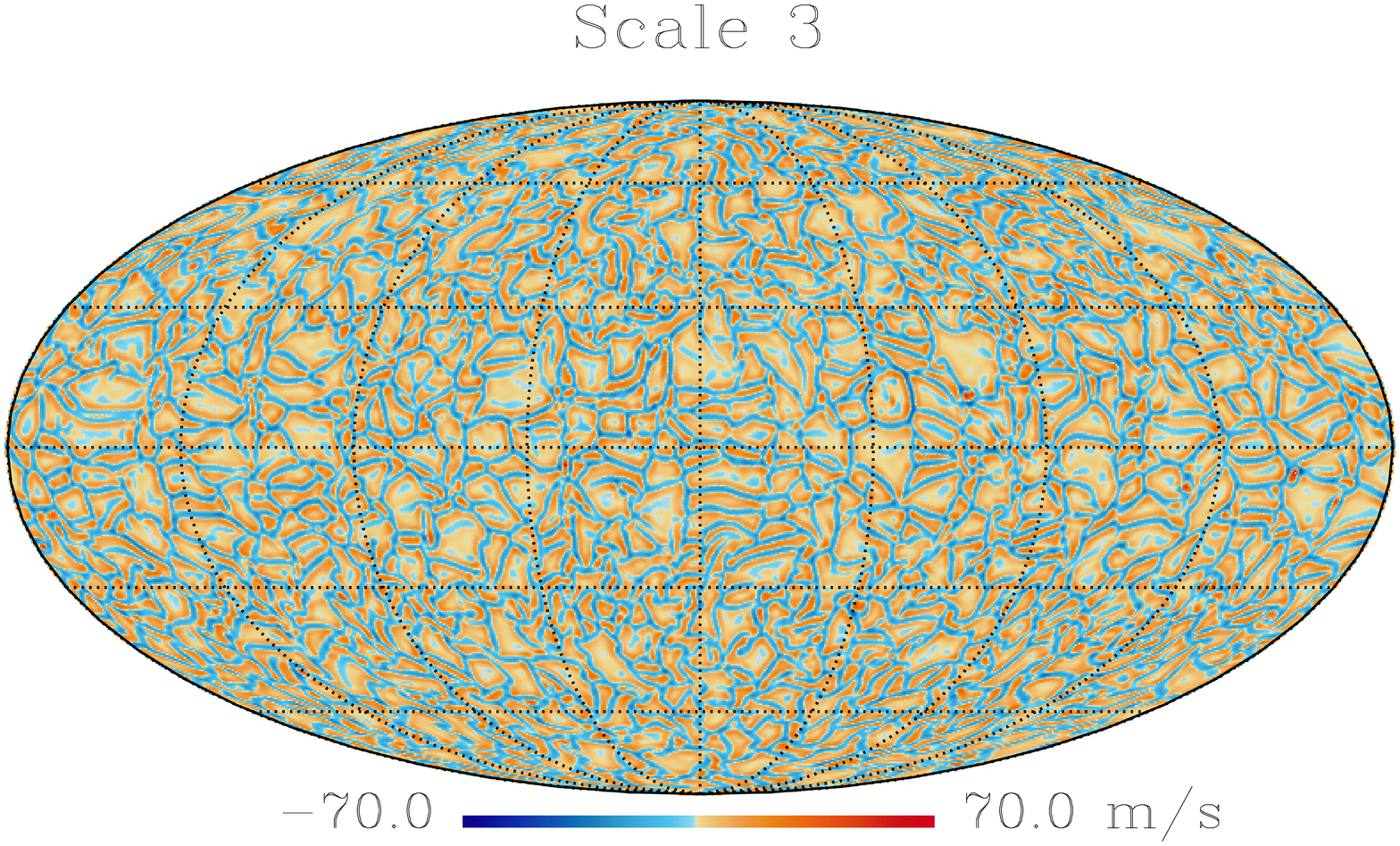}}
\subfigure{\includegraphics[width=4.8cm,angle=90]{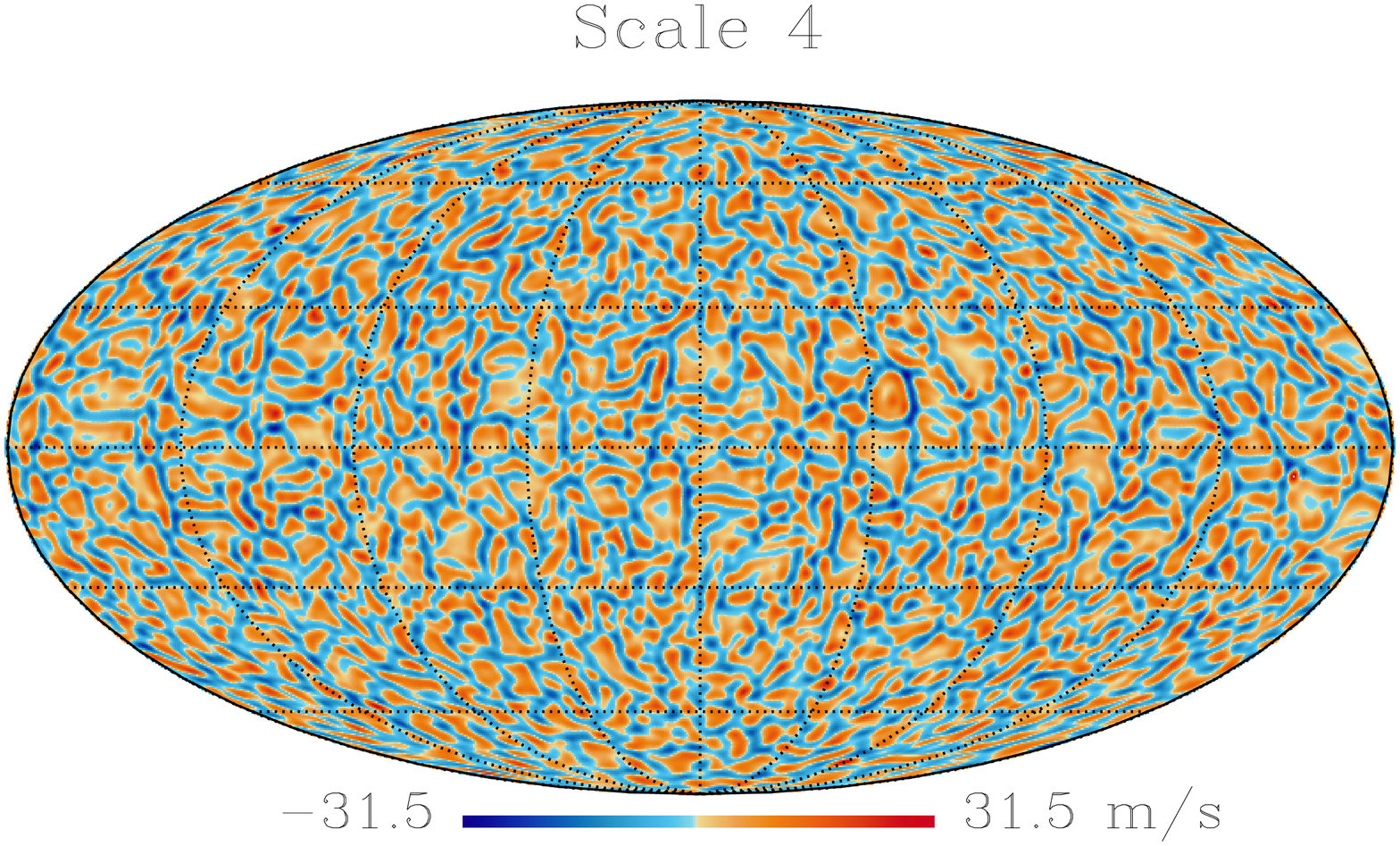}}
\subfigure{\includegraphics[width=4.8cm,angle=90]{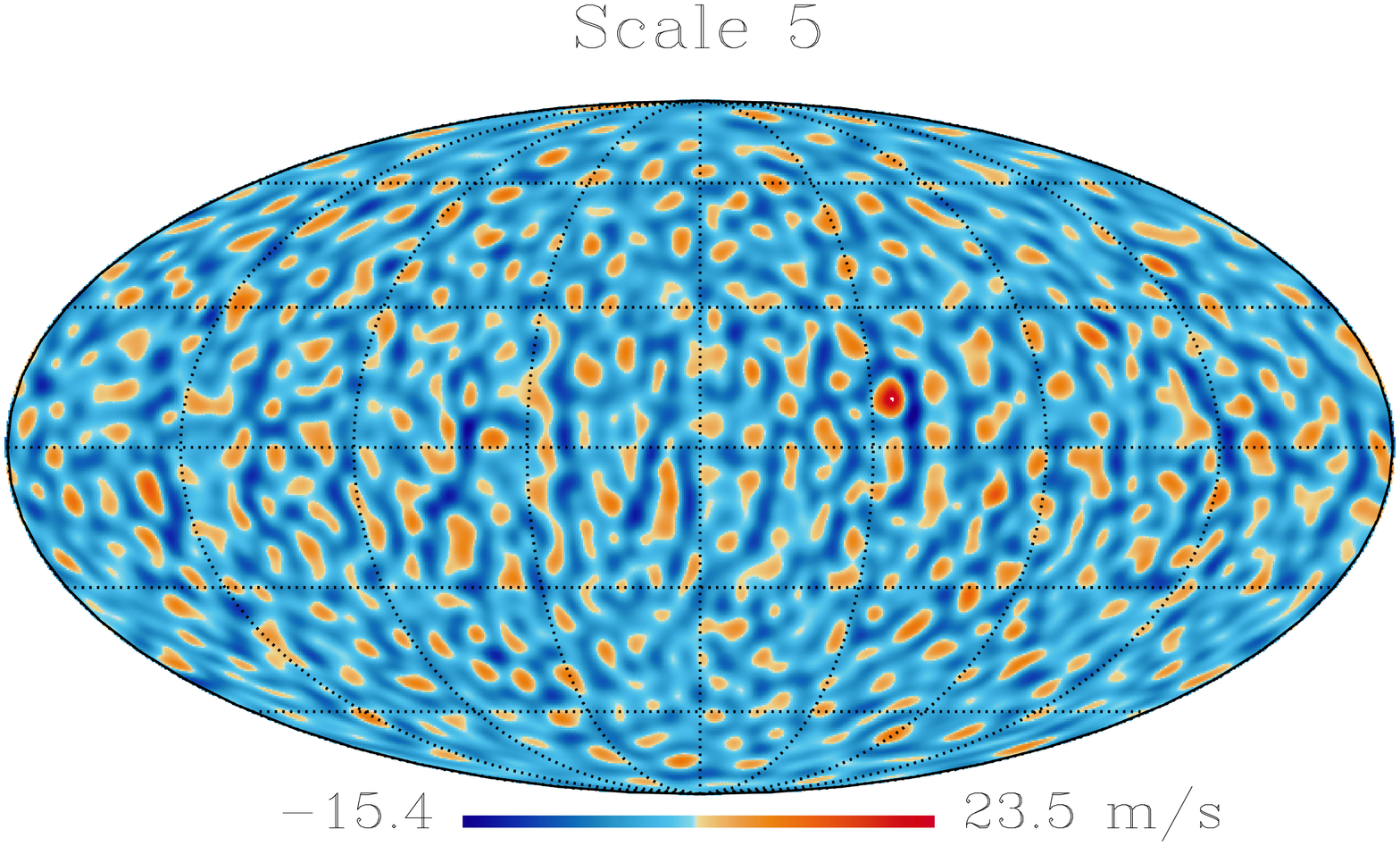}}
\subfigure{\includegraphics[width=4.8cm,angle=90]{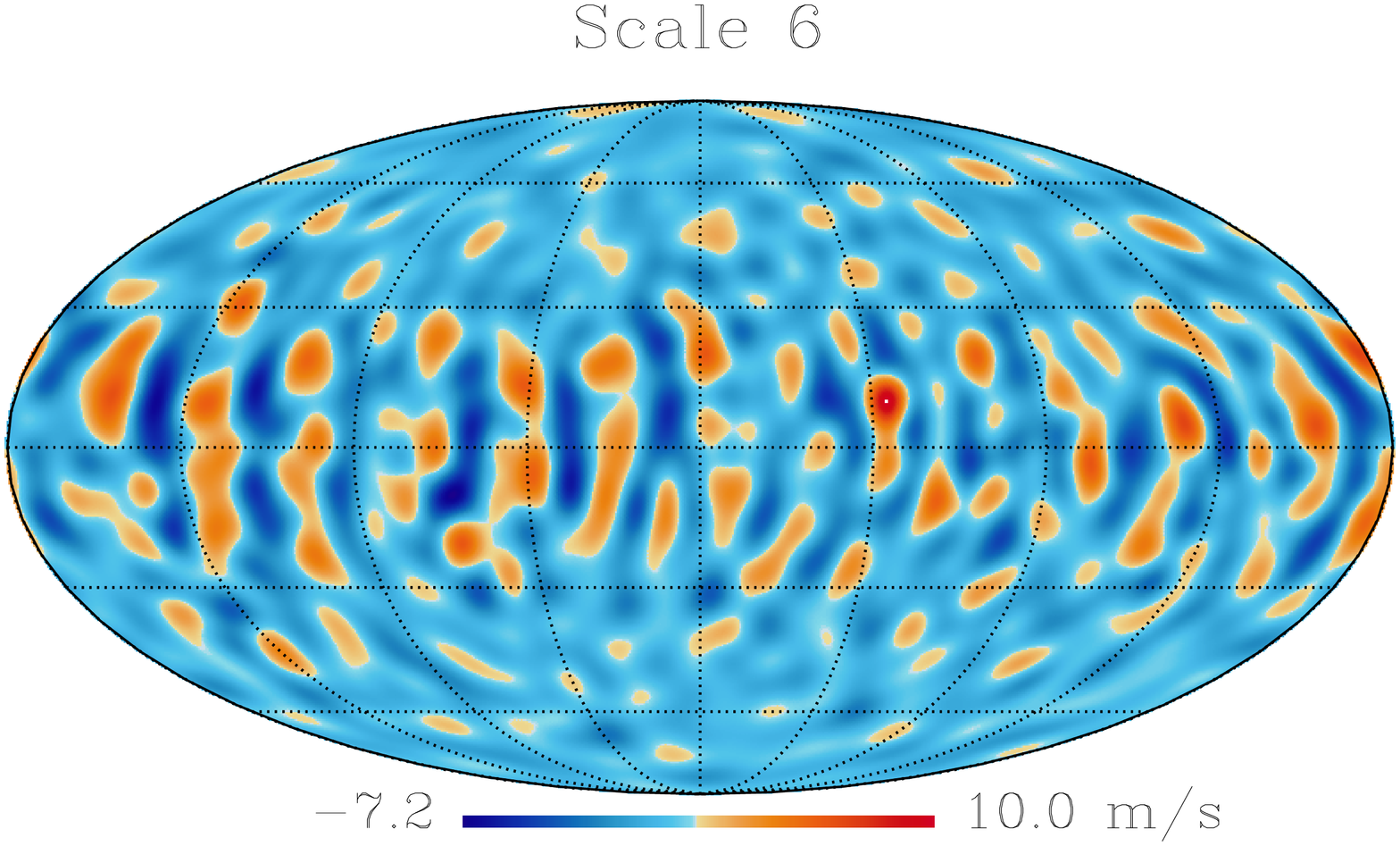}}
\subfigure{\includegraphics[width=4.8cm,angle=90]{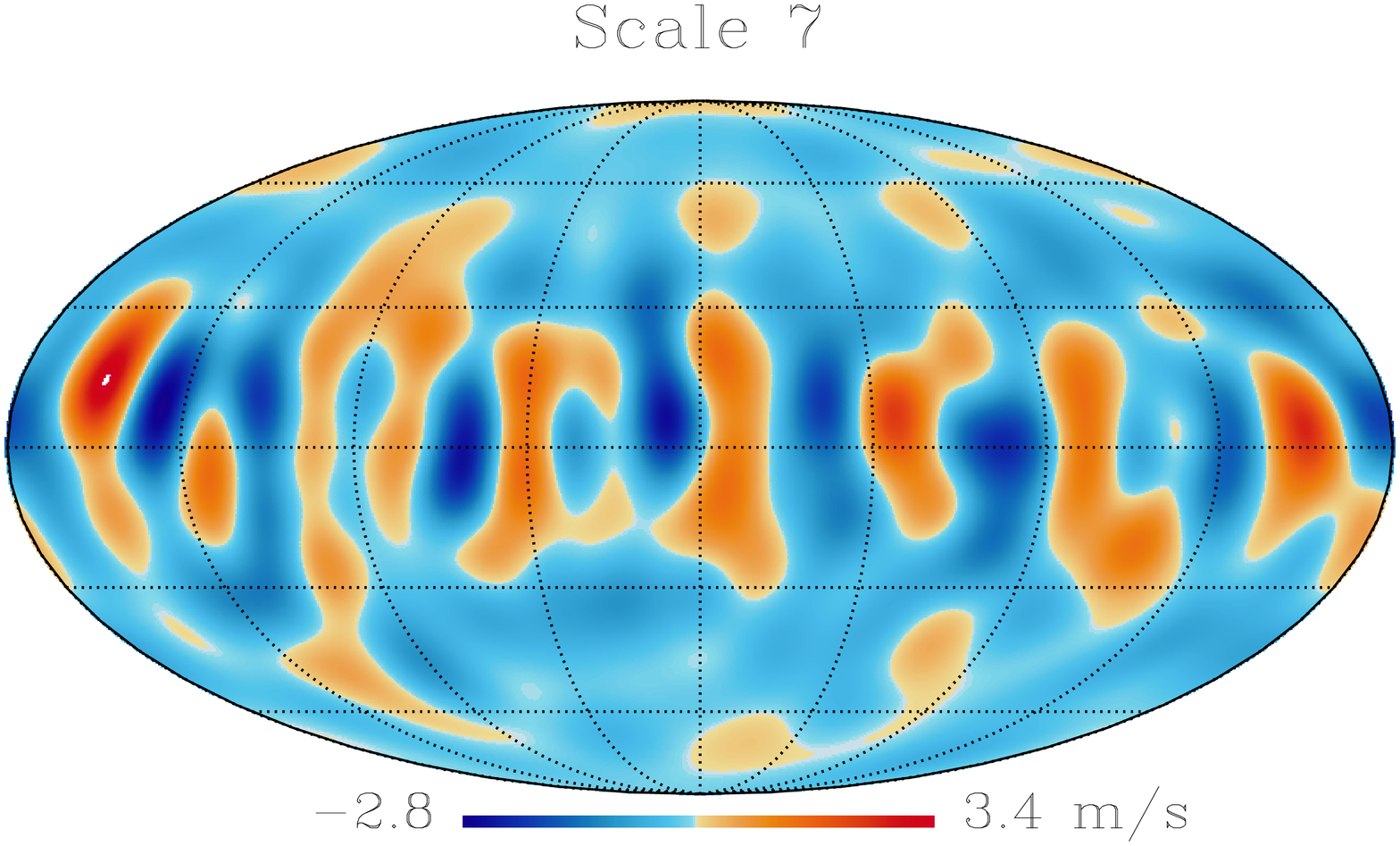}}
\caption{Multi-Resolution wavelet analysis of the radial velocity for the model $\Delta
\rho=100$. The scales 1 and 2 are not represented here because they emphasize
finer structures in the downflows as already shown by the scales 3 and 4. Scales 5 and 6
underline larger structures in the upflows and scale 7 reveals the giant cells signal.}
\label{wt_decompos}
\end{figure*}

\section{The wavelet analysis pipeline}

\subsection{The HEALPIX and MRS package applied to stellar convection} 


Since the simulations that we perform correspond to spherical shells and that 
large scale structures
 are looked for, embedded among complex patches of convection especially near 
 the surface, the use of a wavelet decomposition adapted to this spherical 
 geometry is a powerful tool for detecting giant cells.
The HEALPIX \footnote{developped by NASA initially for the analysis
of the CMB data} pavement of the sphere into equally area blocks at
different resolution with a dyadic description of the different levels is useful 
for developing efficient wavelet decomposition
algorithms and fast statistical analysis for high resolution maps on the 
full-sphere. From our binary output containing different shell slices, conversion
 into FITS format is done using the $\mathbf{ang2pix}$ routine allowing direct 
representation of our data in
the ring format for HEALPIX where data is organized for each iso-latitude circle from
 the North pole to the South one.
Before applying this transformation, an overbinning by a factor 2 is applied to fit to the nine level of resolution of the HEALPIX maps. This corresponds to 
the parameter Nside=$2^9$=512 and do not modify the initial horizontal 
resolution of our simulations which is typically 1024$\times$2048.

Next, the Multi-ReSolution (MRS) package developed at CEA (Starck et al. 2006) is used to 
perform wavelet analysis. The undecimated isotropic wavelet transform is used for its
efficiency since the sum of all the wavelet scales and the coarsest resolution image
provides exactly the original image. 
The analyzing wavelet $\Psi$ for each spatial scale n is defined by the difference between two 
order three box-spline $\Phi$ (whose shape is very similar to a gaussian) with consecutive 
cutoff frequencies having a dyadic distribution and starting
from the Nyquist frequency defined by $l_c=\frac{2}{3} N_{\theta}$ 
for the first scale : 
$\Psi = \Phi_{\frac{l_c}{2^n}}-\Phi_{\frac{l_c}{2^{n-1}}}$.
The interested reader can find all the details of such a wavelet transform in
Starck et al. (2006). Wavelet coefficients for each scale highlight different features in the convective patterns, the finest scale detailing the complex 
downflows structures whereas the coarsest one allowing the detection of large scale flows.   

For each simulation output file and each chosen depth in the convection zone, 
the wavelet analysis is performed for seven scales and for each variable 
studied. An illustration of this wavelet decomposition is presented 
in Figure \ref{wt_decompos} showing the analysis of the radial
velocity for the shell slice near the stellar surface at $r=0.97 R_*$ for the model $\Delta \rho=100$. At the top left hand side, the complex convective patterns of such a high resolution simulation is highlighted with
 asymmetries between fine dowflows and broader upflows. The first two scales correspond to fine structures in the downflow 
plumes. The scale 3 focuses on the downflows network whereas large scale 
structures are stressed in scales 6 and 7. The scales 4 and 5 correspond to intermediate scale structures linked 
to broader upflows of convective cells. Finally, the last scale shows the 
large scale structures which could correspond to giant convective shells.

\subsection{Methodology to detect giant cells}

Various methods are used to detect the signal corresponding to the 
largest scales of convection in our attempt to hunt for giant cells.

\vspace{1mm}

\noindent First, we analyze the strength of the radial velocity variable 
at large scale  by calculating the ratio between the rms speed at 
scale 3 and the rms speed at scale 7 and by distinguishing
both the upflows and downflows.

\vspace{1mm}
 
\noindent Next, we perform a time correlation analysis both on the complete image
and on scale 7 in order to quantify the lifetime of the detected large scale 
structures. 
To do this, we analyse a 60 day long time series of longitude-latitude maps of the radial velocity i.e. about two stellar periods for each model. 
Then, we calculate 
the autocorrelation function (acf in the following) as defined below
and already used in Miesch et al. (2008) for each lag time $\tau$ with respect to the initial snapshot : 

\vspace{-0.6cm}

\begin{eqnarray}
 {\rm acf(r,\Omega_t,\tau) =} \;\;\;\; \;\;\;\;\;\;\;\;\;\;\;\; \;\;\;\; \;\;\;\; \;\;\;\; \;\;\;\;\;\;\;\;\;\;\;\; \nonumber 
\end{eqnarray}
\vspace{-0.4cm}
\begin{equation}
 \frac{\int_{0}^{2\pi} \int_{\theta_1}^{\theta_2}
v_r(\tau=0,\theta,\phi) v_r(\tau,\theta,\phi') sin \theta
d\theta d\phi}{\int_{0}^{2\pi} \int_{\theta_1}^{\theta_2} v_r^2(\tau=0,\phi) sin \theta
d\theta d\phi} 
\end{equation}
where the convective patterns are reprojected in the new reference frame 
 $\phi'=\phi-\Omega_t \tau$ corotating with the chosen tracking velocity 
$\Omega_t$. The optimal tracking velocity $\Omega_{opt}$ partly due to 
the stellar differential rotation is defined for each latitudinal range
[$\theta_1$,$\theta_2$] among a large range of values $\Omega_t$ at depth r 
in order to maximize the acf function.

Then, we define the typical lifetime of large scale structures as the time
$\tau_c$ where the acf function is bigger than a defined threshold 
(typically 0.5 for this study) for each band of latitude.

\begin{table*}
\begin{center}
\begin{tabular}{|c|c|c|c|c|c|c|c|c|}
\hline
\hline
  $\Delta \rho $ & $\nu_{mid},\kappa_{mid} $ & $R_{a}$ &
$T_{a}$ & $R_{o}$ & $Re_{mid}$ & $R_*/L$ &  $\theta_{TC}$   & $N_r, N_\theta, N_\phi$  \\
               & $ (10^{11} cm^2 s^{-1})$ &  &  &  &  & & ($\degree$) &   \\
\hline
13 & 8.45 & $9.9 \times 10^4$ & $3.9 \times 10^4$  & 1.6 & 98 & 14.3 & 68.7 &  $256 \times 1024
\times 2048$  \\ 
37 & 5.44 & $1.89 \times 10^6$ & $1.66 \times 10^6$  & 1.07 & 291 & 7 & 59.4 &  $256 \times 1024
\times 2048$  \\
100 & 3.53 & $4.0 \times 10^7$ & $4.7 \times 10^7$ & 0.92 & 508  & 3.8 & 47.8 &  $256 \times 1024
\times 2048$  \\ 
272 & 4.74 & $1.2 \times 10^8$  & $2.4 \times 10^8$  & 0.71 & 720 & 2.2 & 33.2 
&  $512 \times 1024 \times 2048$  \\ 
\hline
\end{tabular}
\caption{Main characteristics of the performed simulations. The Prandtl number
Pr is 1 for all models. The top values for viscous $\nu$ and thermal $\kappa$ 
diffusivities coefficients are $2 \times 10^{12} cm^2 s^{-1}$. The Rayleigh,
Taylor, convective Rossby numbers and Reynolds number are also reported here 
respectively defined as  
$R_a=-(d\bar{S}/dr)(\partial \rho/\partial S) g L^4/(\bar{\rho} \nu \kappa)$,
$T_a=4\Omega_0^2 L^4/\nu^2$, 
$R_o=\sqrt{Ra/(TaPr)}$ and $Re=v_{rms} L/\nu$ where $L=r_{top}-r_{bottom}$ is the thickness of the convective zone and $v_{rms}$ is the
total vrms velocity at mid-depth. The location of the tangent
cylinder is defined by the colatitude $\theta_{TC}$.  }
\label{sim_prop}
\end{center}
\end{table*}

\vspace{1mm}
 
\noindent Finally, spatial cross-correlations vs depth are 
performed to probe the radial extension of the largest scales. 

\section{Convection properties}

\begin{figure*}
  \centering
  \includegraphics[width=17cm,angle=0]{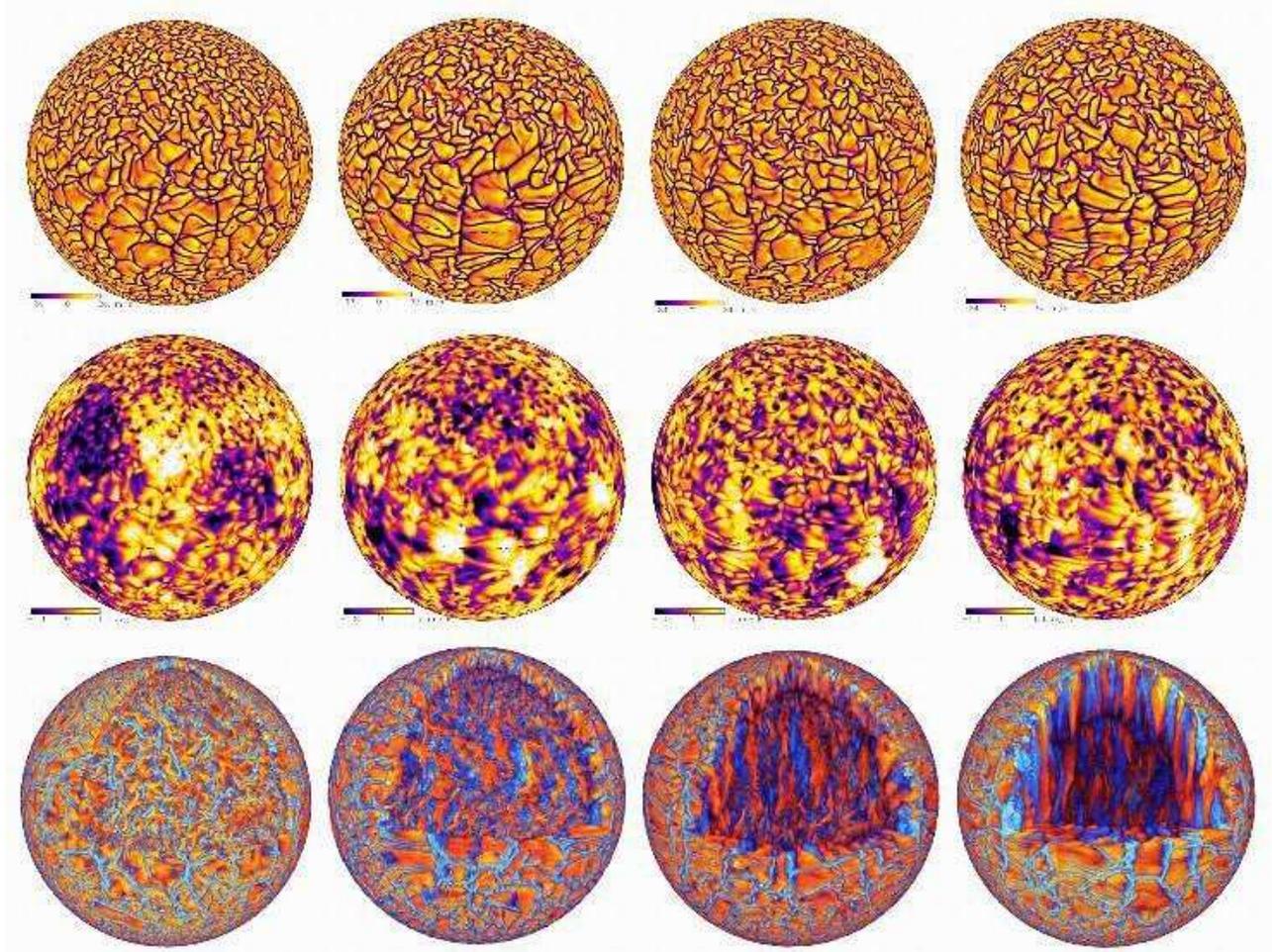} 
\caption{Top : Radial velocity patterns (upper panels) and fluctuating temperature (middle
panels) orthographic maps for all models (by increasing the density contrast from left to right)
at $r=0.97 R_*$. The dashed line corresponds to the stellar equator.
Bottom: 3D rendering of the radial velocity pattern for all models cropped in one quadrant 
in the North hemisphere to highlight the convective structures in depth (done with SDVision).}
\label{conv_T_patterns}
\end{figure*}

\subsection{General characteristics of our simulations}
 
The main characteristics of our convective models are presented in Table
\ref{sim_prop}. These simulations are done with a Prandtl number equal to 
one and
the top dissipative coefficients are chosen to have the same value equal to
$2 \times 10^{12} cm^2 s^{-1}$ for all four models such as to have the same 
level of turbulence near the stellar surface. Indeed, all the other parameters such as the density scale height or the radial velocity profiles (see Fig. 6) are the same in the upper layers of our simulations for all models as we model the same underlying star with different depth for the convective zone. 
The decrease of the convective Rossby number as the convective zone becomes 
deeper indicates that the influence of rotation on convection becomes more 
important. Indeed, thicker convective zones lead to an increase time for convective flows to go up and down hence the Coriolis force has more time to act.  
The main parameter among these models that makes the most difference is the aspect
ratio $R_*/L$ where $L=r_{top}-r_{bottom}$. Its large variation is due to the smooth slope of the density profile of this low
mass young star. This entails a great variation of the Rayleigh and Taylor 
numbers in our simulations for the different models over more than 3 orders 
of magnitude. Finally, we obtain Reynolds numbers calculated from $v_{rms}$ speeds
in the mid layer between 100 and 720. But Reynolds numbers based on peaked
velocities are instead around 1500-2000 and similar for each model, thus 
reaching one of the most turbulent simulations obtained with ASH until now.

\subsection{Turbulent convection structures according to the density contrast}

Even though the models have significantly 
different $R_*/L$, this has in appearance little influence on the convective patterns close to the stellar surface as shown in Figure \ref{conv_T_patterns}. We clearly see the complexity of the convective patterns with a network of narrow downflow lanes surrounding by a mosaic of convective upflows. Such a small 
scale tiling of our models near the surface makes it difficult 
to extract information about the potential underlying giant cells from the surface. We also note that case $\Delta \rho=13$ possesses slightly smaller convective cells at high latitudes than the three other cases. This is due to the thiness of the convective envelope. 
 
Indeed, provided that the level of turbulence is sufficient (i.e. that a fully
nonlinear state has been achieved), the small-scale convection 
patterns are really linked to the local density scale height $H_\rho$ 
which is common to all 
models and equal to 10 Mm. To be more precise, from the conservation of mass in the
anelastic approximation and for typical $H_\rho<<R_*$ , we can deduce that the 
horizontal size $H_{L}$ of the convective patterns is close to $H_{L}=\alpha H_\rho$ 
where $\alpha=v_{horiz}/v_r$ characterize the anisotropy of the flow.
We find that for the $\Delta \rho=13$ model, $\alpha=7.5$ which gives 
$H_{L}=75 Mm$ at $r=0.97R_*$. The spectral energy distribution computed 
in the polar
regions is peaked at a spherical harmonic degree $l=70$
corresponding to the same $H_{L}$ than deduce above. This value is slightly 
larger than the depth of the convective zone (around 60 Mm) for the case 
$\Delta \rho=13$. This relation holds for the other models. We
find that the anisotropy factor increases to 10.3 for the $\Delta \rho=272$ 
model corresponding to $H_{L}=103$ $Mm$. This difference explains the slightly 
smaller size of the convective cells at high latitudes for the $\Delta \rho=13$ 
model.   

By looking at the shape of the convective patterns, it is also important 
to stress the complex deformation of the
convective cells by the differential rotation everywhere and particularly the
strong shearing layer at mid-latitude. There is a clear difference between
the large convective cells near the equator always stretched longitudinally 
and the more horizontally isotropic convective cells at larger latitude. 
The fluctuating temperature structure is also common to all models with a weak
range of variations up to 1.4 K. We also clearly notice the good correlation
between velocity flows and temperature fluctuation (i.e. cold structures sink). In particular, the stronger convective plumes at the downflow
interstice are correlated with intense cold spots in the temperature map. 
This is typical of high resolution turbulent convection.

\begin{figure*}
  \centering
\subfigure{\includegraphics[width=6cm,angle=90]{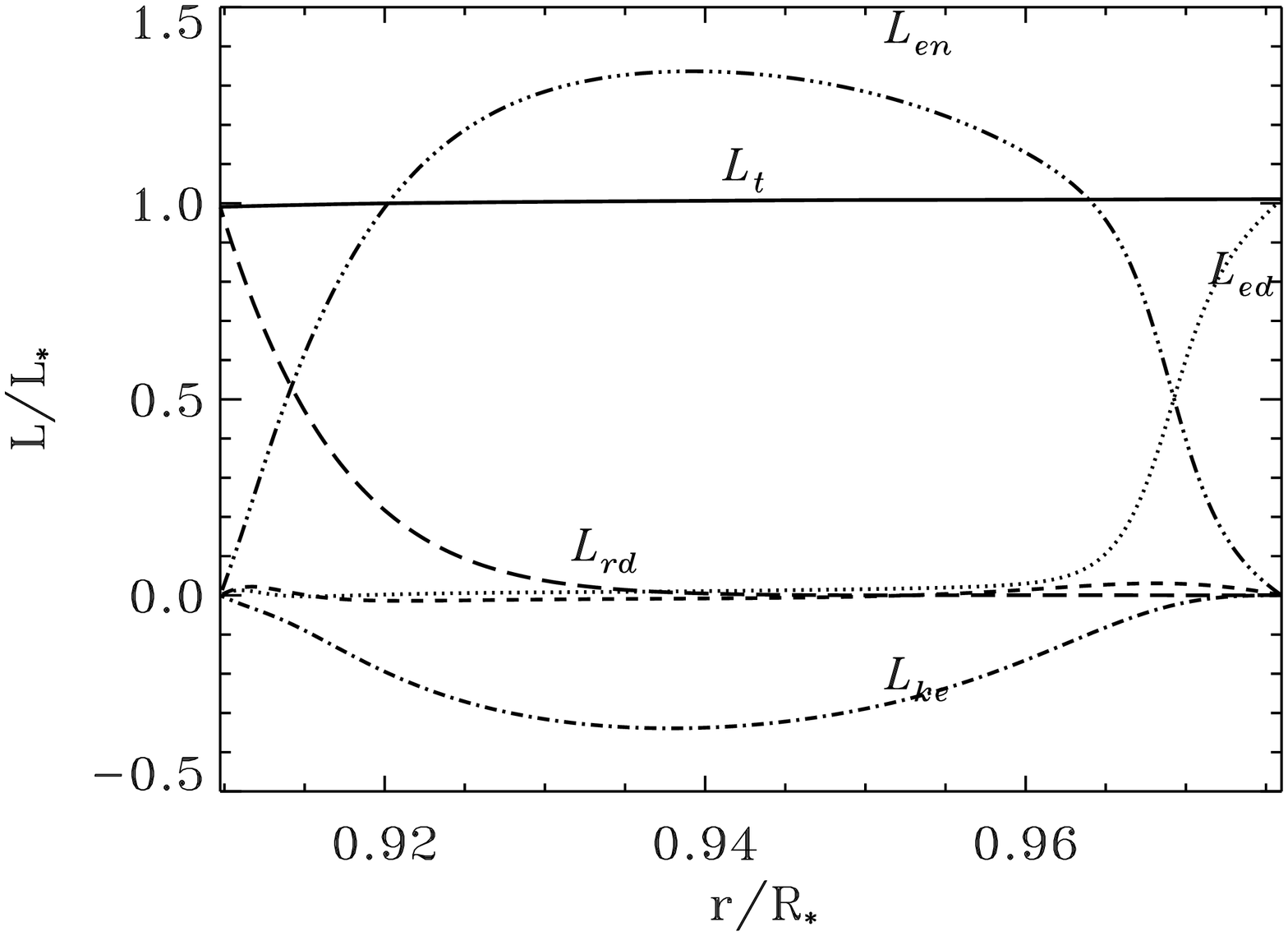}} 
\subfigure{\includegraphics[width=6cm,angle=90]{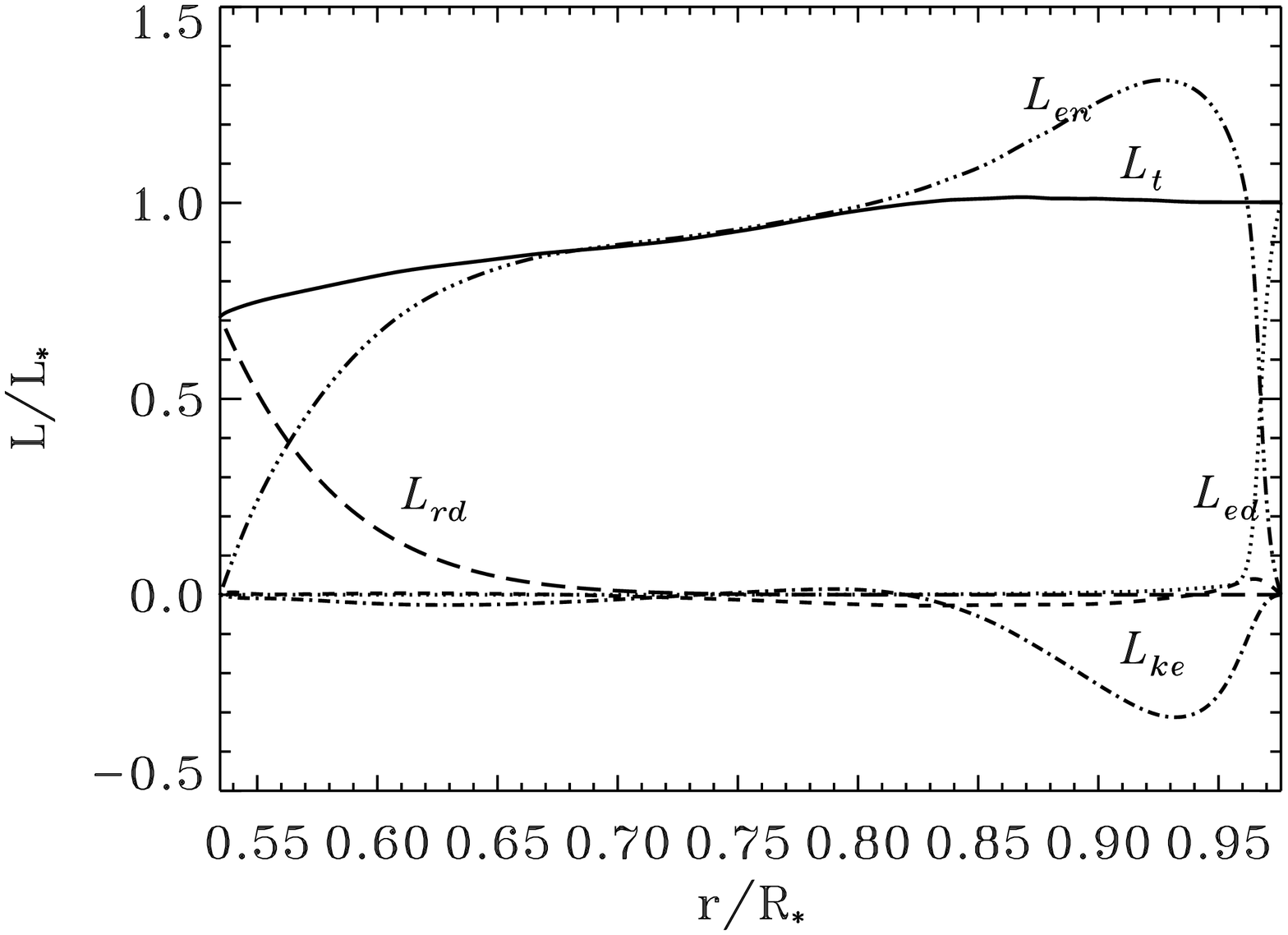}}
\caption{Temporal average of the spherically symmetric 
radial energy flux balance converted to 
stellar luminosity for the $\Delta \rho=13$ (left) and $\Delta
\rho=272$ (right) models. The different fluxes are converted into relative
luminosities with respect to the stellar one. The total flux $L_t$ is the
sum of the enthalpy flux $L_{en}$, the viscous flux $L_\nu$, the unresolved flux
$L_{ed}$, the kinetic energy flux $L_{ke}$ and the diffusive one $L_{rd}$.}
\label{flux_bal}
\end{figure*}

\begin{figure}[!h]
\centering
   \includegraphics[width=5.8cm,angle=90]{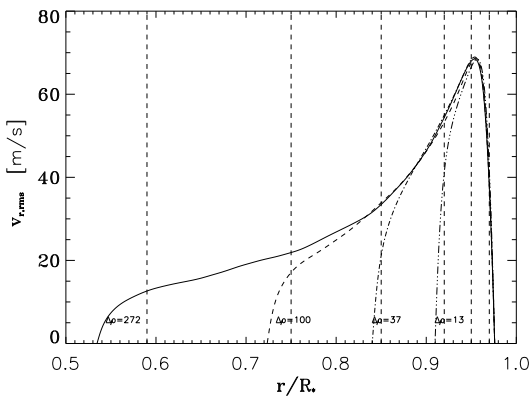}
      \caption{Rms radial velocity profiles for the different models. Each
      horizontal shell slices analyzed in the paper corresponds to one of the 
vertical dashed lines.}
       \label{vr_moy}
\end{figure}

\begin{figure*}[!th]
  \centering
\subfigure{\includegraphics[width=5.5cm,angle=90]{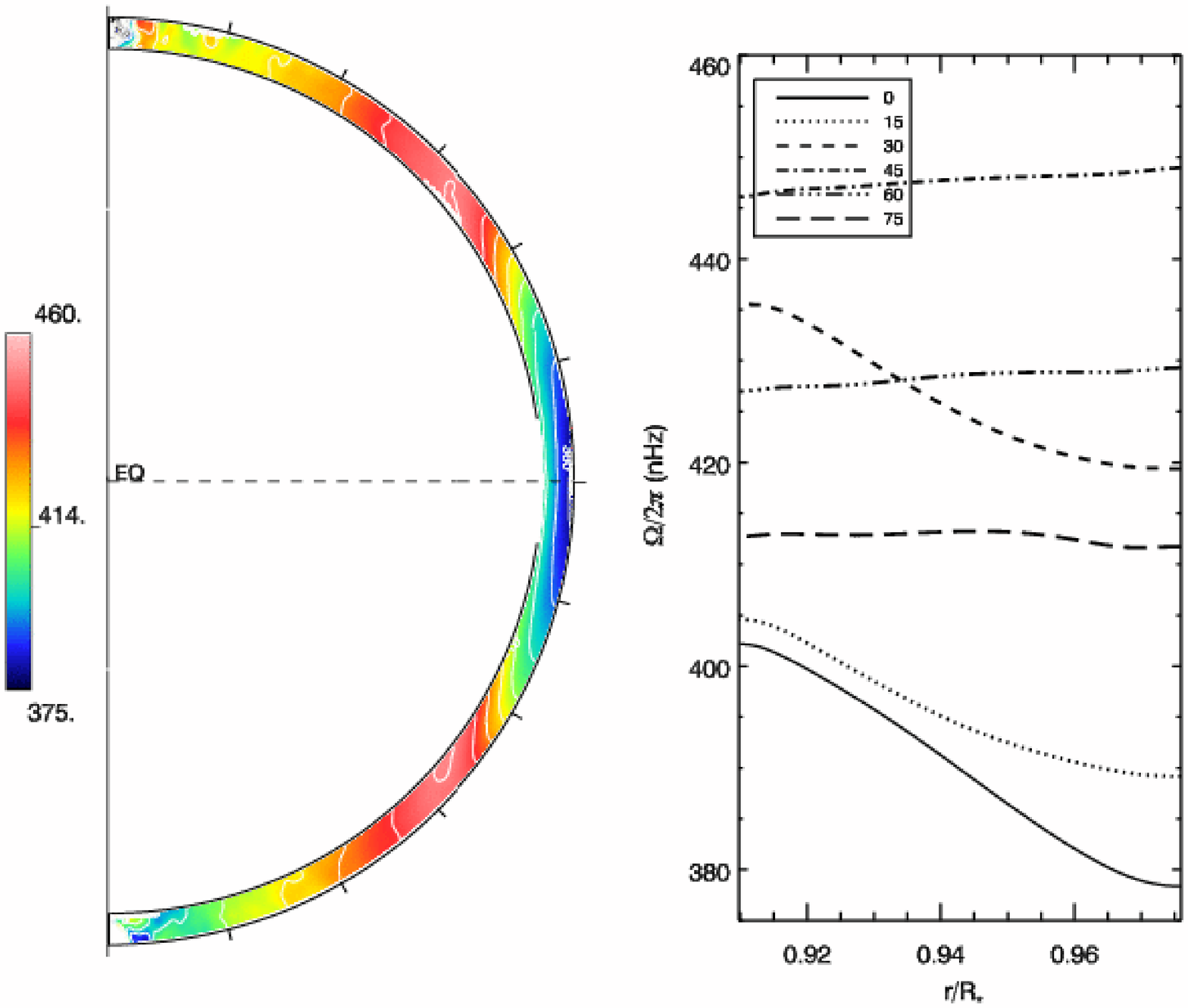}} 
\subfigure{\includegraphics[width=5.5cm,angle=90]{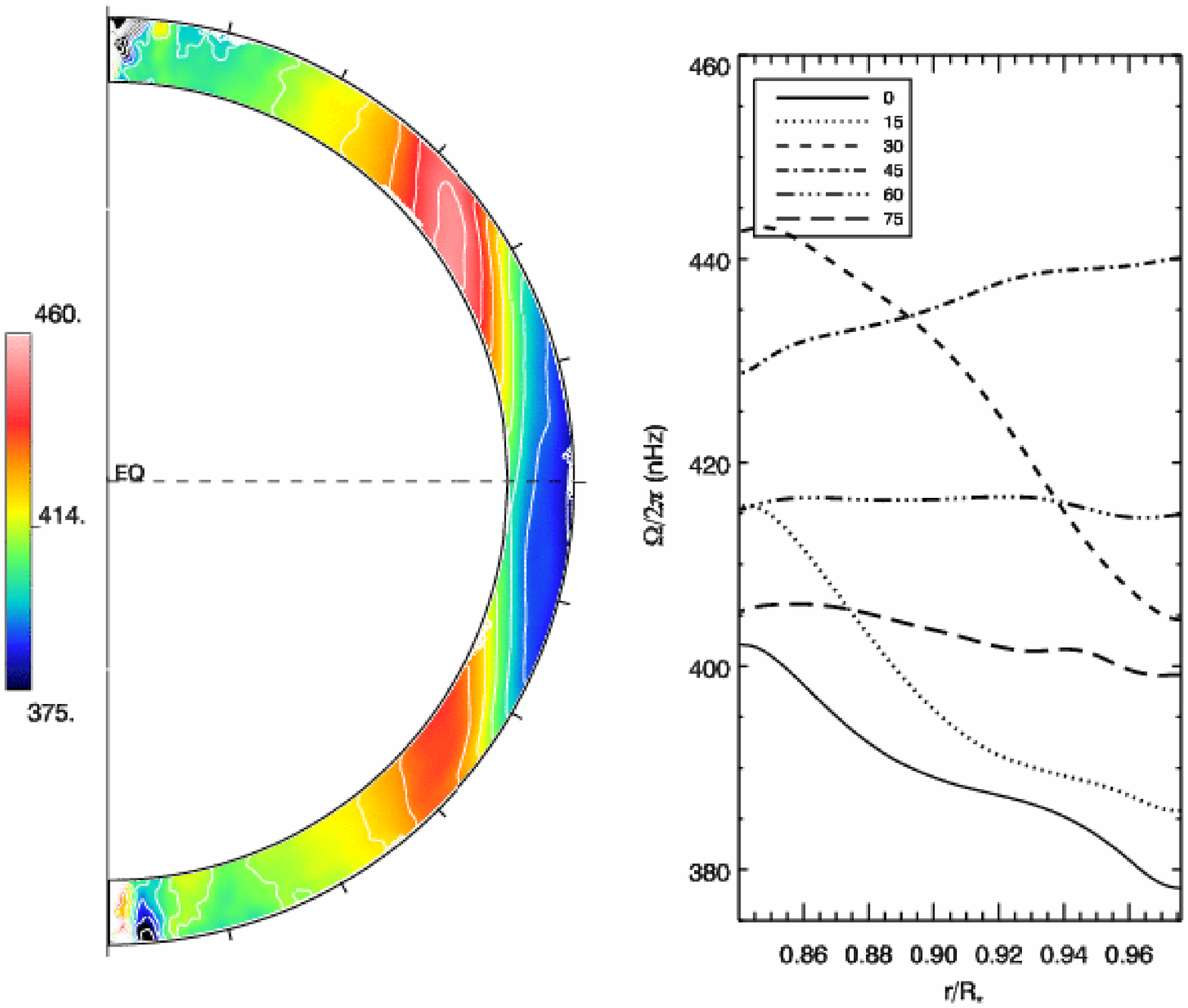}}
\subfigure{\includegraphics[width=5.5cm,angle=90]{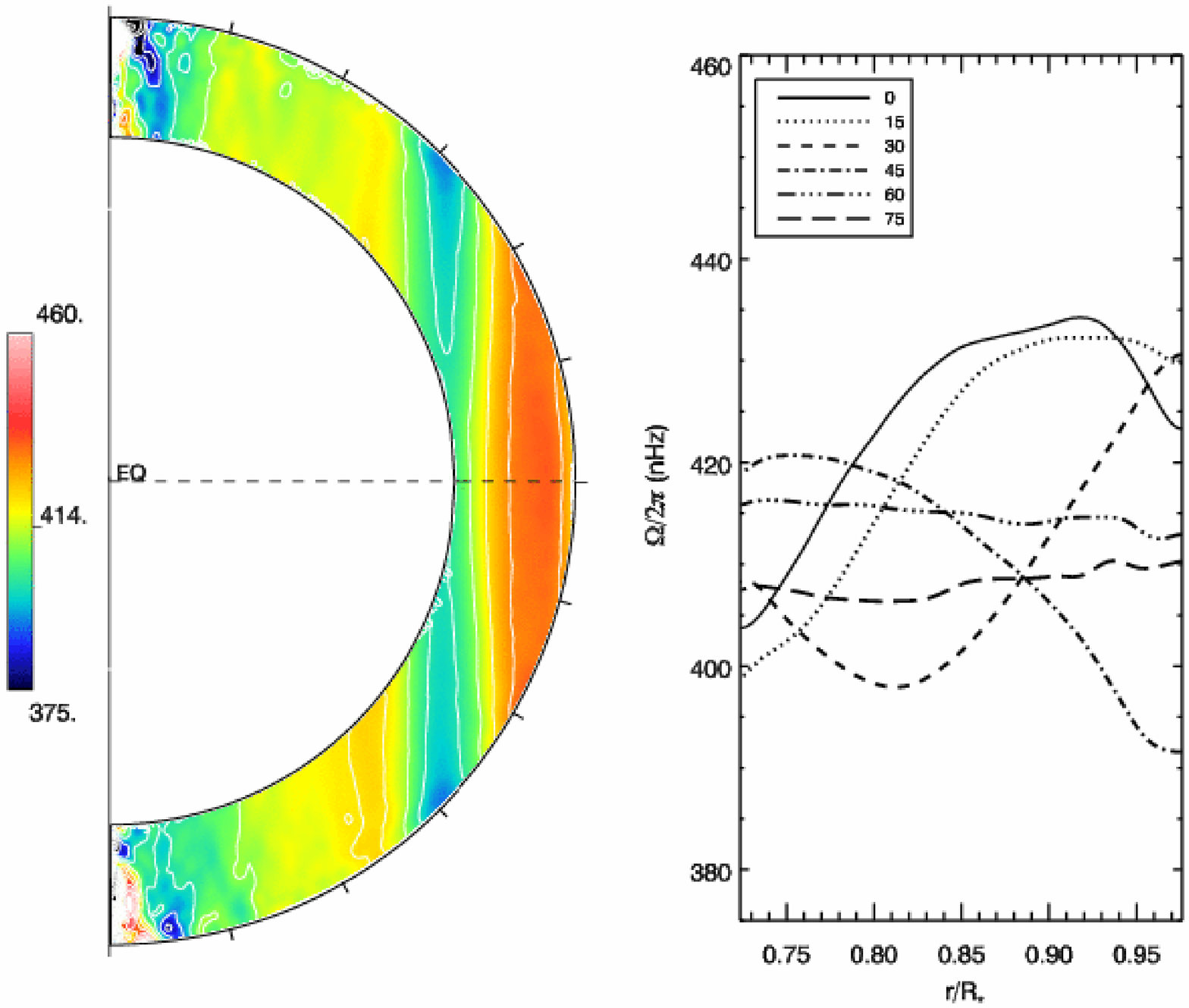}}
\subfigure{\includegraphics[width=5.5cm,angle=90]{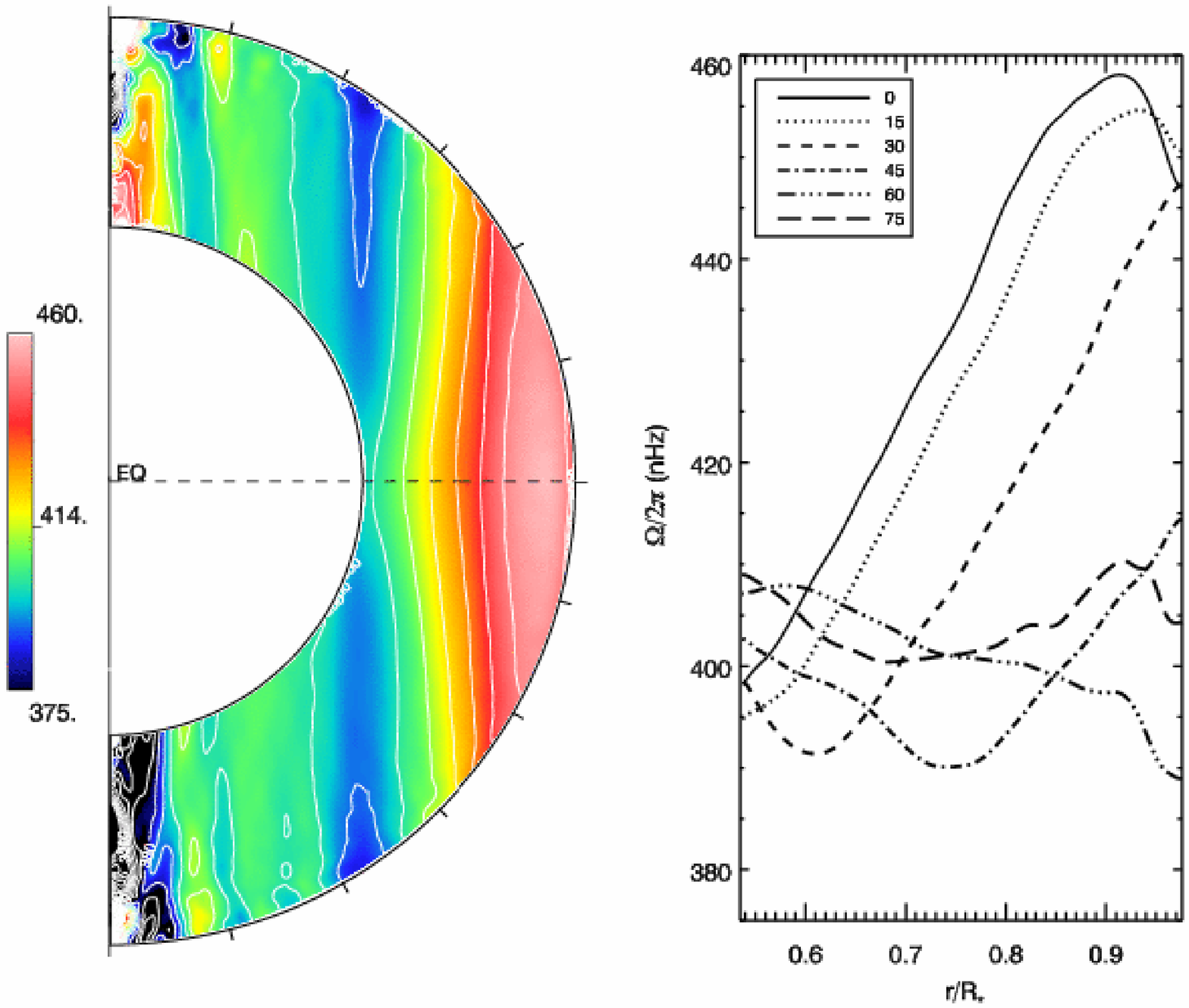}}
\caption{Differential rotation profiles for all models. There is a change of
behaviour in the rotation profiles from retrograde (upper panels) for the low
density contrast cases ($\Delta \rho =13,37$) to prograde (lower panels) for
 the large density cases ($\Delta \rho =100,272$).}
\label{diff_rot}
\end{figure*}

To appreciate the 3D structure of the flow, a 3D
rendering of the radial velocity for each model (done with SDVision, see 
Pomarede et al. 2008) is presented on the last row of Figure
\ref{conv_T_patterns}. These figures clearly show the turbulent nature of 
the flows in our realized convective models. The convective patterns are 
similar whatever the latitude is for the $\Delta \rho=13$ model whereas the 
$\Delta \rho=100,272$ models show
convective cells extended through the whole convective zone and parallel to
the rotation axis at high latitude. Convective structures modify their 
orientation to accomodate for the Coriolis force at the equator. They tend 
to drift at low latitude from being
purely radial (as imposed by gravity) to be aligned with the rotation axis. 
This underlines the influence of rotation in this latter
models where $R_o<1$. The drawing of some stream lines of the velocity 
field in these particular aligned convective cells show flows which  
have a spiraling motion within both downflows and upflows parallel to the 
rotation axis at this high-latitude position really different from the 
simplistic cartoon of convective rolls. We also see the effect of the differential rotation that tilts the convective cells in depth. Convective structures thus  an angle with respect to a pure radial direction both in $r-\theta$ and $\theta-\phi$ planes. This is at the origin of Reynolds stresses and associated angular momentum.

One can also look at how the energy is transferred radially for the differents models after they reach a relaxed state (see Figure 
\ref{flux_bal}). The various expression for each transport process (namely
radiative, kinetic, viscous or unresolved fluxes) are defined 
for instance in Brun et al (2004). We focus on the two extreme cases, the others
having intermediate properties. First, only 70\% of the stellar luminosity is 
reached at the inner edge of the $\Delta \rho=272$ model going much deeper in the
convective zone than the $\Delta \rho=13$ model. This highlights again that the
main source of energy in this pre-main sequence star is provided by the
gravitational contraction as already discussed in Section \ref{model} and is a
consequence of a radially extended heating source.   
We can notice the shape of the unresolved 
flux $L_{ed}$ which is defined such that it increases quickly from 
$r=0.96R_*$ to unity at the outer boundary condition localized at $r=0.98R_*$. 
The transport of energy by viscosity (dashed line) is always negligible in 
these models. The overluminosity of the outward enthalpy flux (between 130 \% and 140 \% of the stellar surface luminosity) characteristing convection
flows is due to the asymmetry between the narrow cooling downflows and the broad warm 
upflows in compressible convection as already illustrated in 
Figure \ref{conv_T_patterns}. This yields a strong inward kinetic
energy flux and force the convection to carry up to 40\% more than the surface stellar radiated flux. This radial convective luminosity resulting from our 3D simulations is really 
different from the initial 1-D model presented in Figure \ref{lum} based on 
mixing-length theory. In our models, turbulent convection is complex, asymetric
and carry an overluminous flux. As explained in Section 2.3, the radiative luminosity is enhanced at the inner boundary to mimic the flux emitted by the unresolved portion of the star.   

Further, studying the azimutally and longitudinally average rms radial velocity
distribution ($v_{rms}$) is also a clue for the global radial convection structure and
 is a mean to check that the inner boundary condition do not disturb and 
 introduce artefacts in our study on the bulk of the convective zone. 
 Figure \ref{vr_moy} show such a radial distribution of $v_{rms}$ 
(also indicating the location of the last shell-slices for each density 
contrast case). 
First, the velocity profile close to the stellar surface is
really common to all models for the shell slices at $r=0.97R_*$ and
$r=0.95R_*$ with a maximum averaged radial velocity of 68 $\mathrm{m.s^{-1}}$. 
Then, there
is a clear difference between models near the inner boundary location with
 respect to the highest density contrast case $\Delta \rho=272$ due to the fact
 that the impenetrable boundary condition demands a cancellation of radial 
 velocity. We see that near the bottom the flow is around 20-30 m/s in most 
cases, except the $\Delta \rho=13$ model where it reaches $\approx 45$ m/s before dropping swiftly to zero.

\begin{figure}[!t]
\centering
\hspace{-1cm}
   \includegraphics[width=8.5cm,angle=0]{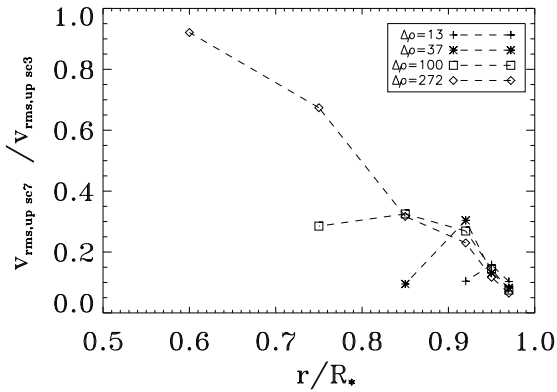}
      \caption{Wavelet analysis of the rms radial velocity amplitude for
      upflows. Similar results is obtained for downflows.}
       \label{wt_vrms}
\end{figure}

After having pointed out the global convection organization and the importance
of rotation, we now study how the interplay between these two ingredients
gives rise to the differential rotation (see Figure \ref{diff_rot}). 
The differential rotation is dominated by band of slow/fast jets. 
They are mostly cylindrical.
The differential rotation is retrograde at low latitudes 
for the low density contrast cases and becomes prograde for the stronger 
density cases. This change of behaviour is linked to the fact that the 
influence of rotation in the non linear regime becomes important for these 
latter models with $R_o<1$ as already seen in Gilman (1979),
Glatzmaier \& Gilman (1982) and Browning, Brun \& Toomre (2004). Actually, 
as the rolls rise, it is more and more influenced by rotation, being tilted
away from purely radial motions. This leads to a gradual change of the
redistribution of angular angular momentum by convection via Reynolds stresses
(see for instance Brun \& Toomre 2002 and Miesh et al. 2008). The rotation 
contrast is similar for all models although a little smaller for 
large density contrast cases. No particular effort has been made to introduce
the influence of a tachocline via a thermal forcing (Miesh et al. 2006) since we are interested mostly in convective
structure and not mean flows profiles (see Ballot et al. 2007 for a discussion of that effect in young stars).

\section{Hunting for giant cells}

After having stressed the existence of large scale structures in our
simulations, we study their spatio-temporal coherence. This should allow us 
to identify them as giant cells and finally to characterize their
3D structure and properties according to the density contrast as well as their 
capacity to transport heat, energy and angular moementum.

\subsection{Extracting the giant cell signal}

\begin{figure*}
  \centering
\subfigure{\includegraphics[width=8cm,angle=0]{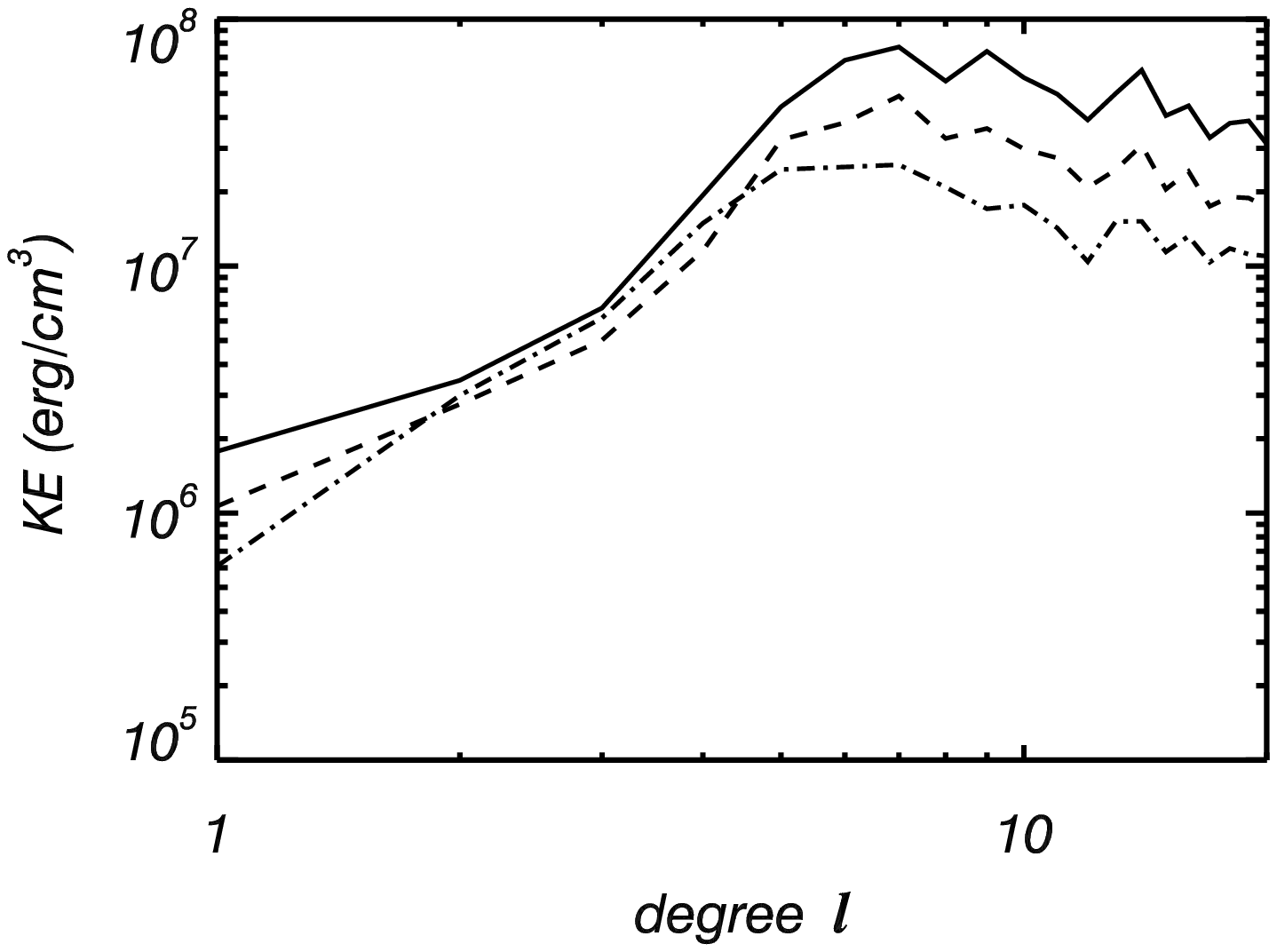}} 
\subfigure{\includegraphics[width=8cm,angle=0]{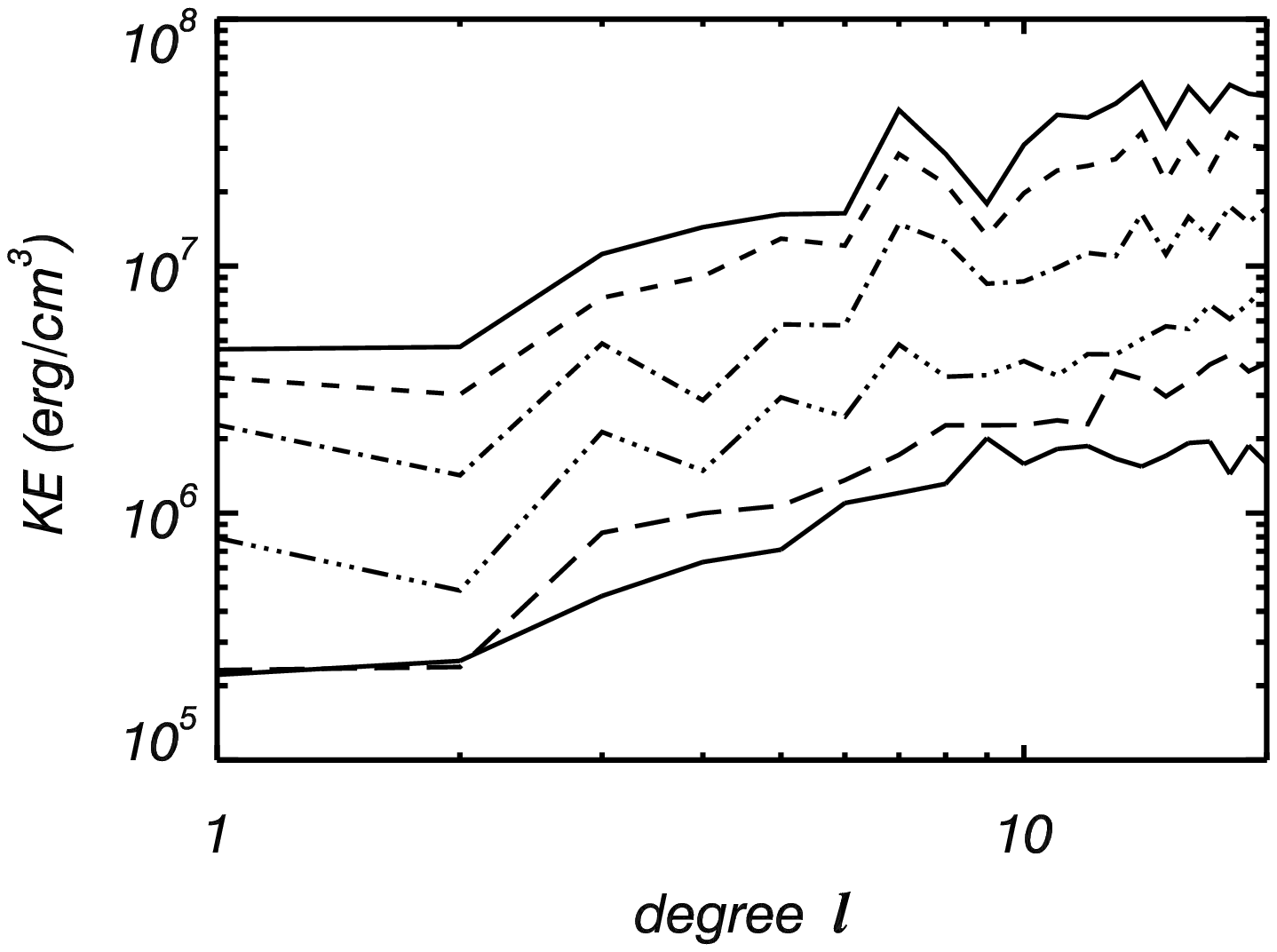}}
\caption{Spectral kinetic energy distribution focused on low l modes
at different depths for the $\Delta \rho=13$ (left) and
 $\Delta \rho=272$ (right) models. The upper solid line corresponds to
 $r=0.97R_*$, the dashed one to $r=0.95R_*$, the triple dotted dashed one to
 $r=0.92R_*$, the dotted dashed one to
 $r=0.85R_*$ and the long dashed one to $r=0.75R_*$.}
\label{sed}
\end{figure*}

Figure \ref{wt_decompos} showed that the large scale structures are localized 
mainly at low latitudes (mainly lower than 30\degree) and that their typical 
width is around 40$\degree$  i.e. 490 Mm. The "potatoe like" shape found here for these giants structures differs from the analysis of solar data by 
Beck et al. (1998) which on the contrary finds longitudinally extended cells 
focused at low latitudes.

The amplitude of the large scale structures increases with the contrast density
(see Figure \ref{wt_vrms})
specially for the shell slices corresponding to the middle depth of the
computed domain of each case, showing the strenghtening of the potential 
underlying giant convective cells with depth. Near the surface, the signal in 
the large scale from both upflows and inflows
stays quite small (typically 5\% of the speed flows at scale 3 as already visible
in Figure \ref{wt_decompos})
since the giant cells are really disturbed by smaller convective
cells there.  We can also notice on Figure \ref{wt_vrms} the effect of the impenetrable inner boundary condition except for the
case $\Delta \rho=272$ where the deepest shell slice is still far from the inner
boundary. By looking at spectral energy distribution (Figure \ref{sed}), we also remark that the
spectral kinetic energy distribution at each depth in the low spherical harmonic
degree l range increases with the density contrast by a
factor of 5 between the $\Delta \rho=13$ model and the $\Delta \rho=272$ case.
This is not only due to faster flow in the more stratified case but also 
to a change of slope in the spectra for this range of $l <10$.
Particularly, there is a maximum at $l=7-10$ for the $\Delta \rho=13$
model which is less pronounced in the stronger density contrast case. In the
analyses of the convection instability done by Chandrasekar (1961), 
Roberts (1972) and Dormy et al. (2004), the most unstable mode for the full sphere is l=1.
As the shell becomes thinner, the most unstable mode drifts towards higher l. 
One can crudely understand this by the fact that the convection instability 
trigger mainly convective rolls with a characteristic scale of the order of 
the aspect ratio of the shell which corresponds to the mode $l=14$ for the 
$\Delta \rho=13$ 
model whereas the characteristic scale for the $\Delta \rho=272$ model 
is instead $l=2$. But we also see that the turbulent nature of our simulations
makes it also almost impossible to pick up a single l mode, that is why our
wavelet analysis is better suited for the purpose of finding giant cells that 
are made of a complex combinaison of l modes.   

Now that we have identified a clear signal at large scale, we wish to analyze its properties. Particularly, we can characterize the temporal coherence of these large-scale 
structures by trying to track them over long timescales at least of the order of the stellar rotation period.

\subsection{Time correlation analysis}

\begin{figure*}
  \centering
\subfigure{\includegraphics[width=3.9cm]{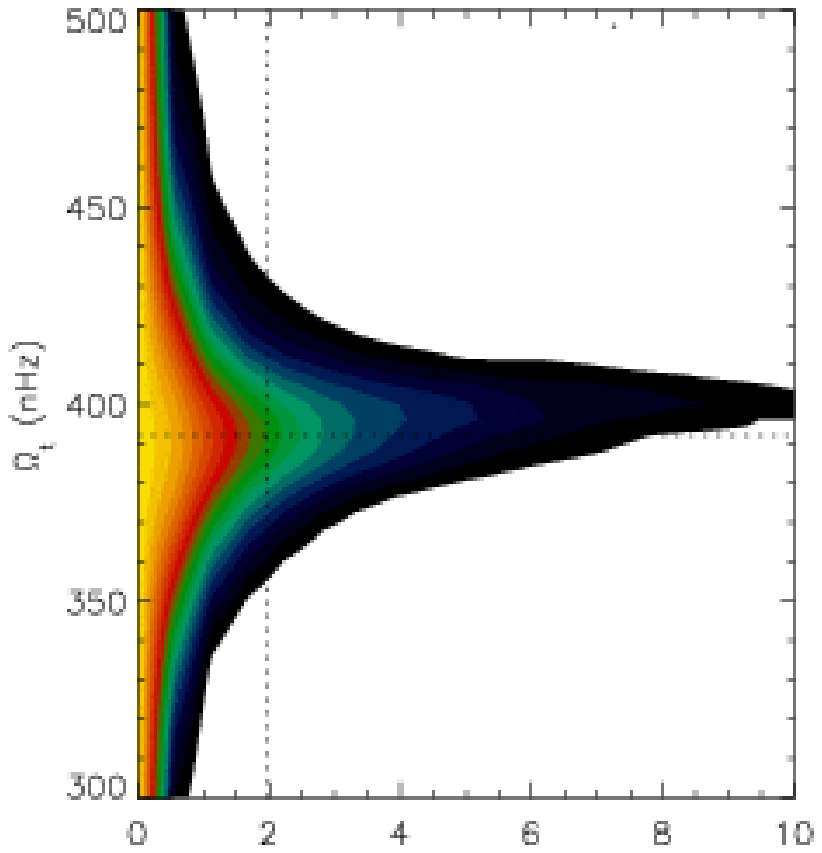}}  
\subfigure{\includegraphics[width=3.9cm]{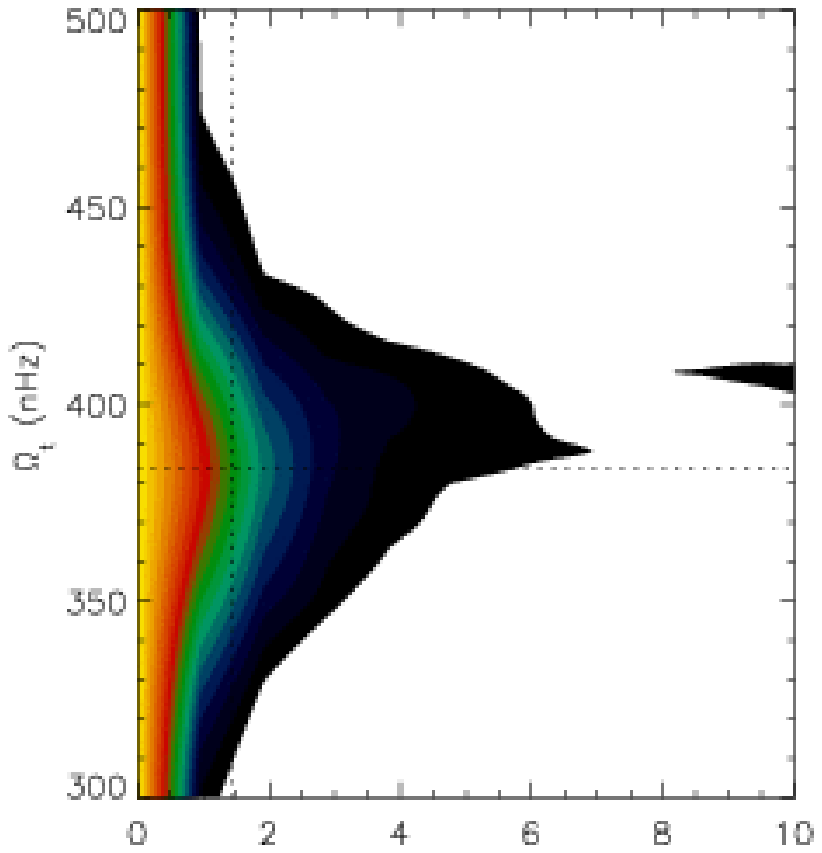}} 
\subfigure{\includegraphics[width=3.9cm]{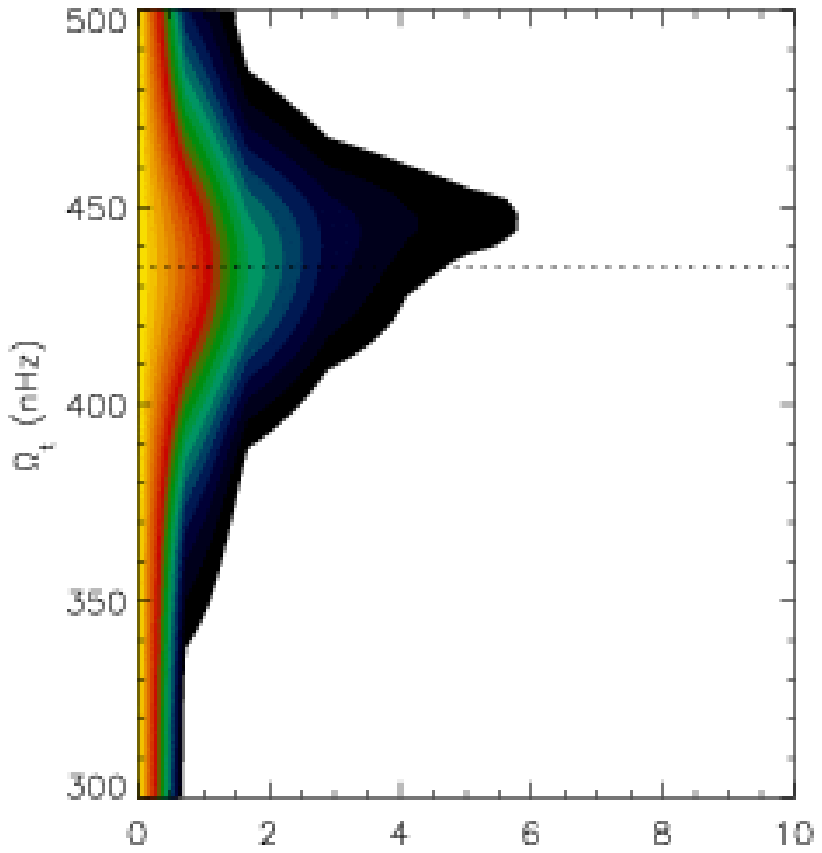}} 
\subfigure{\includegraphics[width=3.9cm]{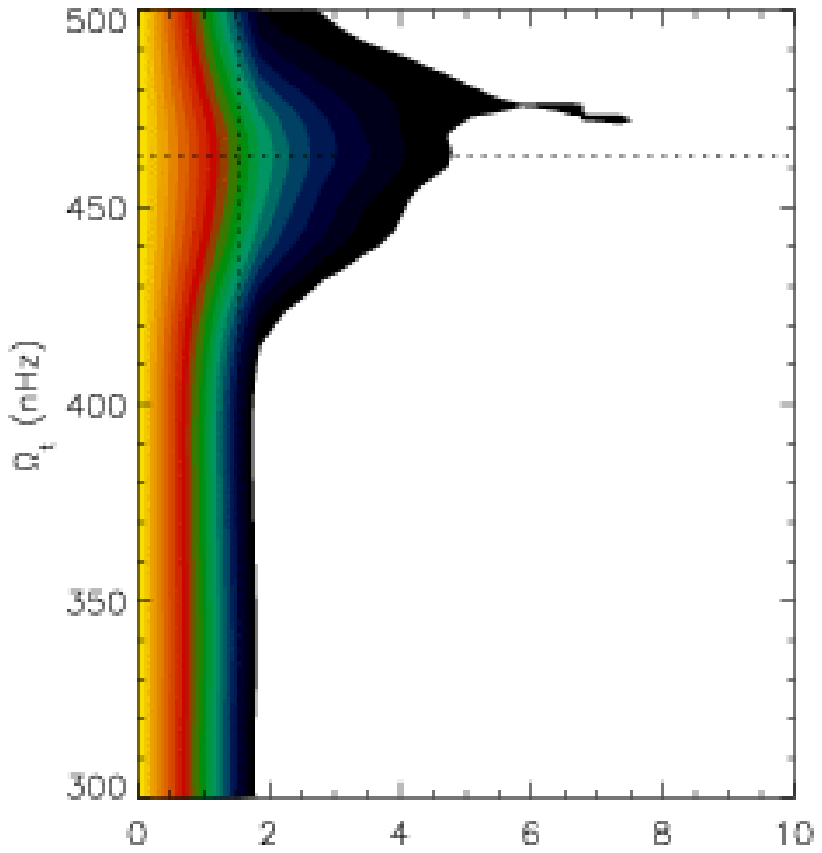}} 
\subfigure{\includegraphics[width=3.9cm]{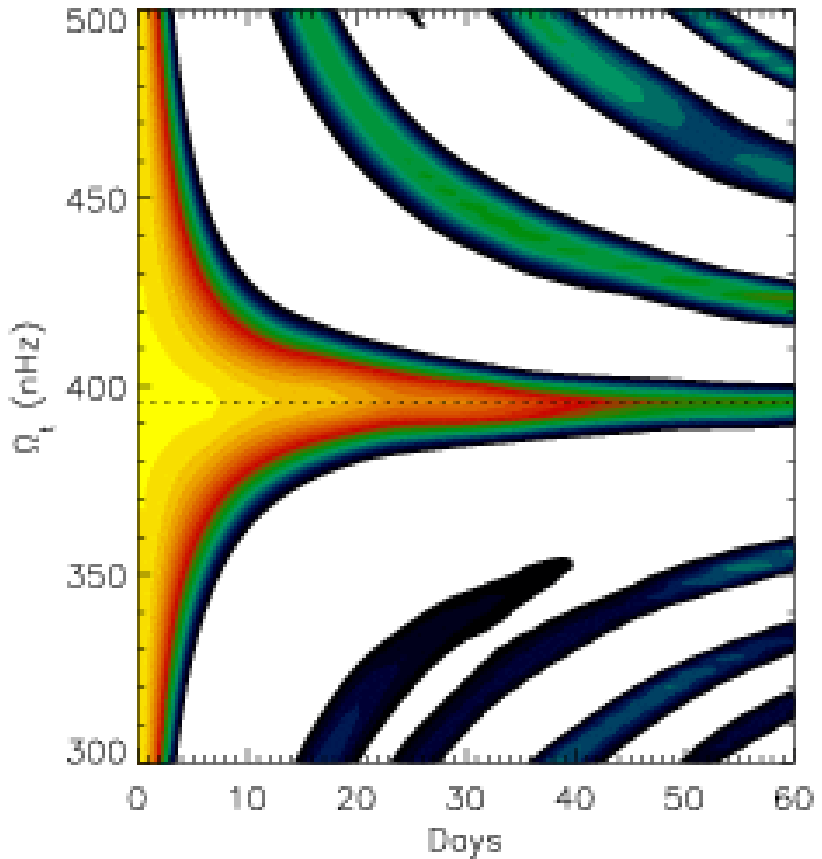}}  
\subfigure{\includegraphics[width=3.9cm]{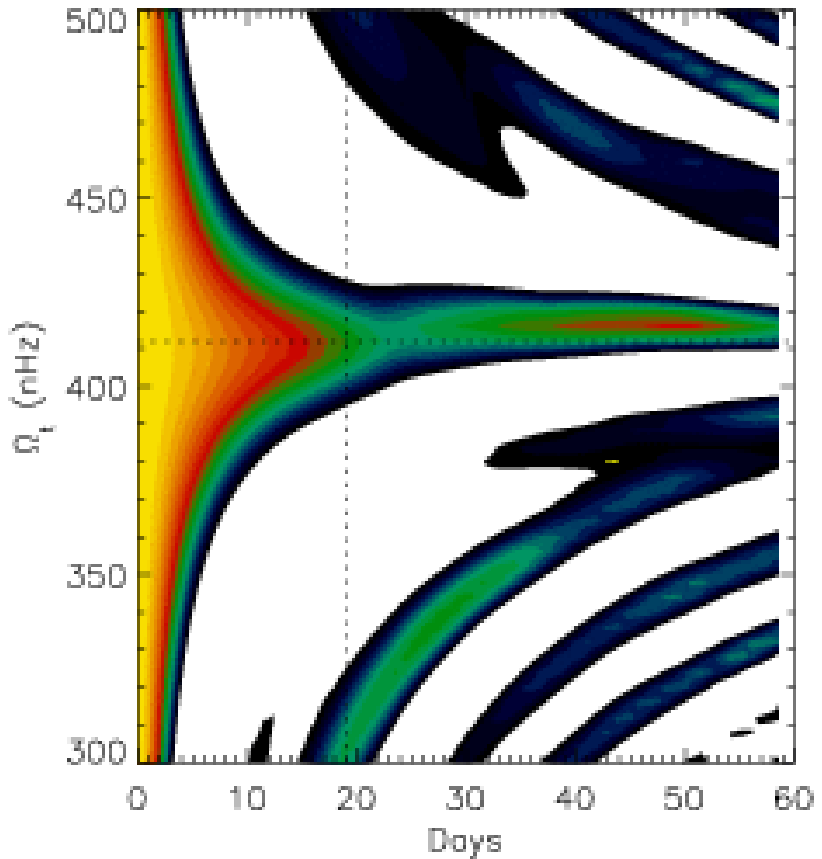}} 
\subfigure{\includegraphics[width=3.9cm]{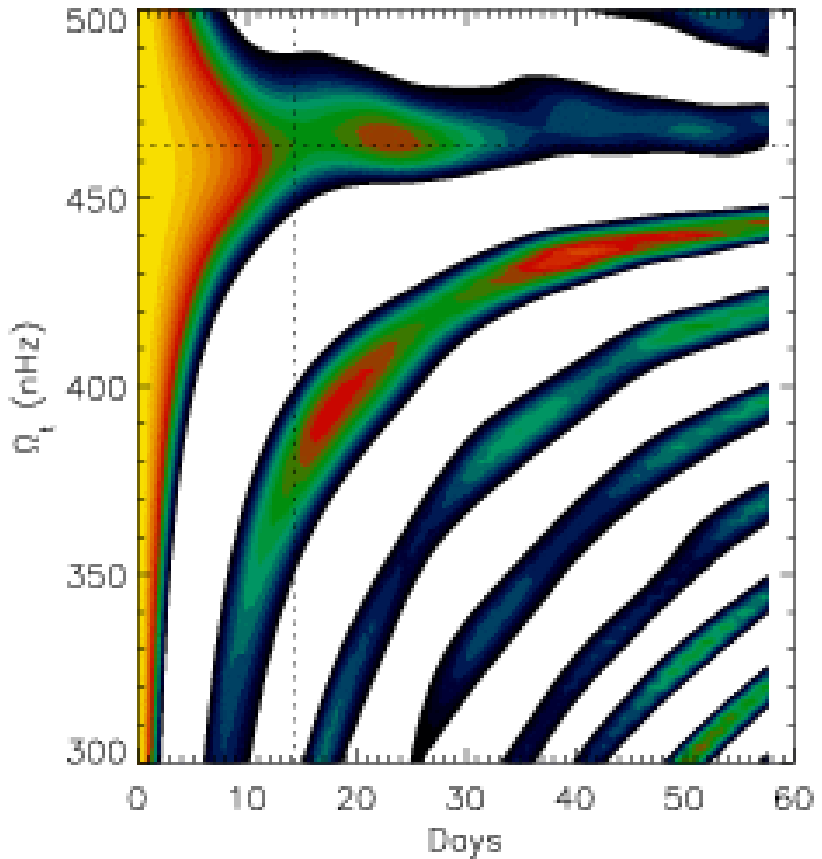}} 
\subfigure{\includegraphics[width=3.9cm]{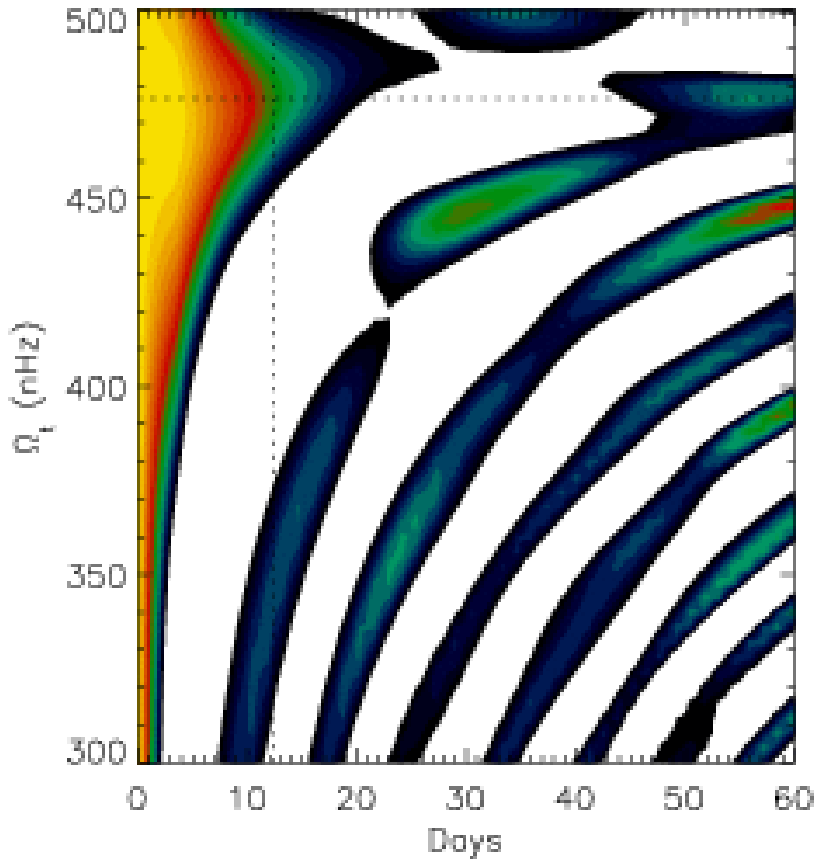}}
\caption{Autocorrelation function (acf, see Eq. 7) computed close to the surface ($r=0.97 R_*$)
and for the latitudinal band [0-20\degree] 
assuming the proper tracking rate $\Omega_t$ for each model. From left to right by increasing the 
density contrast, the acf is computed both for the full (all scales) radial velocity maps 
(upper figures) and only the largest scale (lower figures). Notice the much larger temporal 
range for the large scale analysis.}
\label{corr_vr}
\end{figure*}

We analyse here the time correlation series for all models near the stellar
surface at $r=0.97 R_*$. The result is presented in Figure \ref{corr_vr}.
The threshold for the autocorrelation function is fixed at
0.5 corresponding mainly to the transition between the red and green colors on
the different plots. For the full images composed  of the whole range of scales, 
the lifetime is very similar for our models and equal to 1.5-2 days.
This indicates that the overall density contrast of the models has little influence on the lifetime of convective cells at the surface. This result confirms that the surface convection is well 
controlled by the local density scale height which is common to all models here 
with $H_\rho=10 Mm$. 
By looking at the correlation only for the largest scale (i.e the scale 7), 
we emphasize their much greater lifetime, about a factor of 5 longer with respect 
to that of the small scale convective patterns. 
We can also remark that the other bands of the autocorrelation function visible
only on the analysis of this large scale corresponds to a "stroboscopic" effect
which appears when the chosen tracking velocity differs from the proper
angular velocity of these giant cells and is also another mean to estimate 
their mean longitudinal width. 

\begin{figure*}
\centering
   \includegraphics[width=12cm,angle=90]{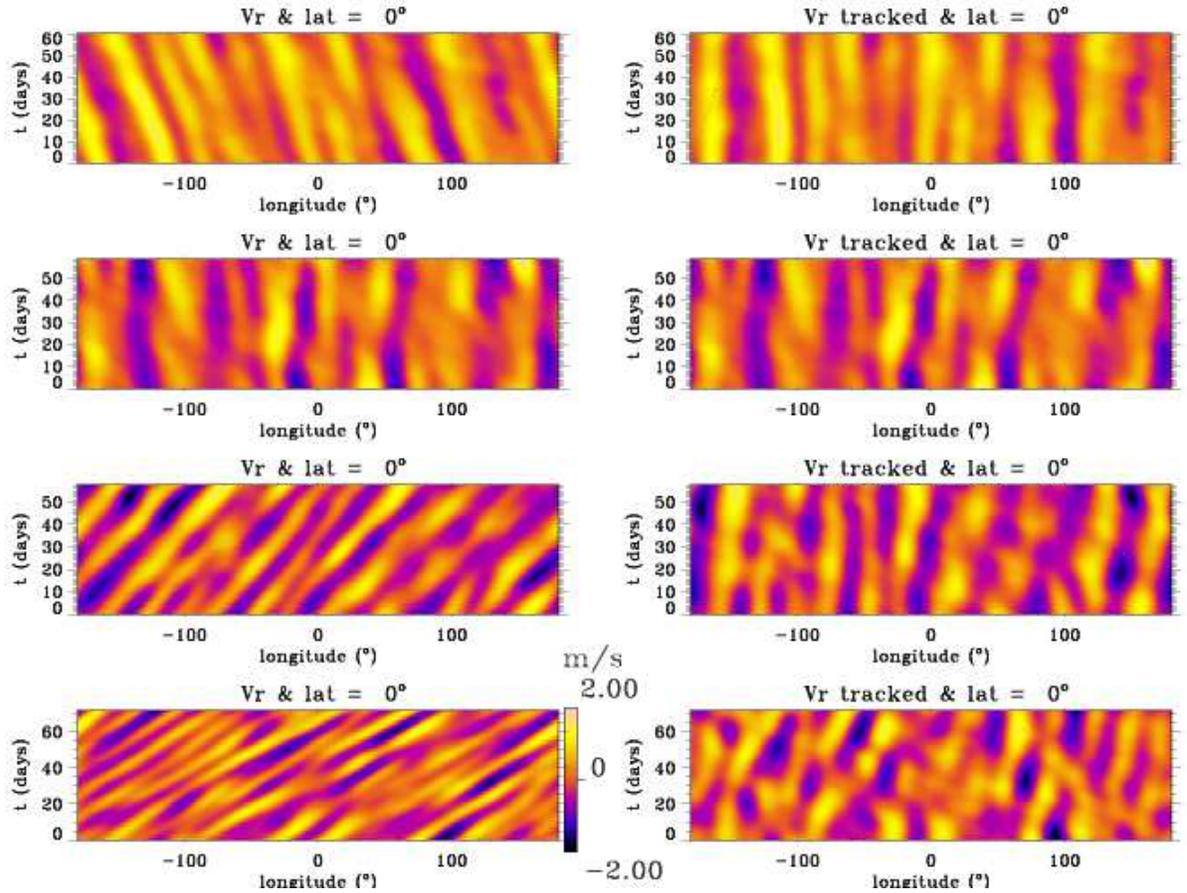}
      \caption{Time longitude diagrams at the equator and 
      close to the stellar surface ($r=0.97 R_*$) for all models with respect to the
      increasing density contrast downwards without (left) and with (right) 
      the optimal tracking rate taken from the time correlation analysis.}
       \label{longtime}
\end{figure*}

We also find that the
lifetime of the large scale structures seems to decrease with the density
contrast from $\tau=60$ days down to $\tau=12$ days. But with a closer view,
specially for the cases $\Delta \rho =37$ and $\Delta \rho =100$, there is a
modulation of the autocorrelation function with respect to the lag time translating into a 
lifetime of $\tau=50$ days if we consider a slightly lower threshold of the autocorrelation
function. 

This conclusion is confirmed by looking at the time-longitude plots
(Figure \ref{longtime}) where we can track these large scale structures in all
models at least for the sixty days studied here while remarking that they 
also undergo much distortions for the high density contrast cases which can 
entail the modulations in amplitude of the acf observed in Figure \ref{corr_vr}. These modulations can be understood by realizing that the stronger 
differential rotation achieved in the high density contrast models can shear 
more easily these large scale structures thus diminishing their
temporal correlation. We can also see that these large scale convective 
patterns have a slight stronger amplitude when 
the density contrast increases even near the stellar surface as already 
found with the previous $v_{rms}$ global method as function of depth.

\begin{figure*}
  \centering
 
\subfigure{\includegraphics[width=3.5cm,angle=90]{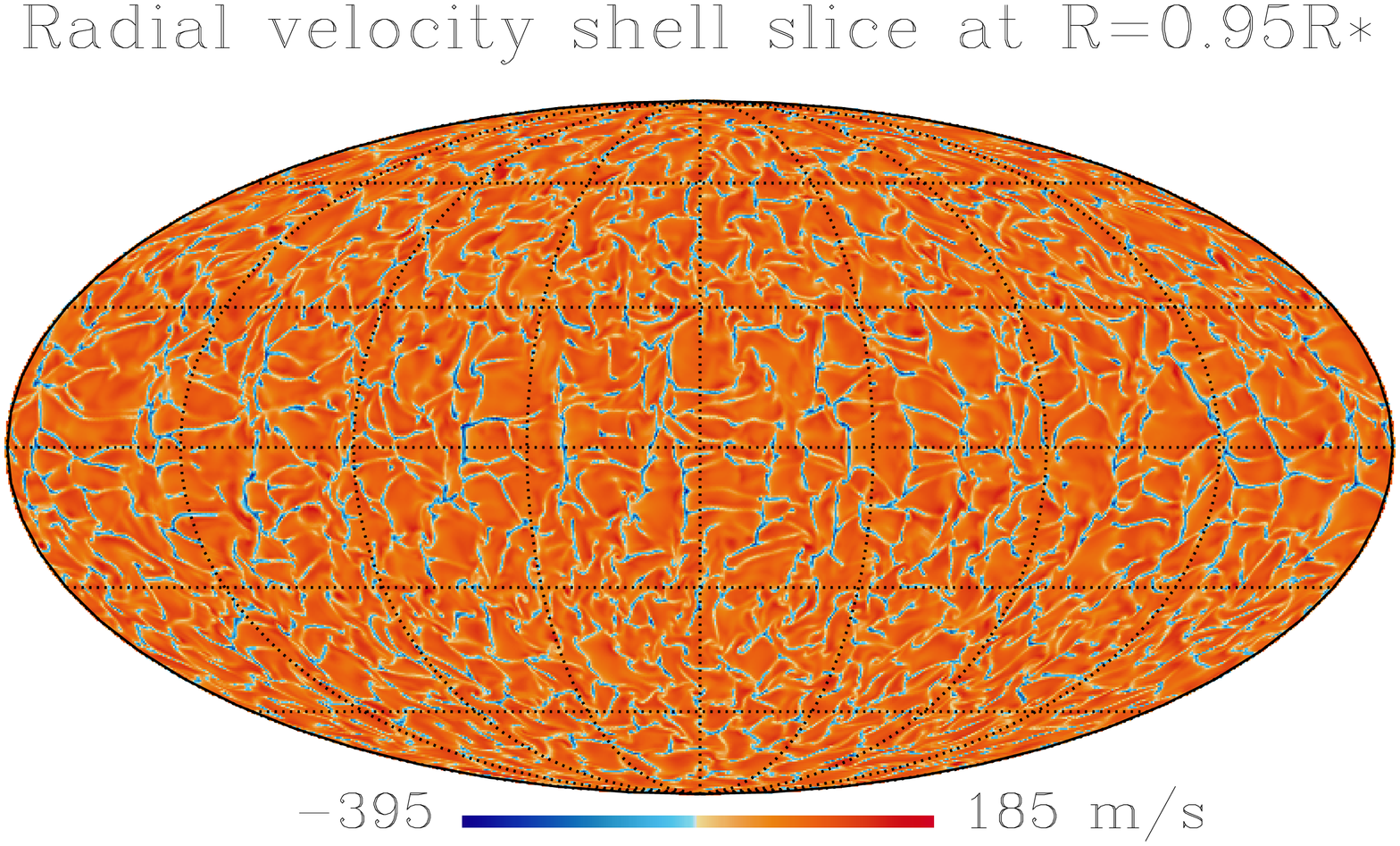}} 
\subfigure{\includegraphics[width=3.5cm,angle=90]{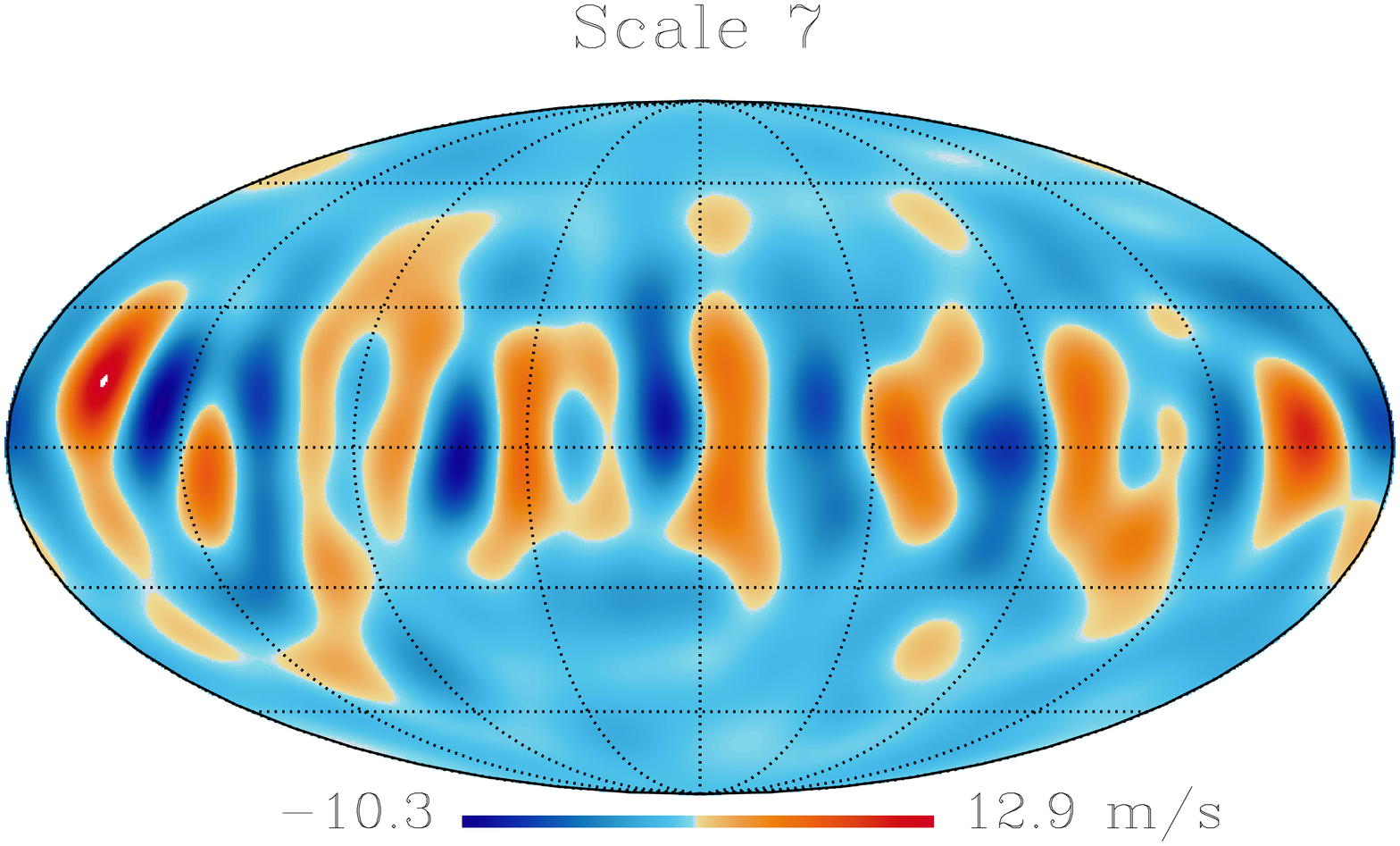}} 
\subfigure{\includegraphics[width=3.5cm]{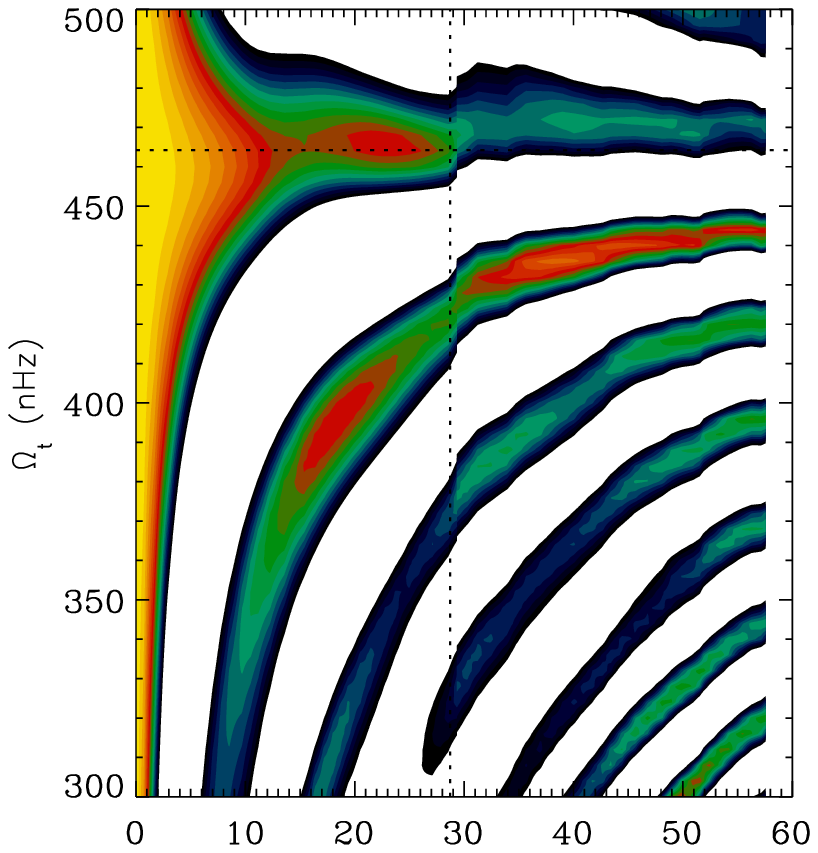}} 
\subfigure{\includegraphics[width=3.5cm,angle=90]{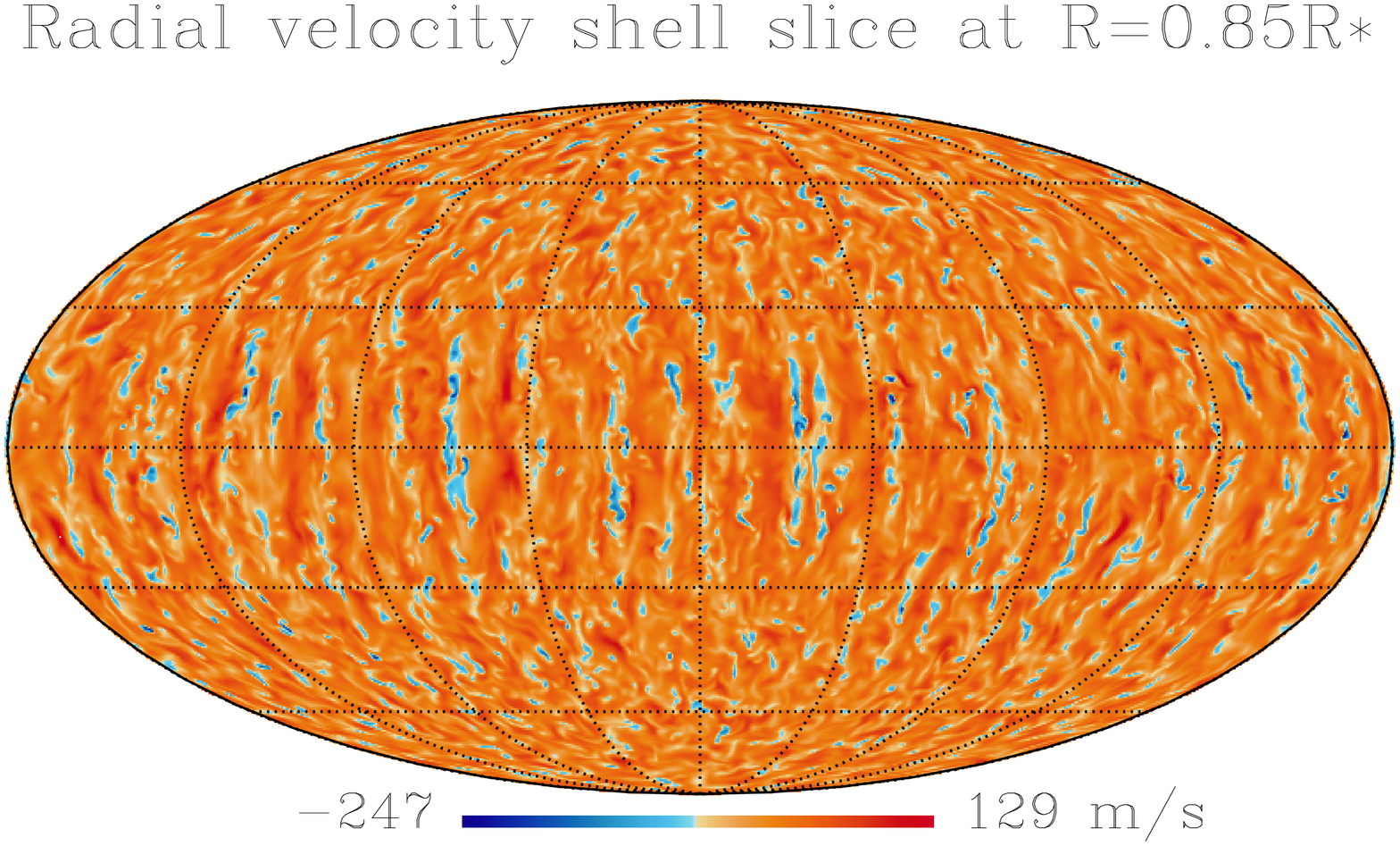}} 
\subfigure{\includegraphics[width=3.5cm,angle=90]{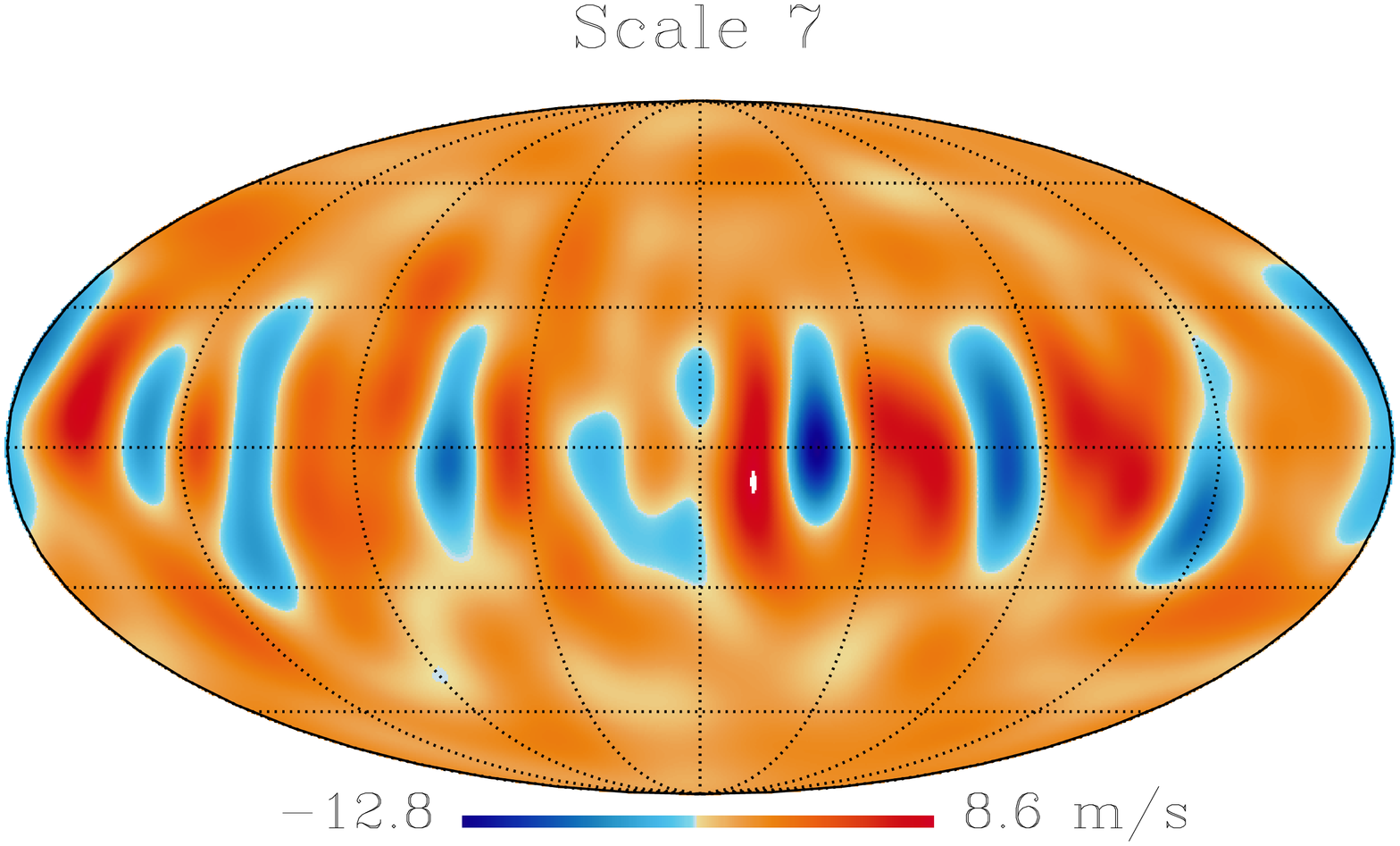}} 
\subfigure{\includegraphics[width=3.5cm]{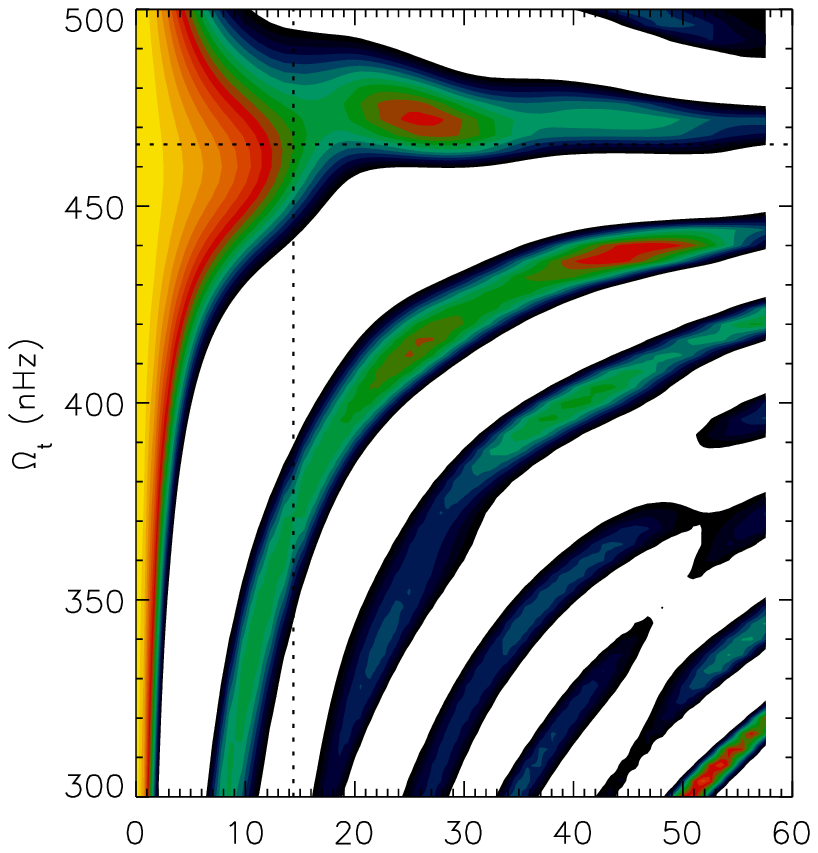}} 
\subfigure{\includegraphics[width=3.5cm,angle=90]{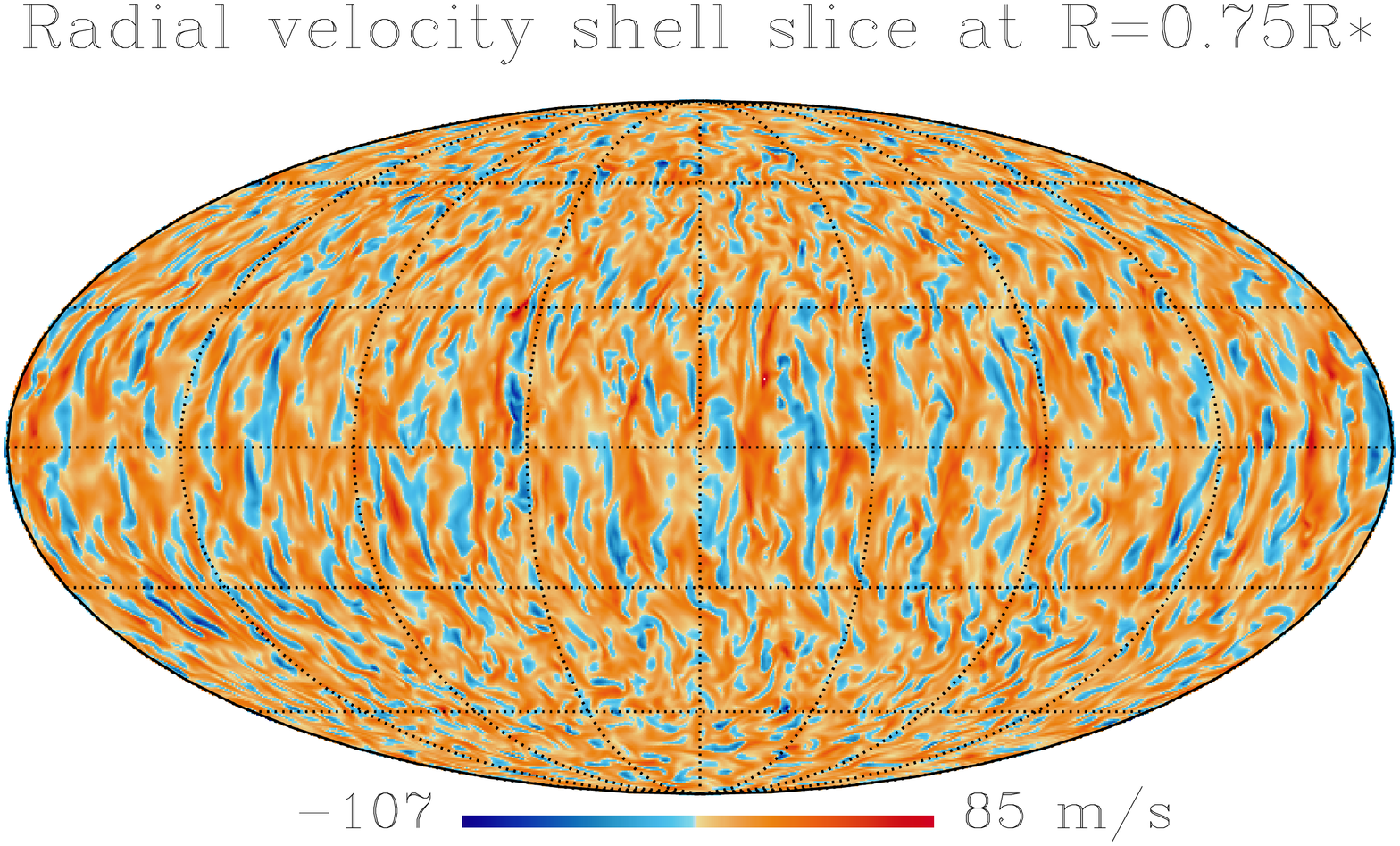}} 
\subfigure{\includegraphics[width=3.5cm,angle=90]{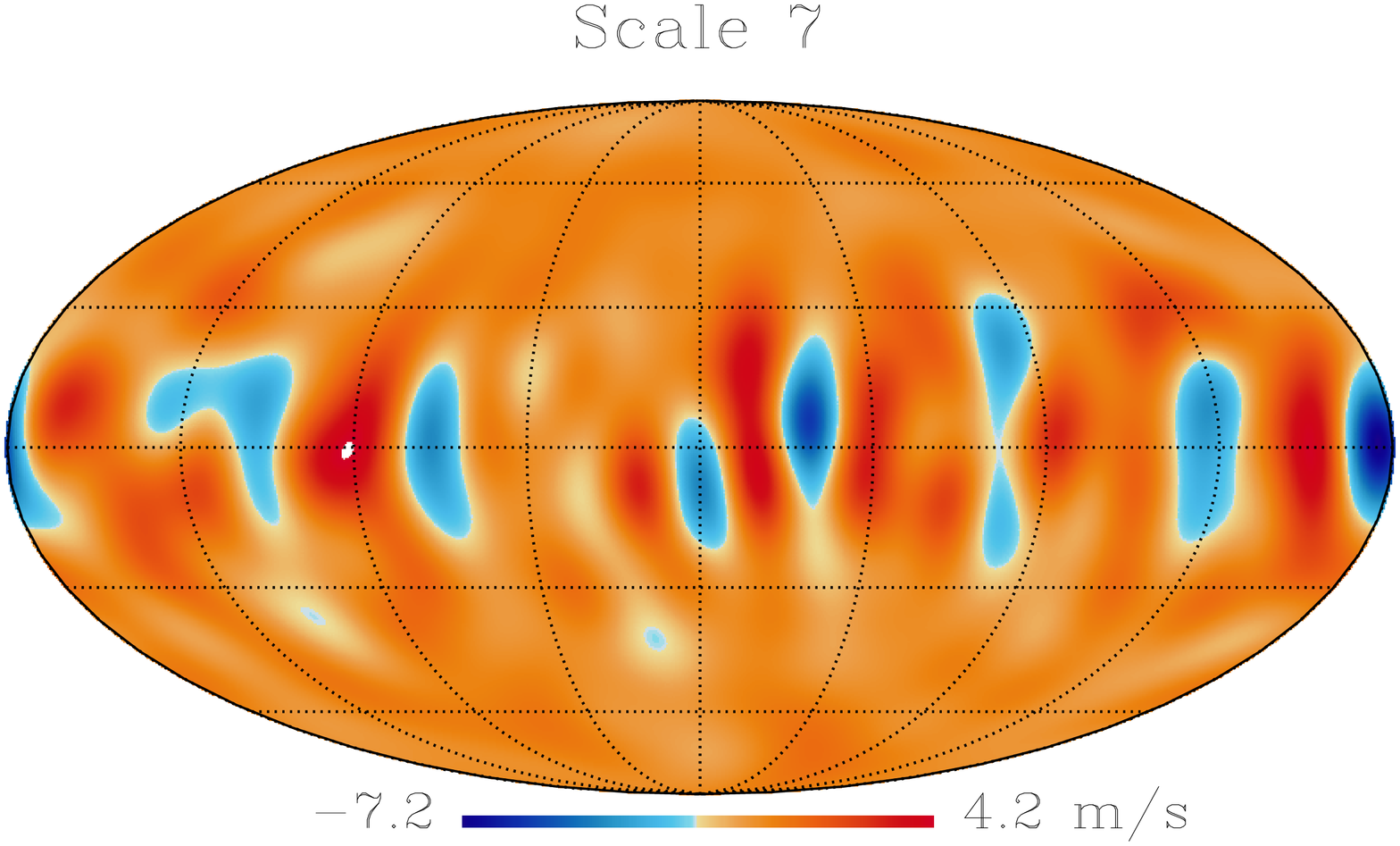}} 
\subfigure{\includegraphics[width=3.5cm]{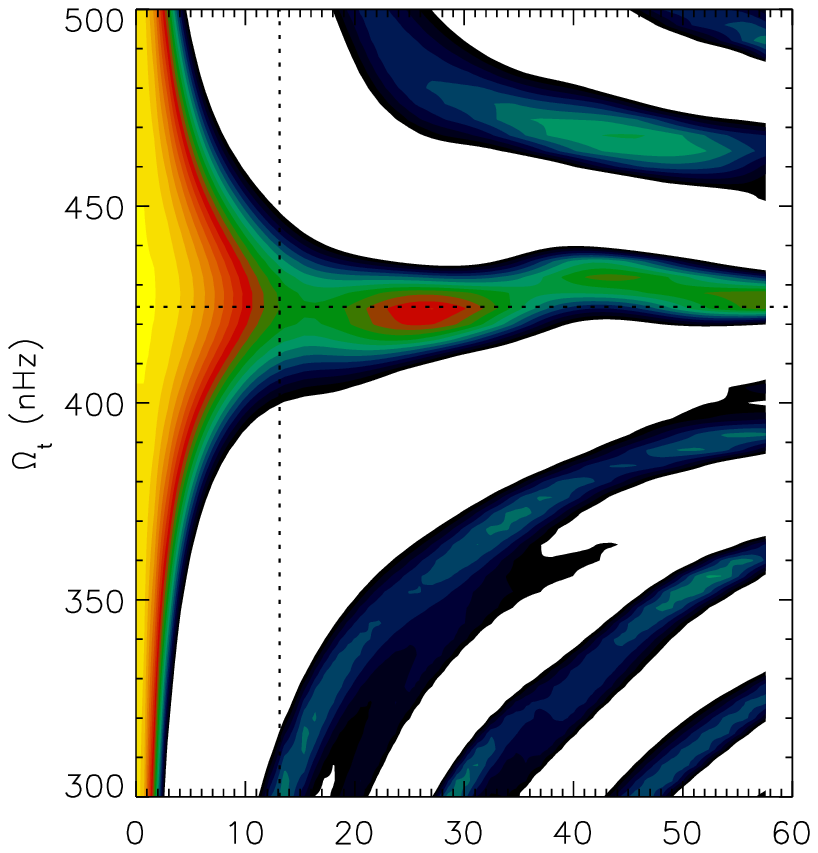}} 
\caption{Radial velocity map both for the full image and only the largest scale 7
deeper in the convective zone at $r=0.95 R_*$, $r=0.85 R_*$ and $r=0.75 R_*$
for the model $\Delta \rho=100$ and the linked
autocorrelation function for the latitudinal band [0-20\degree] 
corresponding to the largest scale.}
\label{corr_depth}
\end{figure*}

Finally, we can have a look at the evolution of the lifetime of these large
structures with respect to depth. Figure \ref{corr_depth} show how the 
flow is structured both at small and large scale for three different depths
 in the $\Delta
\rho=100$ model based on the autocorrellation analysis of the largest
scale. The convective patterns in these deeper shells are mainly dominated by 
the complex distribution of the downflow lanes which have a north-south
orientation and we observe fine structures in the downflow plumes which
penetrate throughout the entire convective zone. We can see that the signal at 
large scale is much stronger than near the stellar surface (see for comparison
 Figure \ref{wt_decompos}) by a factor of 5. We can see
that the lifetime of these large scale patterns increased at $r=0.95 R_*$
showing that the flow is less disturbed by the smaller scales at this depth with
respect to the stellar surface. Otherwise, the optimal tracking rotation rate
obtained for these large scale is the same at each depth except at the deepest one. 
This confirms that their spatio-temporal coherence is maximal from $r=0.98 R_*$ 
down to $r=0.85 R_*$ in this case. Thus, these giant cells seem to propagate at 
their own rate. The spatial correlation in depth becomes
less strong below $r=0.85 R_*$ although the amplitude of the large scale
signal is maximal at this depth as seen in Figure \ref{wt_vrms}.
Actually, we remark on Figure \ref{corr_depth} that at deeper 
depth, the rotation rate is really different indicating that giant cells are not
extended over the whole depth of the convective zone for the deepest case model. One can
understand this behaviour by the fact that the depth $r=0.85 R_*$ corresponds to
the transition between two consecutive giant cells in the radial direction.     
The same behaviour in depth is found for the $\Delta \rho=272$ model whereas
 the giant cells fill in radially the whole depth of the convective zone in the models with
lower density contrast. We thus deduce that giant cells have a radial extension of 0.13 $R_*$
from this analysis.

\subsection{Influence of mean flows and aspect ratio on resulting convective
patterns}

\begin{table}[!h]
\begin{center}
\begin{tabular}{|c|c|c|c|}
\hline
\hline
  $\Delta \rho $ & $\Omega_{diff} $ &  $\Omega$  &  $\Omega_{sc7}$  \\
               & $(nHz)$ & $(nHz)$   &  $(nHz)$  \\
\hline
13 & 378-390 & 391 & 397 \\ 
37 & 378-391 & 390 & 420 \\
100 & 424-430 & 440 & 472  \\ 
272 & 447  & 462 &   485 \\ 
\hline
\end{tabular}
\caption{Differential stellar rotation rates and optimal tracking rates of 
both the small-scale convection and the large scale structures near the 
stellar surface ($r=0.97R_*$) and at low latitudes [0\degree -20 \degree] 
for all models. The models studied here have a mean rotating rate corresponding
to $\Omega_0=414$ nHz.}
\label{omega_prop}
\end{center}
\end{table}

It is interesting to notice that the stronger the density contrast is,
the quicker these giant cells move with respect to the mean stellar differential
rotation of each model. This is illustrated in Table \ref{omega_prop}. 
The small scale structure of convection corresponds to grossly the superganulation scale 
in our simulations, and has also a slightly greater rotation rate with respect
 to the local differential rotation.
This is a well known characteristics as already observed in simulations by
Miesch et al. (2008) and in solar data (see Meunier et al. 2006). However, giant
cells have a very different rate. In most case, they are prograde with respect
to both the global rate and the local differential rotation. For the models 
$\Delta \rho=13,37$, it is striking to notice that they move in a 
counterstreaming direction.   

The latitudinal extension of the large scale structures can also be compared between all models 
discussed in this study, to
see the influence of the domain aspect ratio  which varies between
2 and 14 as reported in Table \ref{sim_prop} and provides an insight on the
3D shape of these cells. An important latitudinal angle is
also reported in Table 1 corresponding to the tangent cylinder for all models, i.e.
the imaginary cylinder tangent to the inner boundary which intersects the outer
boundary sphere at a given colatitude.
First, the graphical analysis of the large scale structures visible from the
scale 7 of our wavelet analysis shows that the longitudinal width of the giant 
cells is nearly constant around 40$\degree$ for all models as the multi-branch of
acf seen in Figure \ref{corr_vr} and  \ref{corr_depth} reveals. It is also important
to notice that the longitudinal extension
is much larger than wavelenght deduced from the number of downflow lanes visible
 at equator on Figure \ref{conv_T_patterns} (more than 20 lanes at equator). 
As far as their latitudinal extent is concerned, there is a clear increase of it
with  the density contrast. Indeed, these cells are well
focused near the equator without reaching latitudes beyond 30$\degree$ for the $\Delta
\rho=13$ case whereas there are extended to high latitudes ($> 60$\degree) for the $\Delta
\rho=272$ case. We thus see that giant cells have different radial and horizontal aspect ratio 
depending on the depth of the convective zone they are embedded into.

\begin{figure}[!h]
\centering
\hspace{-0.7cm}
   \includegraphics[width=8.2cm,angle=0]{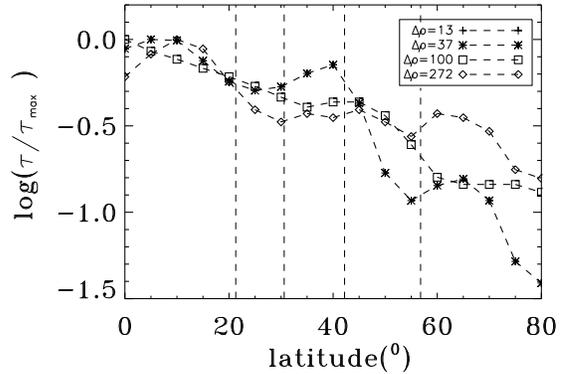}
      \caption{Evolution of the normalized lifetime of giant cells according to
      the latitude. The dashed lines corresponds to the different tangent
      cylinder latitudes as listed in Table \ref{sim_prop}.}
       \label{lat_tau}
\end{figure}

The evolution of the large scale structures lifetime normalized
 to the maximal lifetime found for each model versus the latitude is shown in
Figure \ref{lat_tau}.  For each model, there is a general trend for a steep
decrease of the timescale towards high latitudes which corresponds to the
position of the proper tangent cylinder. However, we can notice that
this decrease is less pronounced when the density contrast increases and this
trend is the strongest at latitudes greater than
$50 \degree$. So the giant cells are quite extended latitudinally for deep
stellar convective zone. Above these clipping latitudes, the giant cells seem 
to be disconnected and correspond to proper polar convection cells as seen in
3D-rendering shown in Figure 4. 

\begin{figure}[!h]
\centering
\subfigure{\includegraphics[width=8cm]{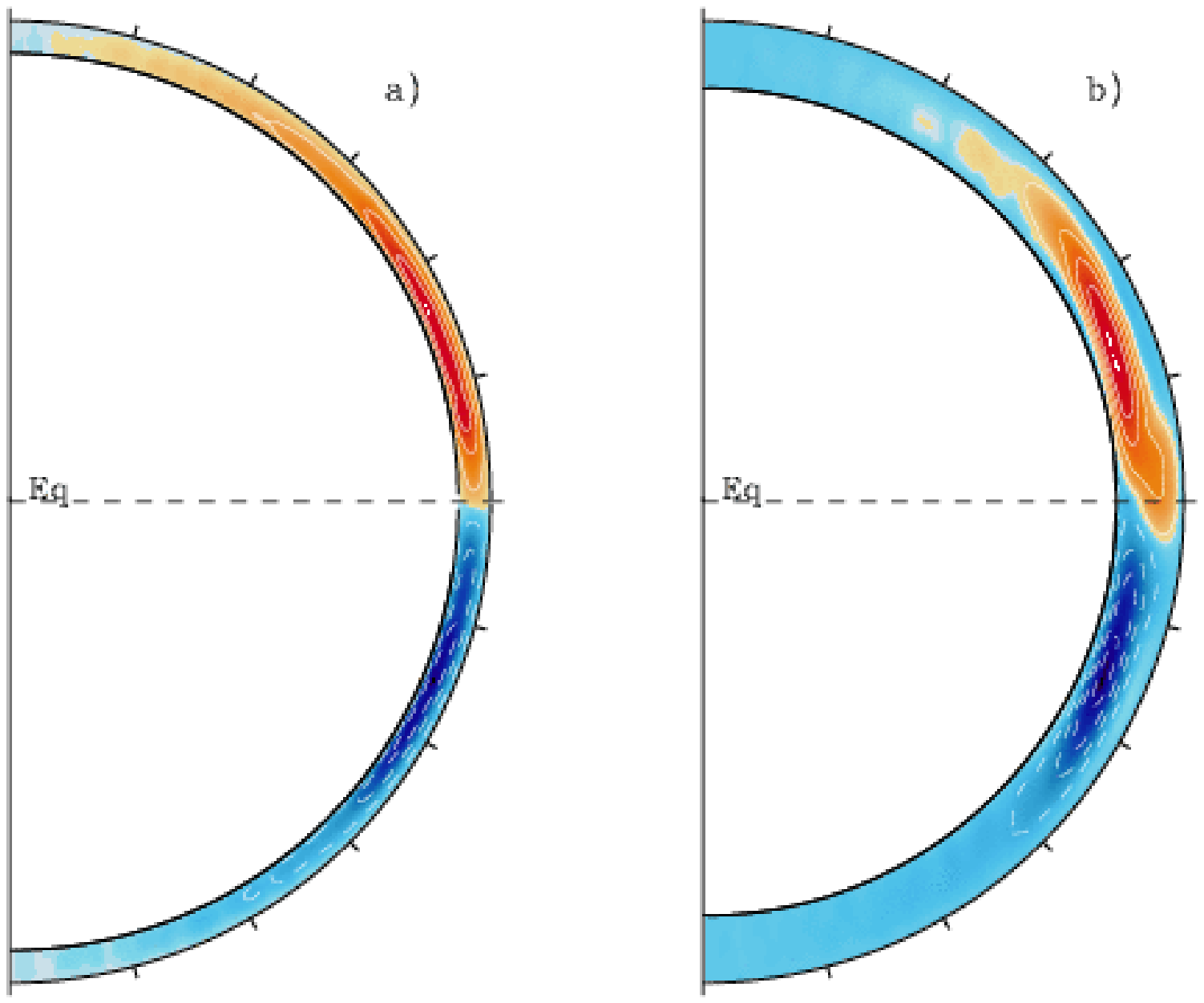}} 
\subfigure{\includegraphics[width=8cm]{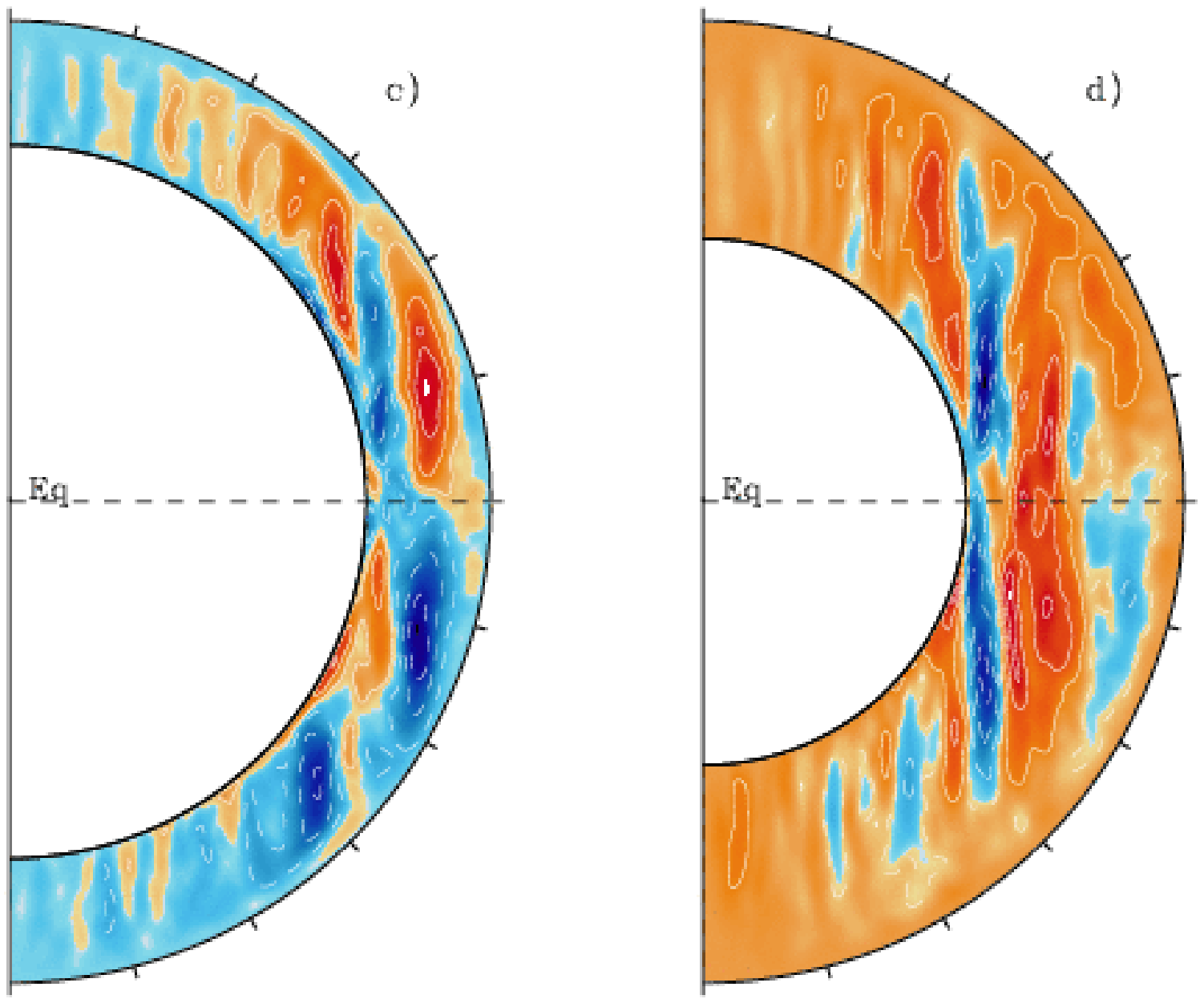}} 
\caption{Meridional circulations for the different models.}
\label{MC}
\end{figure}

The spatial coherence of these giant cells at different depths 
is also studied by cross-correlating the different shell slices of each model.
One finds that the $\Delta \rho=100$ and $\Delta \rho=272$ 
models show a decrease of this coherence for $r<0.85R_*$ contrary to the lower 
density cases as already noticed in Figure \ref{corr_depth} for the $\Delta \rho=100$ 
model. The meridional circulations profiles of theses models 
(see Figure \ref{MC}) give us a hint to understand this property. Whereas the
 low density contrast cases (panels (a) and (b)) present circulations extended
  over the whole depth of the modelized convective zone, larger density cases
  (panels (c) and (d)) present several circulations
cells in the radial direction at low latitudes. However, in these cases, one 
finds thicker circulation cells filling most of the depth that are aligned with 
the rotation axis at larger latitudes corresponding to polar convection cells. 
It is important to stress that the 
meridional circulation profile gives only an indication of the radial extension
since the giant cells are obviously non axisymmetric.   

We can summarize our findings in Table \ref{gc_prop} giving the estimate mean 
aspect ratio of the giant cells for each model by distinguishing their 
properties according to the latitudinal range for the high density contrast models.
We find that they are quite extended horizontally at low latitudes 
($L < R_{\Delta \theta}, R_{\Delta \phi}$) and extend to $\approx 110$Mm in depth 
except at high latitudes for the thicker cases.  

\begin{table}[!h]
\begin{center}
\begin{tabular}{|c|c|c|c|}
\hline
\hline
  $\Delta \rho $ & $R_{\Delta \theta}/R_{\Delta \phi}$ &  $R_{\Delta
  \phi}/{L}$  &  $  R_{\Delta \theta}/ L $  \\
\hline
13 & 0.75 & 7.7 & 5.8 \\ 
37 & 1.25 & 4.2 & 5.2  \\
100 low lat. & 1.5 & 4.5 &  6.8  \\ 
100 high lat. & 1  & 0.5 & 0.5   \\ 
272 low lat.&  2 & 4 & 8.  \\
272 high lat.  & 1  & 0.26 & 0.26  \\ 
\hline
\end{tabular}
\caption{Different mean aspect ratio characterizing the 3D structure of giant cells
for the different models. $R_{\Delta \theta}$, $R_{\Delta \phi}$ and $\Delta
R$ correspond respectively to the latitudinal, longitudinal and radial
extension.}
\label{gc_prop}
\end{center}
\end{table}
  
\subsection{Transport properties of giant cells}

\begin{figure*}
  \centering
 
\subfigure{\includegraphics[width=3.2cm,angle=90]{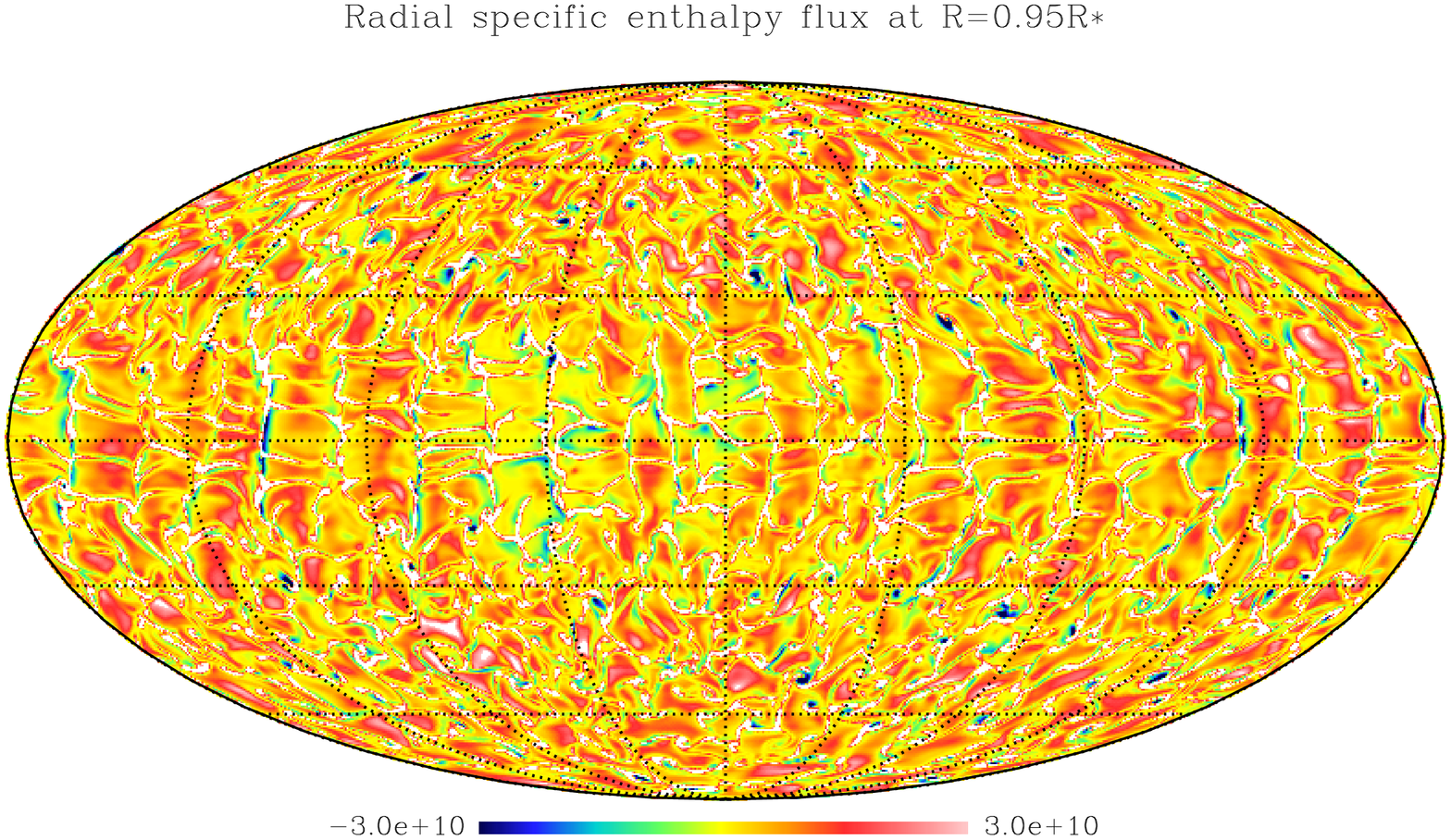}} 
\subfigure{\includegraphics[width=3.2cm,angle=90]{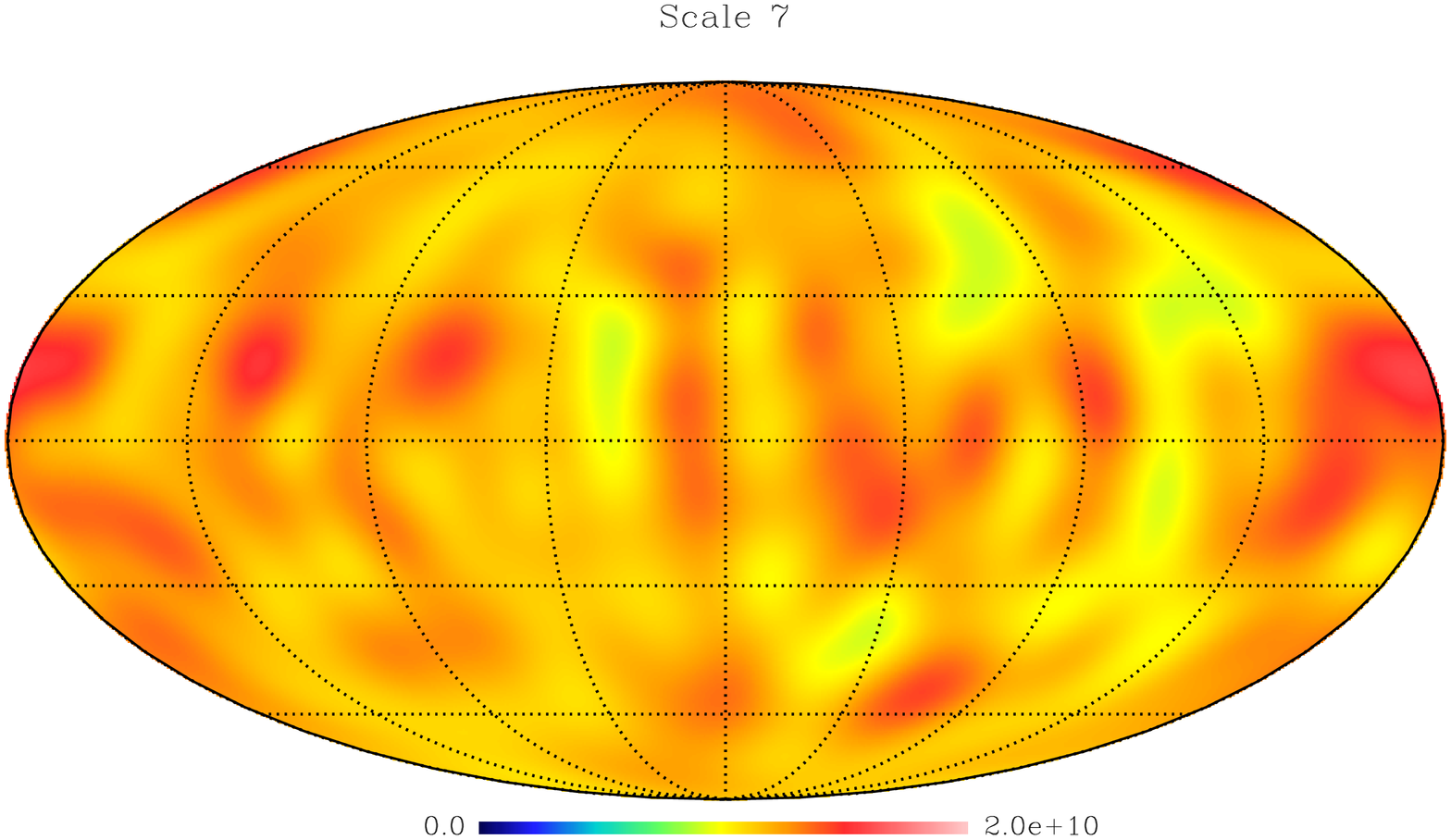}} 
\subfigure{\includegraphics[width=3.2cm,angle=90]{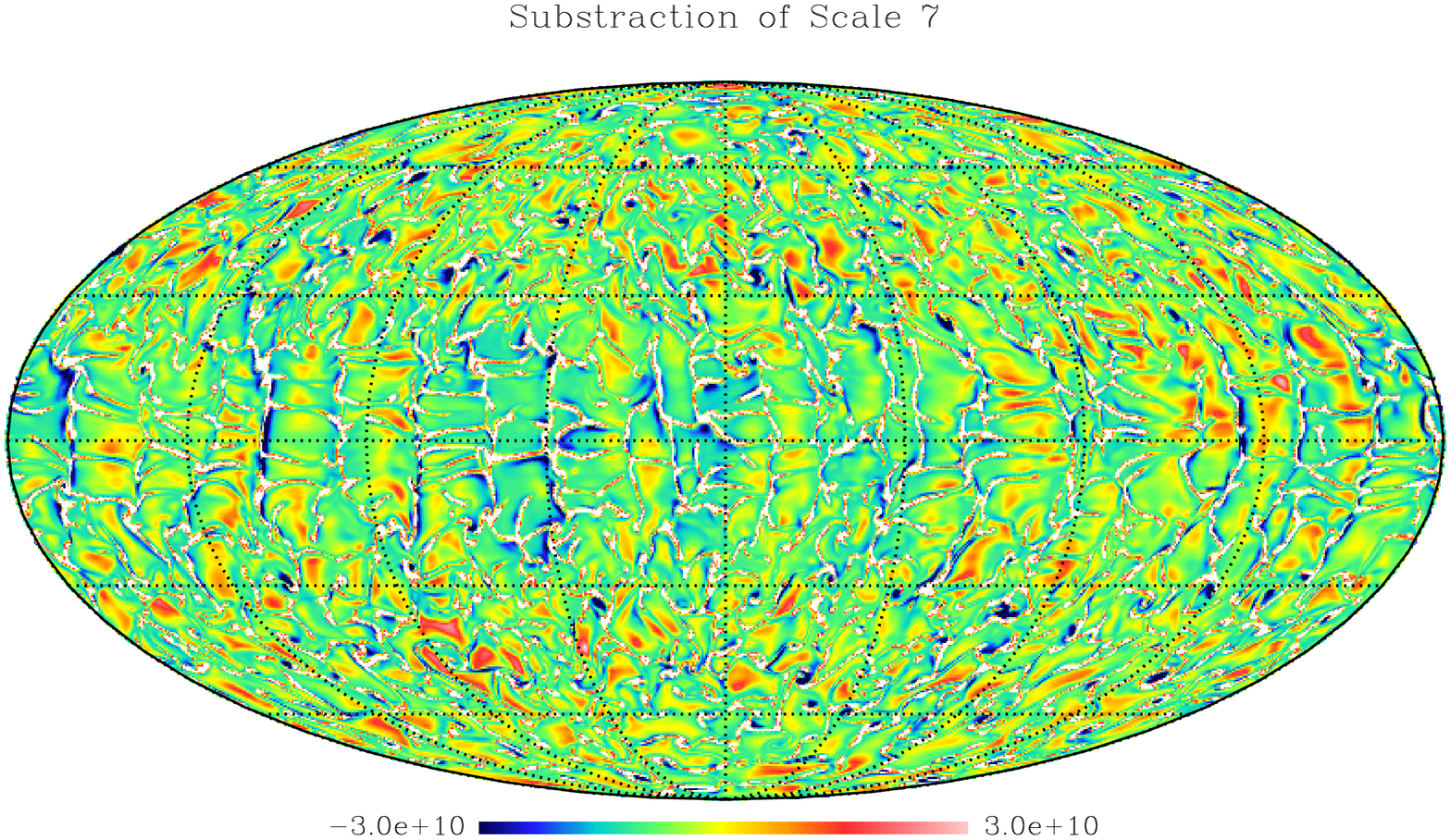}} 
\subfigure{\includegraphics[width=3.2cm,angle=90]{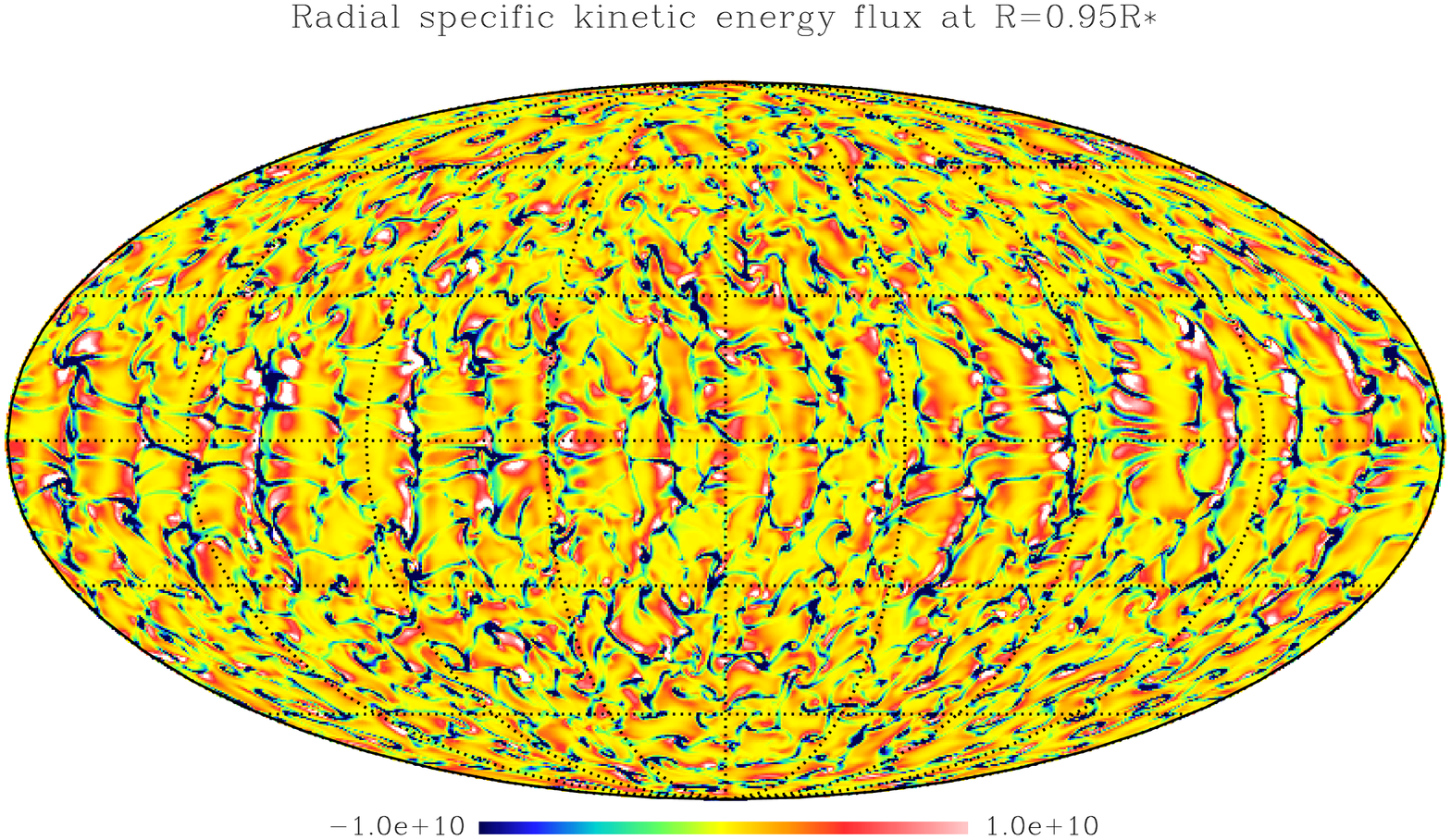}} 
\subfigure{\includegraphics[width=3.2cm,angle=90]{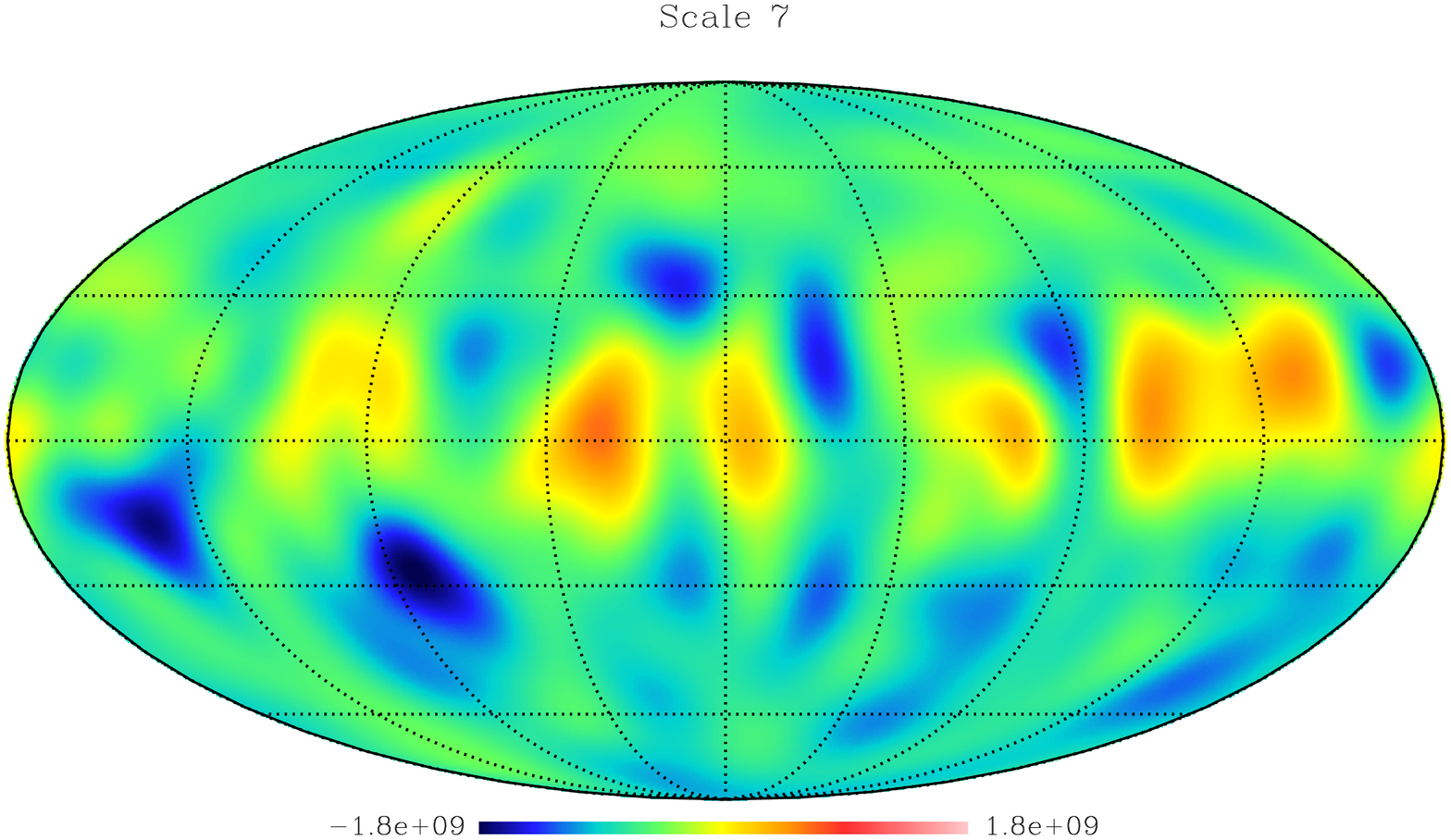}} 
\subfigure{\includegraphics[width=3.2cm,angle=90]{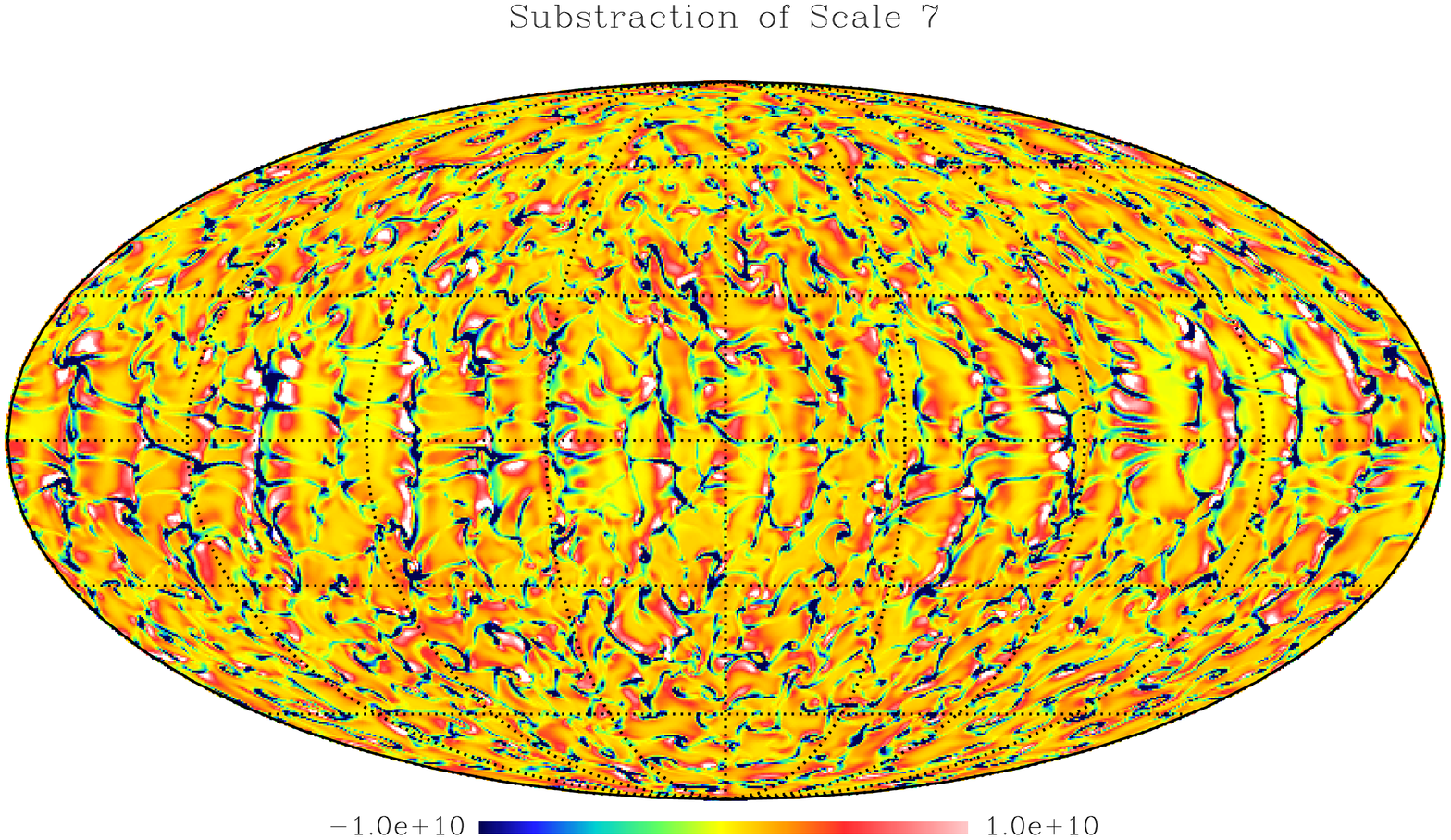}} 
\subfigure{\includegraphics[width=3.2cm,angle=90]{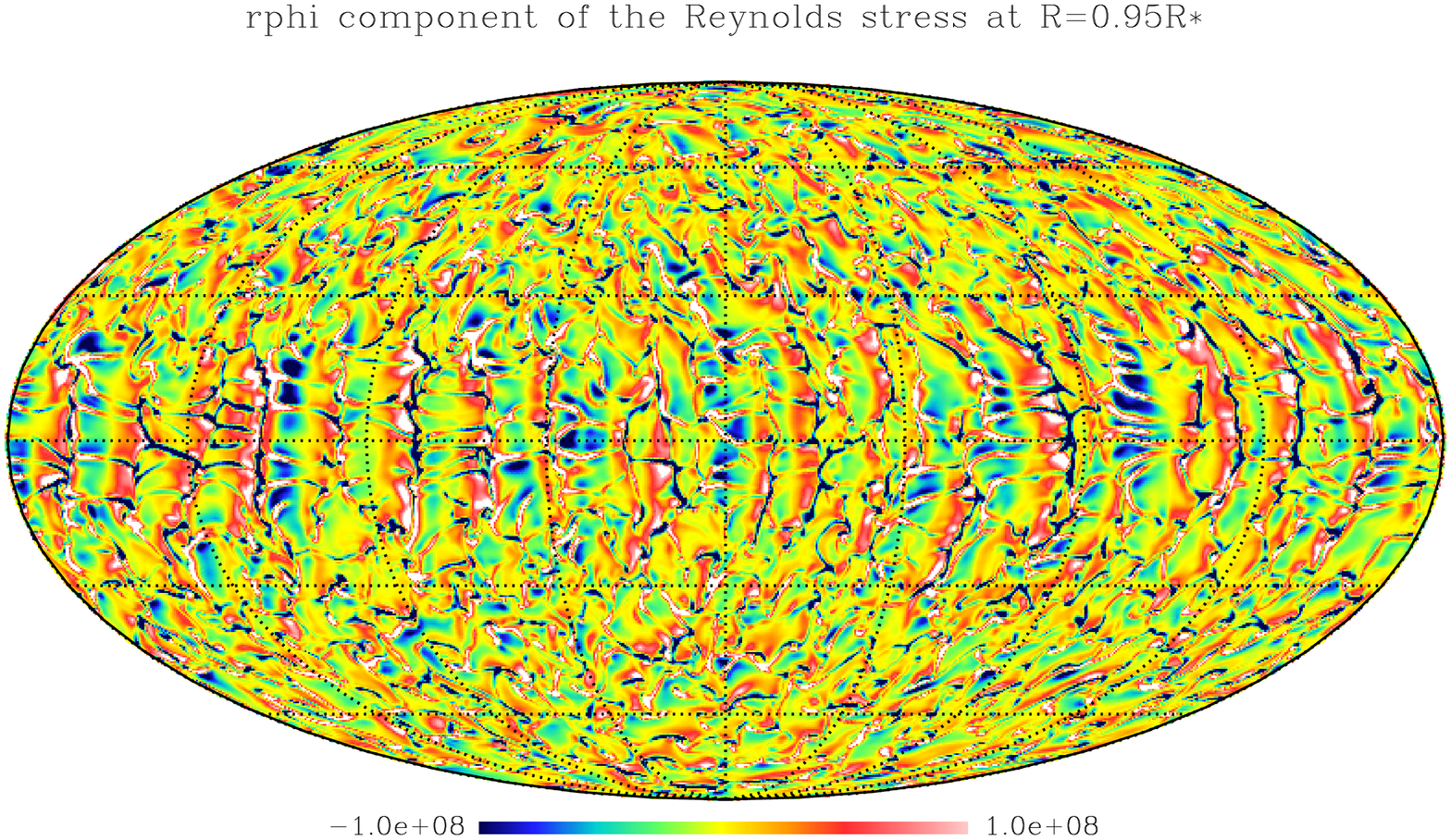}} 
\subfigure{\includegraphics[width=3.2cm,angle=90]{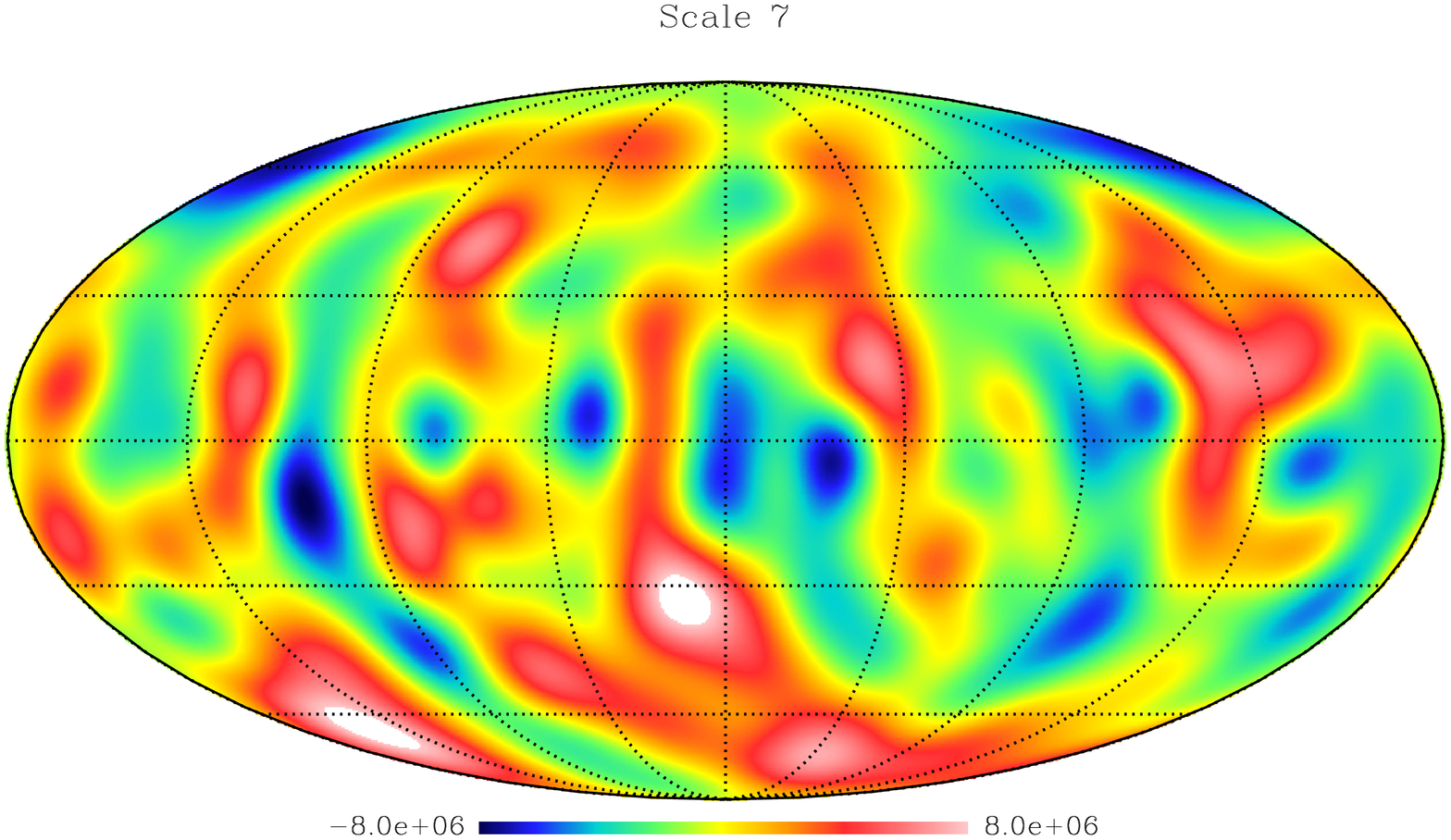}} 
\subfigure{\includegraphics[width=3.2cm,angle=90]{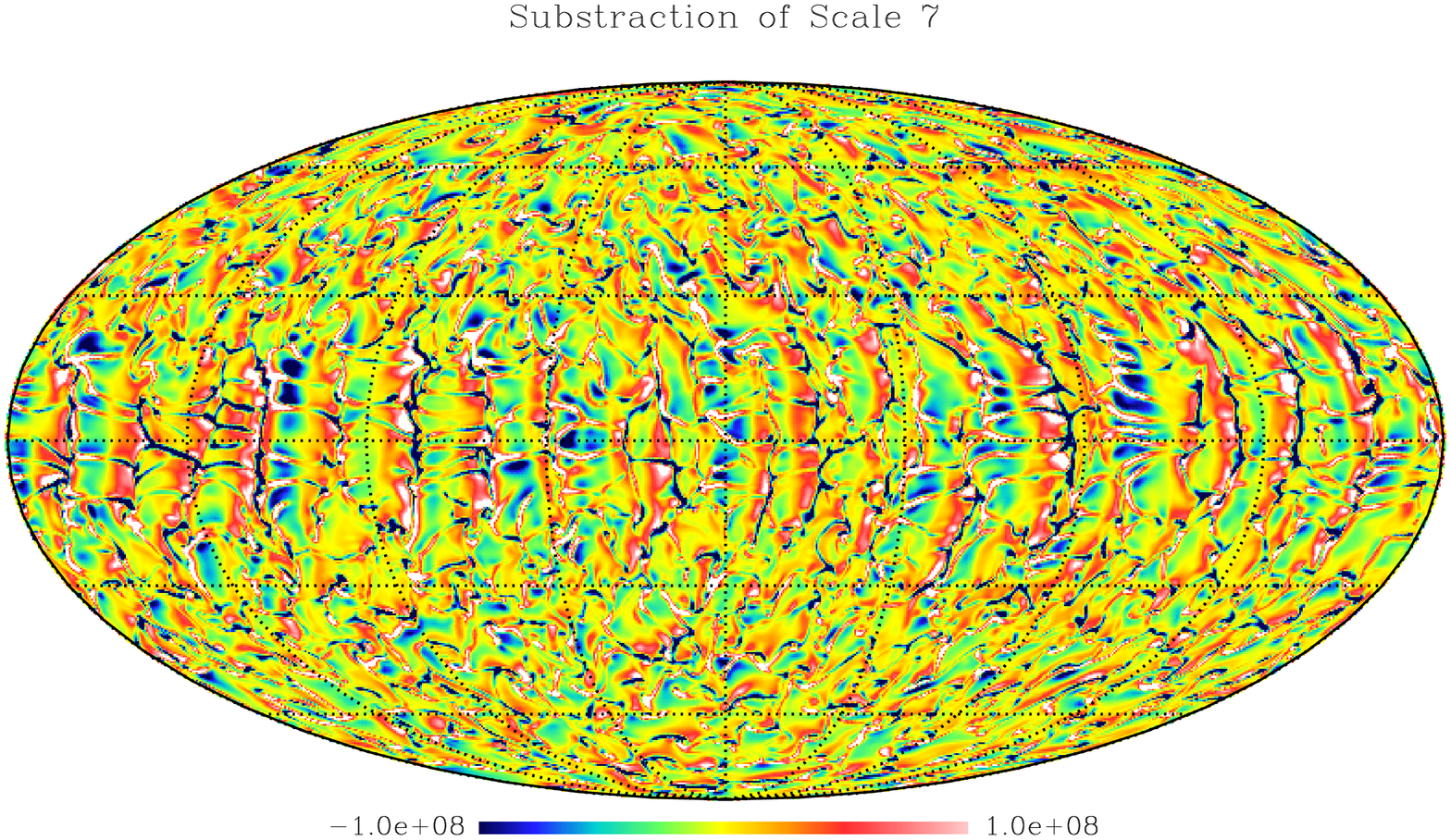}} 
\subfigure{\includegraphics[width=3.2cm,angle=90]{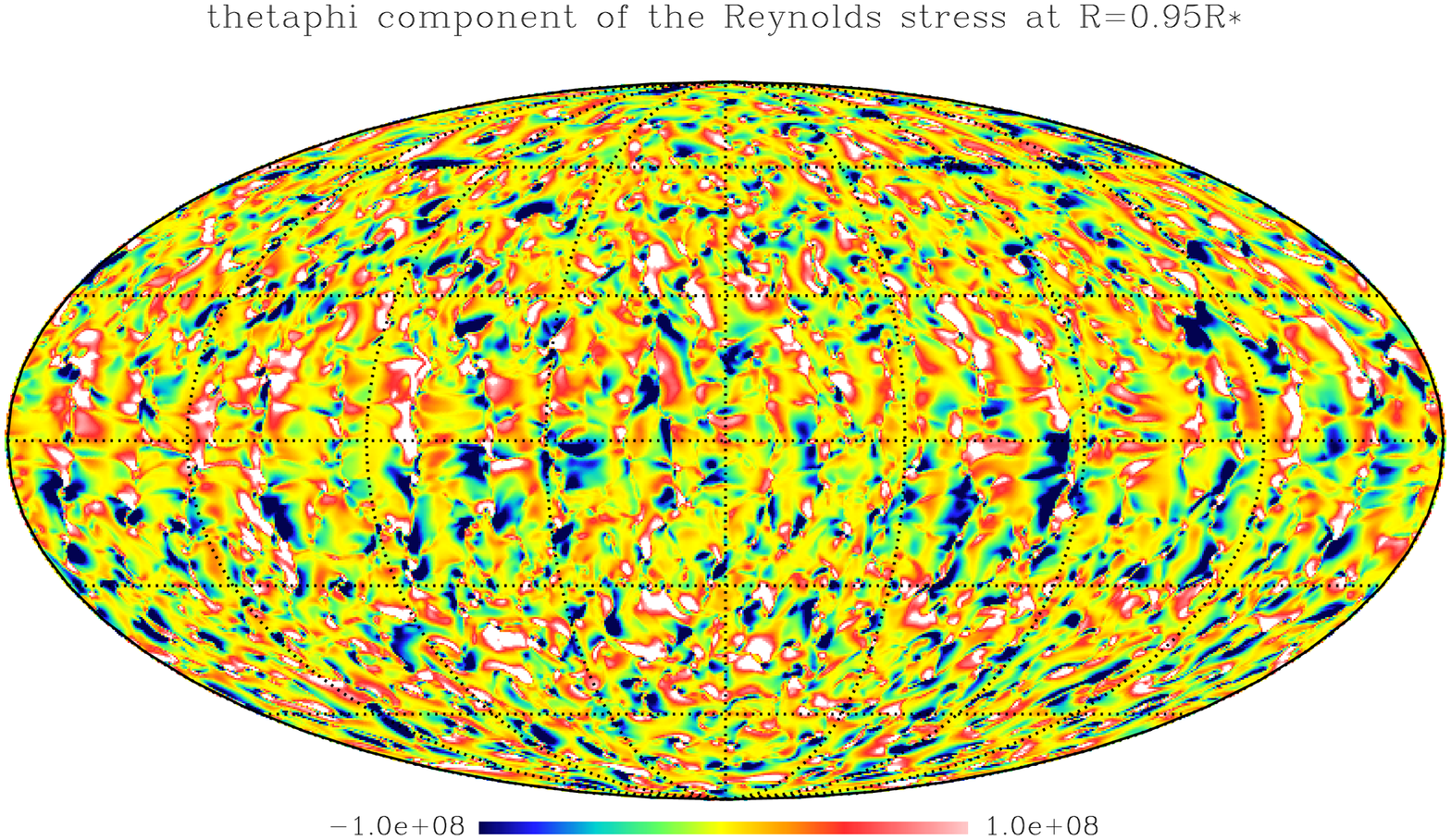}} 
\subfigure{\includegraphics[width=3.2cm,angle=90]{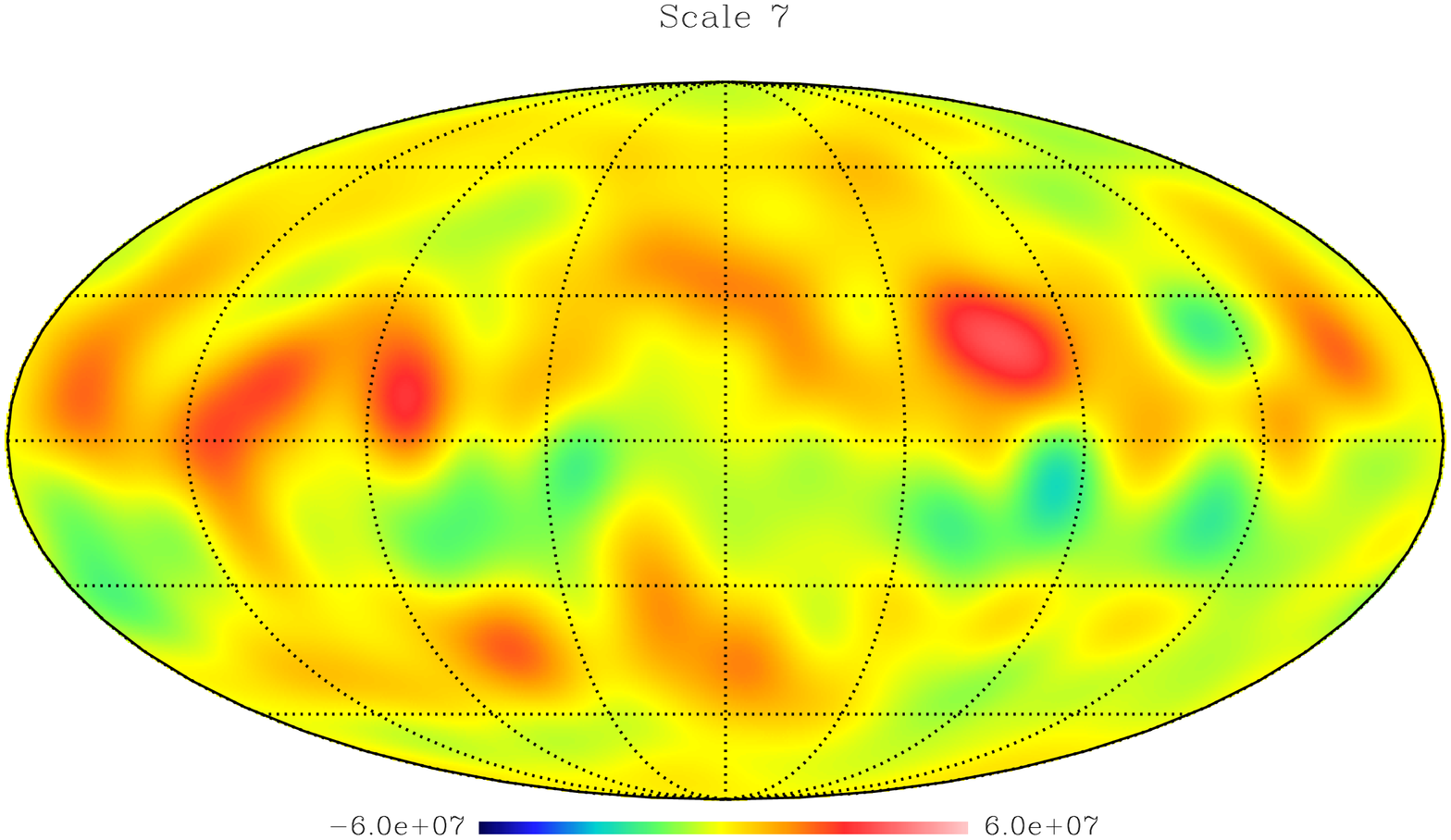}} 
\subfigure{\includegraphics[width=3.2cm,angle=90]{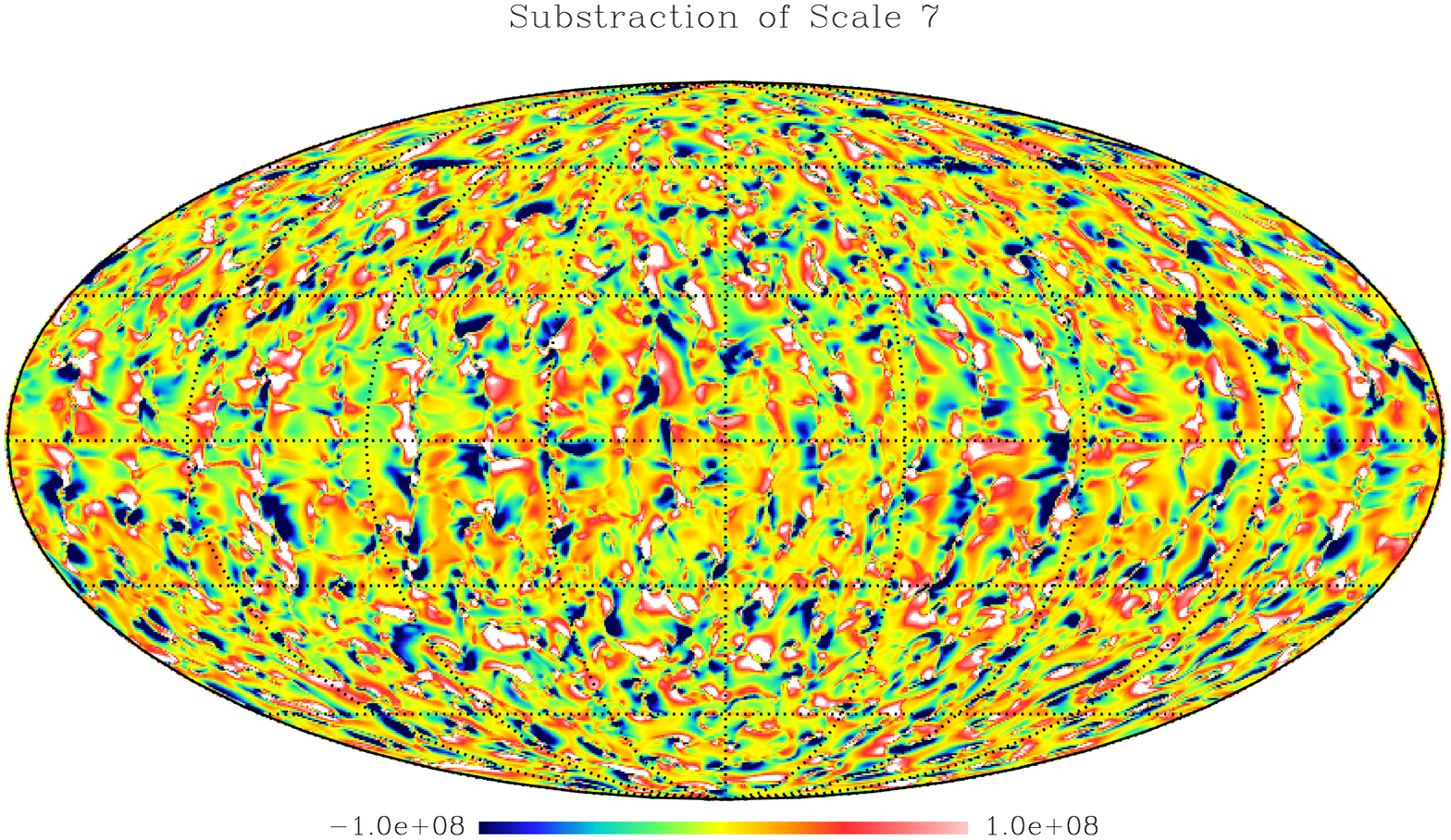}} 
\caption{Specific radial enthalpy and kinetic energy fluxes, and components of the Reynolds stress involved in the angular momentum
fluxes transported both by the entire flows (left) and only by the largest scales (middle) for the $\Delta\rho=272$
model. The right images correspond to the substraction of Scale 7 in the left ones.}
\label{transport_sc7}
\end{figure*}

Using our wavelet analysis pipeline, we can see how these giant cells transport
enthalpy, kinetic energy and angular momentum with respect to smaller scales.
Figure \ref{transport_sc7} shows such an analysis for the specific radial enthalpy, radial kinetic energy and angular momentum
fluxes transported both by the entire flows and only by the largest scales in
 the $\Delta\rho=272$ model. We focus here at one depth corresponding to $r=0.95 R_*$ to have a
 better contrast since for instance the radial specific enthalpy flux tends towards zero near the
 surface due to the chosen unpenetrable boundary conditions. The different
 quantities are computed taking into account only the fluctuating components
 as defined in Brun \& Palacios(2009).  

First, we find that the enthalpy transport by giant cells is always directed 
outwards at scale 7. This represents about 3\% of the total enthalpy flux whereas strong negative enthalpy flux are
found at the edges of convective cells at smaller scales.    

Strong negative kinetic energy fluxes are localized in the fine structures of the 
downflows between the convective cells whereas positive kinetic energy fluxes
correspond to upflows in the full image. At scale 7, there is a good correlation between the giant
cells signal at low latitudes detected via the radial velocity field and the 
kinetic energy fluxes, i.e. that the negative kinetic energy flux is located 
in the downflows of the giant cells. The amplitude of the signal at scale 7
corresponds to 2\% of the signal in the full image.       

The specific angular momentum flux due to the $\theta-\phi$ component of the
Reynolds stress is involved in the latitudinal tranport of angular momentum (see
Brun \& Toomre 2002 for details). 
Whereas this transport is complex at small scales, the scale 7 shows a dominant
transport equatorwards at low-latitudes but also an important transport
polewards particularly in the South hemisphere at mid-latitudes. The transport
by scale 7 represents 5\% of the total signal.  

Finally, the third column of Figure \ref{transport_sc7} shows the reconstruction of the variables without the giant cells' signal (i.e. we have substracted the signal corresponding to Scale 7). This points out the importance of large scale structures to transport enthalpy outwards particularly since the snapshot without this signal leads to a clear lack of transport of enthalpy outwards. The difference is less obvious for the other quantities, but as we have already seen these large scales contribute for few percents to the overall transport.    

Hence we find that these large scale convection structures contribute to the
global distribution of heat, energy and angular momentum in a systematic way,
imposing a large scale trend to the smaller scale flow.

\section{Discussion}

In this paper, we have studied the general properties of multi-scale convection  
within deep stellar convective zones. First, the convection spatial structure
 near the stellar surface is well constrained by the local density scale
 height which is a consequence of the conservation of mass in a statified 
 media. However, this convection organization at small scales depends also on 
 the thickness of the shell to some extent as shown in polar regions for the 
 case $\Delta \rho=13$ that possesses the thinnest shell. Still, the global 
statification has negligible influence on this surface convective patterns 
whenever the thickness of the convective
 zone is sufficient ($> 60$ Mm in our study done with $H_\rho=10$ Mm at the surface).

By using a wavelet analysis, we have stressed that there is a clear signal at
 large scale ($> 200$ Mm) with a typical radial velocity around 2 m/s which is 
 much more elongated in latitude when the density
contrast increases (see Figure \ref{lat_tau}). This ``latitudinal extent'' estimate
 of the giant cells can be a means to determine the depth of the convection zone via correlation 
tracking. This could also be an interesting 
observational method to try preferentially to detect these giant cells at large latitudes.  
These large structures have a proper tracking  rate really greater than the local 
differential rotation and the discrepancy increases with the density contrast. 
It is important to understand that the giant cells detected here, even they have
several characteristics similar to the linear convection equatorial modes, 
are clearly different in nature.
Gilman (1975) shows that these equatorial modes are the most unstable in the
linear regime for high m modes ($>$20) when the Taylor number increases ($Ta>10^6$) for a quite thin 
convective zone ($R/ L =5$) but with a decrease in the most unstable m
mode for thicker convective zones as studied in Glatzmaier \& Gilman (1981). 
However, the longitudinal extent of our giant cells is really similar for all 
models despite the great range of Taylor numbers covered in these simulations.
Gilman (1975) also shows that these equatorial modes are well confined at low latitudes 
($<35$\degree) which is not the case for our high density contrast models.            
These giant cells are also different from the basic convective rolls since 
they are not extended over the full depth of the convective zone particularly 
up to mid latitudes for the much realistic high density contrast cases with a 
really turbulent flow as shown in Figure \ref{conv_T_patterns}. 

We thus believe that the signal seen at large scale is genuine and instructive to reveal observationally stellar giant cells. As in Lisle et al. (2004), we find that giant cells are predominantly oriented along the North-South direction. 
More recently, Hanasoge et al. (2010) have looked for the existence of giant cells using a 
time distance analysis of numerical simulations of the solar convection with the ASH code. After a calibration of MDI solar surface observations from these simulations, they obtain an 
upper limit for the longitudinal velocities around 
10 m/s for $l < 10$ corresponding to the range of spatial scales identified in 
our simulations. In Figure 12, we have 
found a maximal radial velocity for the giant cell signal of 12.9 m/s. A wavelet analysis on $v_{\phi}$ and $v_\theta$ gives 
respectively maximal velocities of 11 m/s and 7 m/s consistent with the error bars of Hanasoge et al. (2010) as reported on their Figure 5. 
However, the direct comparison of their results with ours is not so straigthforward since we model a young star with a density stratification and velocity profiles that differ from the Sun. On the other hand, it is likely that simulations done with the ASH code possess a different energy spectra in the large scales with respect to reality since the range of scales available spans only of the order of 3 decades. This could lead to an over estimation of the amplitude of the giant cell signal even if our simulations clearly show evidence of their existence.

The lifetime for these giant cells is much greater than the stellar rotation rate.
These giant cells take part in a systematic way to the transport of heat, energy
and angular momentum although their contribution do not exceed a few percent of
the transport carried out by the whole range of flows. There are actually two types of giant cells in the deepest convective zones corresponding to
$\Delta\rho=100,272$ models and separated roughly by the tangent cylinder. 
The giant cells localized at low
latitudes do not span the entire convection zone due to the strong low latitude
radial shear in these models and we find mainly two of them in the 
radial direction (i.e. around 110 Mm in thickness)
contrary to the giant cells at higher latitudes which 
are aligned with the rotation axis and coherent trough the whole thickness 
of the convective zone. This two different modes of convection, with very different aspect ratio for the convective cells, can be understood by the 
different orientation of gravity and Coriolis force  with respect to the flow
combined with the increase influence of rotation in these deep convective zones.
We have thus found that giant cells are likely to be present in deep convective
zones with aspect ratio of 4-5 and that they are more easily detected at
mid-latitudes. 

The presence of giant cell flows in our purely hydrodynamic simulations suggest
that some large scale organisation of the stellar turbulent convection
is possible. In the presence of magnetic field these large scale ordering of
 the flow may contribute to the global scale dynamo observe in most solar-type 
 stars. We intend to study in the near future (Bessolaz \& Brun, in preparation)
the subtle interplay between multi-scale convection, rotation and magnetic field
in young solar type stars that possess deep convection zone and possibly giant
cells.



\acknowledgments{}
The simulations were carried out on the national super computing centers
such as the CCRT at CEA, JADE at CINES and also Vargas at IDRIS via GENCI
project 1623. 
This work is part of the STARS$^2$ project (http://www.stars2.eu) funded by a
grant \#207430 of the European Research Council awarded to A.S.B. 
We would like to thank J. Toomre and  M.S. Miesh for useful discussions, and J.L. Starck and 
S. Pires for their advice on our wavelet analysis.


\end{document}